\documentclass[twocolumn]{jpsj3}
\usepackage{txfonts}
\newcommand{\lsim}
 {\ \raise.35ex\hbox{$<$}\kern-0.75em\lower.5ex\hbox{$\sim$}\ }
\newcommand{\gsim}
 {\ \raise.35ex\hbox{$>$}\kern-0.75em\lower.5ex\hbox{$\sim$}\ }
\def\journal #1#2#3#4{#1 {\bf #2}, #3 (#4)}

\def\APNY{Ann.\ Phys.\ (New York)}

\def\IJQC{Int.\ J.\ Quantum\ Chem.}

\def\JPCM{J.\ Phys.:~Condens.~Matter}

\def\JPSJ{J.\ Phys.\ Soc.\ Jpn.}
\def\JPSCP{JPS Conf.\ Proc.}

\def\NJP{New J.~Phys.}
\def\NC{Nat.~Commun.}
\def\NP{Nat.~Phys.}

\def\PP{Phys.~Proc.}

\def\PR{Phys.\ Rev.}

\def\PRB{Phys.\ Rev.\ B}

\def\PRL{Phys.\ Rev.\ Lett.}

\def\SST{Supercond.~Sci.~Technol.}

\def\EPJB{Eur.\ Phys.\ J.\ B}
\def\EPL{Europhys.\ Lett.}
\def\RPP{Rep.~Prog.~Phys.}
\usepackage{color}
\definecolor{Green}{rgb}{0,0.7,0}

\title{Staggered Flux State in Two-Dimensional Hubbard Models} 
\author{
Hisatoshi Yokoyama\thanks{E-mail, yoko@cmpt.phys.tohoku.ac.jp},$^{1}$ 
Shun Tamura$^{2}$, and 
Masao Ogata$^{3}$ 
} 

\inst{
$^{1}$Department of Physics, Tohoku University, Sendai 980-8578, Japan \\ 
$^{2}$Department of Applied Physics, Nagoya University, Nagoya 464-8603, 
Japan \\
$^{3}$Department of Physics, University of Tokyo, Bunkyo-ku, Tokyo 113-0033.
Japan 
}

\abst{ 
The stability and other properties of a staggered flux (SF) state or a 
correlated $d$-density wave state are studied for the Hubbard ($t$-$t'$-$U$) 
model on extended square lattices, as a low-lying state that competes with 
the $d_{x^2-y^2}$-wave superconductivity ($d$-SC) and possibly causes 
the pseudogap phenomena in underdoped high-$T_{\rm c}$ cuprates and organic 
$\kappa$-BEDT-TTF salts. 
In calculations, a variational Monte Carlo method is used. 
In the trial wave function, a configuration-dependent phase factor, which is 
vital to treat a current-carrying state for a large $U/t$, is introduced 
in addition to ordinary correlation factors. 
Varying $U/t$, $t'/t$, and the doping rate ($\delta$) systematically, we show 
that the SF state becomes more stable than the normal state (projected 
Fermi sea) for a strongly correlated ($U/t\gtrsim 5$) and underdoped 
($\delta\lesssim 0.16$) area. 
The decrease in energy is sizable, particularly in the area where Mott 
physics 
prevails and the circular current (order parameter) is strongly suppressed. 
These features are consistent with those for the $t$-$J$ model. 
The effect of the frustration $t'/t$ plays a crucial role in preserving 
charge 
homogeneity and appropriately describing the behavior of hole- and 
electron-doped cuprates and $\kappa$-BEDT-TTF salts.
We argue that the SF state does not coexist with $d$-SC and is not a `normal 
state' from which $d$-SC arises. 
We also show that a spin current (flux or nematic) state is never stabilized 
in the same regime.
}

\begin{document}
\maketitle

\section{Introduction\label{sec:intro}}
%
Superconductivity (SC) in underdoped high-$T_{\rm c}$ cuprates should be 
understood through the relationship to the pseudogap phase observed 
for $T_{\rm c}<T<T^*$, where $T_{\rm c}$ [$T^*$] is the superconducting 
(SC) transition [pseudogap] temperature.\cite{Yoshida12,Uchida12} 
Because the pseudogap phase appears in the proximity of half filling, 
it is probably related to Mott insulators\cite{OF,AtoZ} 
(precisely, charge-transfer insulators\cite{Zaanen}).
Experimentally, the pseudogap phase presents various features distinct from 
an ordinary Fermi liquid.\cite{OF,AtoZ}
(1) A large gap different from the $d_{x^2-y^2}$-wave SC ($d$-SC) gap opens 
in the spin degree of freedom near the momenta of $(\pi,0)$ and $(0,\pi)$. 
(2) However, the material is conductive and does not have a charge gap. 
(3) Fragmentary Fermi surfaces, i.e., Fermi arcs\cite{Fermi_arc1,Fermi_arc2} 
or hole pockets,\cite{hole_pocket1,hole_pocket2} appear in the zone-diagonal 
direction near ($\pi/2,\pi/2$). 
\par

The origin of the pseudogap has often been studied as a linkage to $d$-SC, 
although it will not be related to SC 
fluctuation.\cite{note-PG,Nernst1,Nernst2,Kondo,Dubroka,Uykur} 
On the other hand, recent experimental studies argued that the pseudogap 
phase is accompanied by some symmetry-breaking phase transitions at 
$T^*$.\cite{Uchida12} 
(1) Time-reversal symmetry breaking\cite{T*,mag-order1,mag-order2,mag-order3} 
is claimed from polarized neutron scattering signals at the momentum $(0,0)$ 
as well as from the appearance of the Kerr effect.
(2) Rotational symmetry breaking (or nematic order) similar to the stripe 
phase is observed, and the oxygen sites between copper atoms are 
involved.\cite{Uchida12,Yamada12}
(3) Charge orders or charge density waves are observed in resonant X-ray 
scattering experiments.\cite{Keimer1,Keimer2,Keimer3}
(4) ($\pi,\pi$)-folded (shadow) bands appear in ARPES spectra, and so 
forth.\cite{SB1,SB2,SB3,SB4} 
Note, however, that the thermodynamic properties such as specific heat and 
spin susceptibility have not provided any evidence of the phase transition.
It is also important to study whether a pseudogap and other orders coexist 
or are mutually exclusive.\cite{Uykur,coexist1,coexist3,Yoshida} 
\par

In this context, we study a symmetry-breaking state---a staggered flux (SF) 
state (sometimes called a $d$-density wave state)---as a possible pseudogap 
state for the Hubbard model. 
We should understand such a state in the context of a doped Mott 
insulator.\cite{OF,AtoZ} 
To respect the strong correlation, we use a variational Monte Carlo (VMC) 
method,\cite{t-J-d} which deals with the local correlation factors exactly 
and has yielded consistent results for many aspects of 
cuprates.\cite{YO,YTOT,YOT,YOTKT,Paramekanti,Tahara} 
If the pseudogap phenomena are generated by a symmetry-breaking state, 
it should be more stable than the (symmetry-preserved) ordinary normal 
state. 
Also, when a predominant antiferromagnetic (AF) or $d$-SC state is suppressed 
for some reason, features of the symmetry-breaking state will manifest 
themselves. 
Note that a recent VMC calculation with a band-renormalization effect showed 
that an AF state is considerably stabilized compared with the $d$-SC state 
in a wide region of the Hubbard model.\cite{BR}
\par

Since the early years of research on cuprate SCs, the SF state has been 
studied by many groups from both weak- and strong-correlation sides. 
In the early studies,\cite{Affleck,Schulz,Lederer,Zhang,Liang,TKLee,OgataRice} 
the main aim was to check whether the SF state becomes 
the ground state, but it was shown mainly using the $t$-$J$ model that 
the SF state yields to other ordered states (AF and $d$-SC) for any 
relevant parameters. 
Later, the SF state was mainly studied as a candidate for a normal 
state that causes the pseudogap phenomena and underlies $d$-SC in underdoped 
cuprates.\cite{LW01,Kishine,Wang,Tsuchiura1,Tsuchiura2} 
\par

At half filling, for the Heisenberg model, owing to the SU(2) symmetry, 
the SF state is equivalent to the $d$-wave BCS state,\cite{Affleck-SU2,ZGRS} 
which has a very low energy\cite{t-J-d,YO} comparable to that of the AF 
ground state.\cite{t-J-AF,AF-QMC,AF-QMC2} 
The $t$-$J$ model with finite doping was studied using U(1) and SU(2) 
slave-boson mean-field theories\cite{Ubbens,Wen-Lee,Hamada} and 
a perturbation theory of Hubbard $X$ operators,\cite{Cappelluti} which 
revealed that the SF phase exists in phase diagrams but is restricted 
to very small doping regions.\cite{Hamada}
As a more reliable treatment, VMC calculations\cite{Liang,TKLee,YO,Ivanov} 
showed that the SF state has lower energy than the projected Fermi sea, 
although the $d$-SC state has even lower energy. 
It was pointed out that the $\delta$ dependence of the SC condensation 
energy using 
the SF state as a normal state becomes domelike\cite{Ivanov} but that 
the SF state tends to be unstable toward phase separation.\cite{Ivanov-PS} 
These VMC results claim that the strongly correlated Hubbard model should 
have the same features. 
\par

For the Hubbard model, SF states have been studied using a phenomenological 
theory,\cite{Chakravarty} 
mean-field theories,\cite{Nersesyan,Ozaki,Normand} and 
more refined renormalization group methods\cite{RG-Binz,Honerkamp} 
from the weak-correlation side. 
These studies obtained various knowledge of the SF state, but it is still 
unclear whether or not the SF state is stabilized in the weakly as well as 
strongly correlated regions. 
On the other hand, a Gutzwiller approximation study\cite{Normand} claimed 
that the SF state is not realized in the Hubbard model. 
A study using a Hubbard operator approach\cite{Phillips} showed the absence 
of SF order for a large $U/t$ ($=8$) unless an attractive intersite 
interaction is introduced. 
A study using a dynamical cluster approximation for a $2\times 2$ 
cluster\cite{Jarrel} argued that the circular-current susceptibility increases 
in the pseudogap-temperature regime but does not diverge, and there is no 
qualitative change as $U/t$ and $t'/t$ are varied. 
A study using a variational cluster approach\cite{Arrigoni} concluded that 
the SF phase is not stabilized with respect to the ordinary normal state for 
a strongly correlated region ($U/t\gtrsim 4$). 
An extended dynamical-mean-field approximation showed that although the SF 
susceptibility is enhanced, it is dominated by $d$-SC and an inhomogeneous 
phase for $t'/t=0$.\cite{Otsuki} 
Thus, it is still unclear whether the results in the Hubbard model are 
consistent with those in the $t$-$J$ model.
\par

The purpose of this paper is to show that the SF state becomes considerably 
stable with respect to the projected Fermi sea (an ordinary normal state) 
in the underdoped regime for large values of $U/t$ and $t'/t\sim -0.3$ 
in the Hubbard ($t$-$t'$-$U$) model, and to clarify various properties of 
this state on the basis of systematic VMC calculations. 
It is essential to introduce a configuration-dependent phase factor to 
treat a current-carrying state such as the SF state in the regime of 
Mott physics.\cite{Drude} 
Without it, the SF state is never stabilized in models permitting double 
occupation such as the Hubbard model. 
We change the model parameters $U/t$, $t'/t$, and the doping rate $\delta$ 
($=1-N/N_{\rm s}$) in a wide range, with $N$ and $N_{\rm s}$ being the 
numbers of electrons and sites, respectively. 
Additionally, we study the spin-current flux phase (sometimes called the 
spin-nematic phase) using the same method. 
\par

Besides cuprates, we consider a model for layered organic conductors, 
$\kappa$-(BEDT-TTF)$_2$X, [henceforth, abbreviated as $\kappa$-(ET)$_2$X] 
with X being a univalent anion.\cite{Organic12,O3,O4} 
In these compounds, SC arises for $T_{\rm c}\lesssim 12$~K, and a pseudogap 
behavior similar to that of cuprates has been observed.
Therefore, we need to check whether its origin is identical to that of 
cuprates. 
Various low-energy properties of $\kappa$-(ET)$_2$X are considered to be 
described by the Hubbard model\cite{O6} on an anisotropic two-dimensional 
triangular lattice. 
The value of $U/t$ can be controlled by applying pressure. 
$U$ is estimated as $U\sim W$--$2W$ with $W$ being the 
band width.\cite{Organic12} 
The degree of frustration $t'/t$ can be varied by substituting X or applying 
uniaxial pressure. 
$t'/t$ is estimated by ab initio calculations as $0.4$--$0.7$ for 
weakly frustrated compounds and $\sim 0.8$ for the highly frustrated 
compound $\kappa$-(ET)$_2$Cu$_2$(CN)$_3$.\cite{Organic12,O7} 
Among the former compounds, deuterated 
$\kappa$-(ET)$_2$Cu[N(CN)$_2$]Br ($t'/t\sim 0.4$) 
under applied pressure has been shown to exhibit pseudogap behavior such as 
a steep 
decrease in the NMR spin-lattice relaxation time ($1/T_1T$) in the metallic 
phase ($T>T_{\rm c}$). 
On the other hand, $\kappa$-(ET)$_2$Cu$_2$(CN)$_3$, which has a spin liquid 
state in the insulating phase under ambient pressure, exhibits 
the Korringa relation ($1/T_1 T=$ const.) in the metallic phase under 
pressure, 
namely, pseudogap behavior is absent.\cite{O8} 
Furthermore, similar pseudogap behavior was observed in a hole-doped 
$\kappa$-ET salt [$\kappa$-(ET)$_4$Hg$_{2.89}$Br$_8$],\cite{O9} in which the 
doping rate is $0.11$ and $t'/t\sim 0.8$. 
With these experimental results in mind, we study the SF state 
on an anisotropic triangular lattice in the framework applied to the 
frustrated square lattice for cuprates. 
\par

This paper is organized as follows. 
In Sect.~\ref{sec:formalism}, we introduce the model and method used in this 
paper.
In Sects.~\ref{sec:square} and \ref{sec:doped cases}, we discuss the results 
mainly for the simple square lattice ($t'=0$) at half filling and in doped 
cases, respectively, to grasp the common properties of the SF state.
Section \ref{sec:t'} is assigned to the effect of the diagonal 
hopping term $t'$ for the frustrated square lattice and anisotropic 
triangular lattice. 
In Sect.~\ref{sec:discussions}, we discuss the results. 
In Sect.~\ref{sec:conclusions}, we recapitulate this work.
In Appendix\ref{sec:nonintSF}, we summarize the fundamental features of the 
noninteracting SF state.
In Appendix\ref{sec:t-J}, we briefly review the stability of the SF phase 
for $t$-$J$-type models with new accurate data.
In Appendix\ref{sec:SSC}, we show that the spin current (flux) state is 
unstable 
toward the projected Fermi sea in $t$-$J$-type models for any $J$ ($>0$) 
and $\delta$. 
Preliminary results on the effect of $t'$ terms have been reported in two 
preceding publications.\cite{SCES,ISS}
\par

\bigskip
\section{Model and Wave Functions\label{sec:formalism}}
In Sects.~\ref{sec:model} and \ref{sec:wf}, we explain the model and 
variational wave functions used in this paper, respectively. 
In Sect.~\ref{sec:phase}, we introduce a phase factor essential for 
treating a current-carrying state in a strongly correlated regime. 
In Sect.~\ref{sec:VMC}, we describe the numerical settings of our VMC 
calculations. 
\par
\subsection{Hubbard model\label{sec:model}}

\begin{figure}[htb]
\begin{center}
\vskip -0mm
\includegraphics[width=8cm,clip]{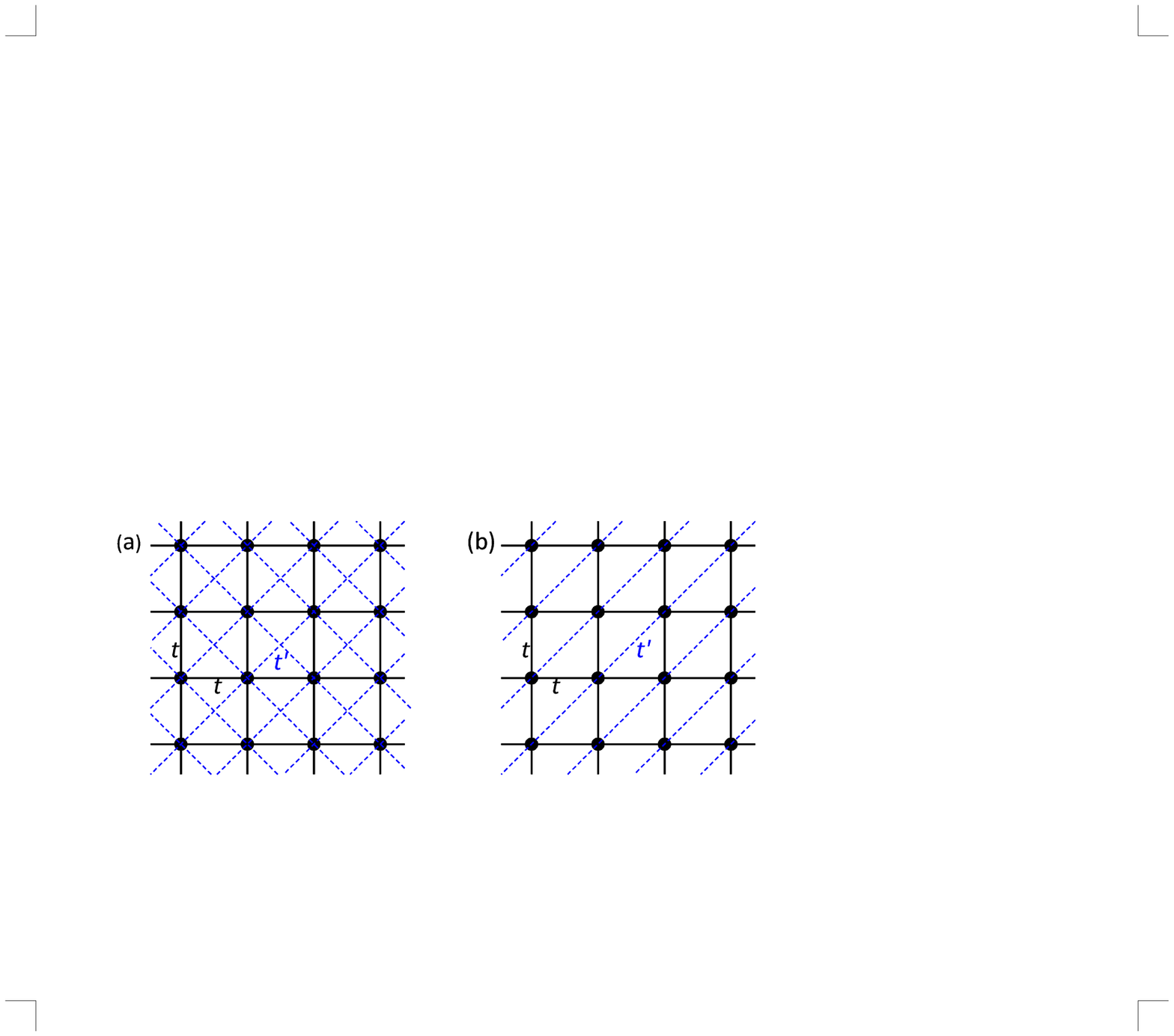}
\end{center} 
\vskip -3mm 
\caption{(Color online) 
Lattice connectivity or hopping paths in extended square lattice 
addressed in this study. 
(a) Frustrated square lattice for cuprate SCs. 
(b) Anisotropic triangular lattice for organic $\kappa$-ET 
salt SCs.
At lattice points (solid circles), onsite repulsion $U$ acts. 
}
\vskip -5mm 
\label{fig:model} 
\end{figure}

As models of cuprates and $\kappa$-ET organic conductors, we consider the 
following Hubbard model ($U\ge 0$) on extended square lattices 
(Fig.~\ref{fig:model}):
\begin{eqnarray}
{\cal H}&=&{\cal H}_{\rm kin}+{\cal H}_U \nonumber \\
        &=&-\sum_{(i,j),\sigma}t_{ij}\left( 
        c^\dagger_{i\sigma}c_{j\sigma} + \mbox{H.c.}\right)
         +U\sum_j n_{j\uparrow}n_{j\downarrow}, 
\label{eq:Hamil}
\end{eqnarray} 
where 
$n_{j\sigma}=c^\dag_{j\sigma}c_{j\sigma}$ 
and $(i,j)$ indicates the sum of pairs on sites $i$ and $j$. 
In this work, the hopping integral $t_{ij}$ is $t$ for nearest 
neighbors ($\ge 0$), $t'$ for diagonal neighbors, and $0$ otherwise 
(${\cal H}_{\rm kin}={\cal H}_t+{\cal H}_{t'}$) for the two lattices shown 
in Fig.~\ref{fig:model}. 
The bare energy dispersions are
\begin{equation}
\epsilon_{\bf k}=\left\{
\begin{array}{ll}
-2t\left(\cos k_x+\cos k_y\right)
                      -4t'\cos k_x\cos k_y, & \mbox{(a)} \\
-2t\left(\cos k_x+\cos k_y\right)
                      -2t'\cos\left(k_x+k_y\right). & \mbox{(b)}
\end{array}
\right.
\label{eq:disp-t'}
\end{equation}
In the following, we use $t$ and the lattice spacing as the units of energy 
and length, respectively.
\par

We refer to the former (latter) lattice as a frustrated square 
(anisotropic triangular) lattice for convenience. 
The effective values of $t'/t$ are considered to be $-0.4$--$-0.1$ 
($\sim 0.3$) 
in hole-doped (electron-doped) cuprates.\cite{OF,Tohyama}
For the organic compounds, $t'/t$ is $0.4$--$0.8$. 
Hubbard models have been extensively studied, and we have shown that 
a first-order Mott transition occurs at $U=U_{\rm c}\sim W$ at half filling 
for nonmagnetic cases and that a doped Mott insulator is realized 
for $U\gtrsim W$ near half filling.\cite{YTOT,YOT,YOTKT}
In this paper, we show that similar Mott physics appears in the SF states.
It has been shown that, in a wide range of the parameter space of concern, 
a $d$-SC state becomes stable compared with the projected Fermi sea (ordinary 
normal state).\cite{YOTKT} 
\par

\subsection{Trial wave functions\label{sec:wf}}
We follow many-body variation theory using Jastrow-type trial wave 
functions: $\Psi={\cal P}\Phi$, where ${\cal P}$ indicates a product of 
many-body projection (Jastrow) factors discussed later and $\Phi$ is 
a mean-field-type one-body wave function. 
\par 

As a normal (paramagnetic) and reference state, we use a projected Fermi sea, 
$\Psi_{\rm N}={\cal P}\Phi_{\rm N}$, with 
\begin{equation}
\Phi_{\rm N}(t_1/t)
=\prod_{{\bf k}\in {\bf k}_{\rm F}(t_1/t),~\sigma} 
c_{{\bf k}\sigma}^\dag|0\rangle, 
\end{equation}
where $t_1/t$ is a band-adjusting variational parameter independent of 
$t'/t$ in ${\cal H}$ and ${\bf k}_{\rm F}(t_1/t)$ denotes a Fermi surface 
obtained by replacing $\epsilon_{\bf k}(t')$ in Eq.~(\ref{eq:disp-t'}) 
with $\epsilon_{\bf k}(t_1)$. 
It was shown for $\Psi_{\rm N}$\cite{BR} that the band-renormalization 
effect through $t_1/t$ owing to the electron correlation (${\cal P}$) 
is sizable for $U\gtrsim U_{\rm c}$, a finite $t'/t$, and $\delta\sim 0$.  
\par 

\begin{figure}[htb]
\begin{center}
\vskip -0mm 
\includegraphics[width=8cm,clip]{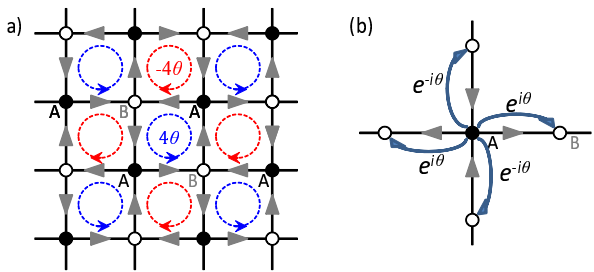} 
\end{center} 
\vskip -0mm 
\caption{(Color online) 
(a) Schematic figure of local current in staggered flux state. 
Arrows denote the directions of currents. 
(b) Peierls phase factors attached to hopping terms in ${\cal H}^{\rm SF}$
(a case of sublattice A); 
the sign of the phase depends on the relative directions of the hopping and 
current. 
}
\vskip -3mm 
\label{fig:SF-fig} 
\end{figure}
%
As a candidate for the pseudogap state, we study a correlated SF state, 
$\Psi_{\rm SF}={\cal P}\Phi_{\rm SF}$. 
Here, $\Phi_{\rm SF}$ is the one-body SF state, namely, the ground state of 
the noninteracting SF Hamiltonian ${\cal H}^{\rm SF}$ [shown in 
Eq.~(\ref{eq:HamilSF})], given as 
\begin{equation}
\Phi_{\rm SF}=\prod_{{\bf k}\in{\bf k}_{\rm F},\sigma}
\frac{1}{\sqrt{N_{\rm s}}}
\left[\sum_{i\in{\rm A}}
\Gamma_{\theta,{\bf k}}
e^{i{\bf k}\cdot{\bf r}_i}c^\dag_{{\rm A}i\sigma}
     +\sum_{i\in{\rm B}}
e^{i{\bf k}\cdot{\bf r}_i}c^\dag_{{\rm B}i\sigma}
\right]|0\rangle,
\label{eq:SF}
\end{equation}
with
\begin{equation}
\Gamma_{\theta,{\bf k}}=
\frac{e^{i\theta}\cos{k_x}+e^{-i\theta}\cos{k_y}}
{{\cal S}_{\theta,{\bf k}}}, 
\end{equation}
\begin{equation}
{\cal S}_{\theta,{\bf k}}
=\sqrt{\cos^2{k_x}+2\cos{2\theta}\cos{k_x}\cos{k_y}+\cos^2{k_y}}. 
\label{eq:sk}
\end{equation}
In ${\cal H}^{\rm SF}$, a Peierls phase $\theta$ is added to the hopping 
integrals so that circular current flows in alternate directions in
each plaquette as shown in Fig.~\ref{fig:SF-fig}(a). 
In the present variational theory, $\theta$ is a variational parameter to be 
optimized together with the other parameters.
Because ${\cal H}^{\rm SF}$ breaks time-reversal and lattice-translational 
symmetries, $\Phi_{\rm SF}$ does not have these symmetries. 
The lower-band energy dispersion of ${\cal H}^{\rm SF}$ is given as
\begin{equation}
E^{\rm SF}_-({\bf k})=-2t{\cal S}_{\theta,{\bf k}}=
-\sqrt{\frac{1+\cos2\theta}{2}}\sqrt{\varepsilon_{\bf k}^2
+\Delta_{\bf k}^2}, 
\label{eq:SF-disp}
\end{equation}
with 
\begin{eqnarray}
\varepsilon_{\bf k}&=&-2t(\cos k_x+\cos k_y), 
\label{eq:dispersion} \\
\Delta_{\bf k}&=&\Delta_\theta(\cos k_x-\cos k_y), 
\label{eq:SFgap}
\end{eqnarray}
and
$\Delta_\theta=2t\sqrt{(1-\cos2\theta)/(1+\cos 2\theta)}$. 
Equation (\ref{eq:SF-disp}) is similar to the quasiparticle dispersion of the 
$d$-wave BCS wave function. 
Note that some important features of the bare $\Phi_{\rm SF}$ 
(summarized in Appendix\ref{sec:nonintSF}) survive in 
$\Psi_{\rm SF}={\cal P}\Phi_{\rm SF}$. 
We do not consider Band-renormalization effects on $\Psi_{\rm SF}$ because 
those due to diagonal currents or hopping are known to raise the variational 
energy for typical cases.\cite{SO} 
\par

The correlation factor ${\cal P}$ is defined as 
\begin{equation}
{\cal P}=
{\cal P}_\phi\ 
{\cal P}_{\rm Q}\ 
{\cal P}_{\rm G}.
\end{equation}
Here, ${\cal P}_{\rm G}$ is the fundamental onsite (Gutzwiller) projection 
${\cal P}_{\rm G}=
\prod_j[1-(1-g)n_{j\uparrow}n_{j\downarrow}]$\cite{Gutz} 
and ${\cal P}_Q$ is an asymmetric projection between a nearest-neighbor 
doubly occupied site (doublon) and an empty site 
(holon),\cite{D-H1,D-H2,YOTKT}
\begin{equation}
{\cal P}_Q=\prod_j\left[1
-\zeta_{\rm d}d_j\prod_\tau\left(1-h_{j+\tau}\right)
-\zeta_{\rm h}h_j\prod_\tau\left(1-d_{j+\tau}\right)
\right],
\label{eq:D-H}
\end{equation}
where $d_j=n_{j\uparrow}n_{j\downarrow}$, 
$h_j=(1-n_{j\uparrow})(1-n_{j\downarrow})$, and $\tau$ runs over the 
nearest-neighbor sites of site $j$. 
$g$, $\zeta_{\rm d}$, and $\zeta_{\rm h}$ are variational parameters. 
As shown before,\cite{YOT,YTOT} the doublon-holon (D-H) binding effect is 
crucial for appropriately treating Mott physics. 
At half filling, $\zeta_{\rm d}$ and $\zeta_{\rm h}$ become identical 
because of the D-H symmetry. 
In addition to ${\cal P}_{\rm G}$ and ${\cal P}_Q$, it is vital 
for the SF state to introduce a phase-adjusting factor ${\cal P}_\phi$, 
which we will explain in the next subsection. 
\par

\subsection{Configuration-dependent phase factor\label{sec:phase}}
A current-carrying state is essentially complex, because the current is 
proportional to $|\Psi|^2\nabla\Theta$, if we represent the state as 
$\Psi({\bf r})=|\Psi({\bf r})|e^{i\Theta({\bf r})}$. 
It is natural that when electron correlation is introduced, the phase part 
$\Theta({\bf r})$ varies accordingly. 
However, the conventional correlation factors, ${\cal P}_{\rm G}$ and 
${\cal P}_Q$, are real and do not modify the phase in $\Phi_{\rm SF}$. 
Therefore, we need to introduce an appropriate phase-adjusting factor into 
the trial wave function. 
Such a phase factor was recently introduced for calculating the Drude and 
SC weights in strongly correlated regimes; \cite{Drude} thereby, 
a long-standing problem proposed by Millis and 
Coppersmith\cite{Millis}---D-H binding wave functions yield finite (namely 
incorrect) Drude weights even in the Mott insulating regime--- was solved. 
\par

\begin{figure}[htb]
\begin{center}
\vskip -2mm
\includegraphics[width=8.5cm,clip]{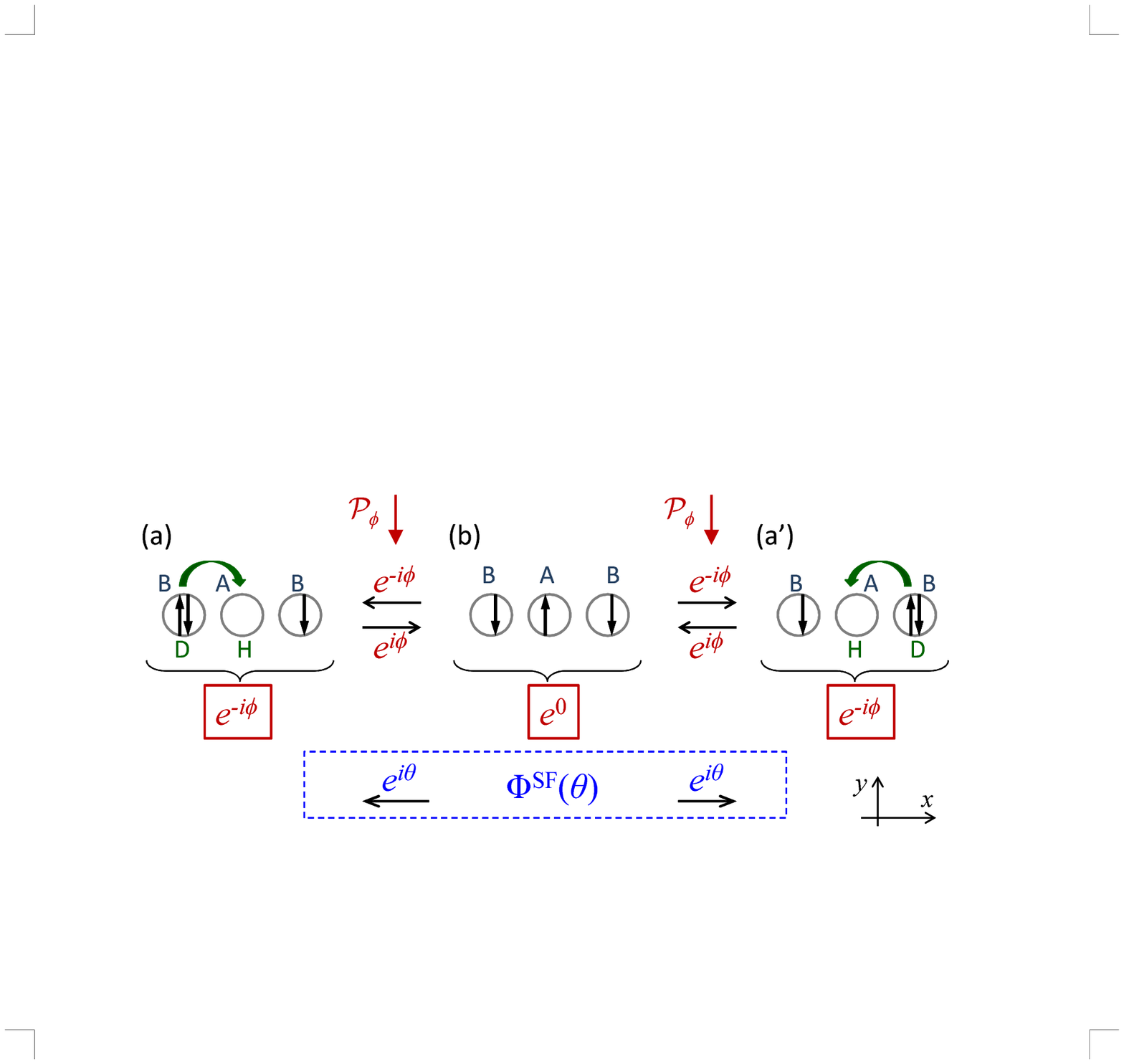} 
\end{center} 
\vskip -2mm 
\caption{(Color online) 
Illustration for assigning configuration-dependent phase $\phi$ in 
${\cal P}_\phi$. 
Here, we assume that an electron hops in the $x$ direction. 
For the hopping in the $y$ direction, the signs of $\theta$ and $\phi$ have 
to be reversed. 
The (Peierls) phase factor assigned by $\Phi_{\rm SF}$ in hopping 
[Fig.~\ref{fig:SF-fig}(b)] is shown in the blue dashed box. 
The values in the red boxes are the phase factors in ${\cal P}_\phi$ 
corresponding to the three-site parts shown. 
The ratio ($e^{\pm\phi}$) indicated by red arrows is produced by 
${\cal P}_\phi$ in hopping.
}
\vskip -3mm 
\label{fig:SF-scheme} 
\end{figure}
%
We show that this type of phase factor also plays a vital role in the 
correlated SF state. 
In $\Phi_{\rm SF}$, a phase $\theta$ or $-\theta$ is added when an 
electron hops to a nearest-neighbor site depending on the direction and 
position (sublattice), as shown in Fig.~\ref{fig:SF-fig}(b).
In the noninteracting case, such hopping occurs equally in all directions. 
On the other hand, in the strongly correlated regime, the probability of 
hopping depends on the surrounding configuration 
(see Fig.~\ref{fig:SF-scheme}).
For example, when a D-H pair is created [configurations (a) and (a$'$)], 
the next hopping occurs probably in the direction in which the singly occupied 
configuration is recovered [configuration (b)]. 
This hopping process does not contribute to a global current in the Mott 
regime ($U\gsim U_{\rm c}$).\cite{YOTKT} 
According to a previous study,\cite{Drude} to reduce the energy, it is 
important to cancel the phase attached in this type of hopping ($\pm\theta$) 
by introducing a phase parameter. 
\par

This hopping process can be specified by its local configurations and, 
correspondingly, we can attach a phase-adjusting variational factor to the 
trial wave function. 
To be more specific, ${\cal P}_\phi$ gives $e^{-i\phi}$ as shown by the 
solid boxes (red) in Fig.~\ref{fig:SF-scheme}, with $\phi$ being a 
variational parameter. 
This phase assignment can be written as 
\begin{eqnarray}
{\cal P}_\phi&=&\exp\Bigg[
i\phi\sum_{\lambda=1}^2(-1)^{\lambda+1}\sum_j d_{\lambda,j} 
\nonumber \\ 
&&\times\left(h_{\lambda,j+{\bf x}}+h_{\lambda,j-{\bf x}}
             -h_{\lambda,j+{\bf y}}-h_{\lambda,j-{\bf y}}\right)
\Bigg],
\qquad
\label{eq:phase}
\end{eqnarray}
where ${\bf x}$ and ${\bf y}$ indicate the lattice vectors in the $x$ and $y$ 
directions, respectively,  $\lambda=1$ ($\lambda=2$) indicates sublattice 
A (B), and $j$ runs over all the lattice points in sublattice $\lambda$. 
By ${\cal P}_\phi$, a phase factor $e^{\pm i\phi}$ is assigned to a D-H 
creation or annihilation process, in which $e^{\mp i\theta}$ is yielded 
by $\Phi_{\rm SF}$ as shown in the dashed box (blue) in 
Fig.~\ref{fig:SF-scheme}. 
Therefore, when the relation $\phi=\theta$ holds, the total phase shift in 
a D-H process vanishes.\cite{note-phase}
This phase cancelation is acceptable since a phase shift does not appear 
in an exchange process in the $t$-$J$ model.
On the other hand, the phase is not canceled in the hopping processes 
unrelated to doublons (or of isolated holons).\cite{note-free-holon} 
\par

The configuration-dependent phase factor ${\cal P}_\phi$ is conceptually 
distinct from position-dependent phase factors used in various 
contexts.\cite{Liang,Weber,Paramekanti} 
Note that, without ${\cal P}_\phi$, the energy of the SF state is never 
reduced from that of $\Psi_{\rm N}$ for any model parameters, but 
$\Psi_{\rm SF}$ with ${\cal P}_\phi$ has lower energy than $\Psi_{\rm N}$, 
as we will see below.\cite{note-real} 
Incidentally, this type of phase factor was also recently shown to be crucial 
for SF states in a Bose Hubbard model\cite{Toga} and a $d$-$p$ 
model.\cite{Tamura-dp}
In the regime of Mott physics, the D-H binding affects not only the real part 
but also the phase in the wave function. 
\par

\subsection{Variational Monte Carlo calculations\label{sec:VMC}}
To estimate variational expectation values, we adopt a plain VMC 
method.\cite{VMC1,VMC2,VMC3,VMC4} 
In this study, we repeat linear optimization of each variational parameter 
with the other ones being fixed, typically for four rounds of iteration. 
The linear optimization is convenient for obtaining an energy that is 
discontinuous in some parameters ($\theta$ in this case). 
After convergence, we continue the same processes for more than 16 rounds 
and estimate the optimized energy by averaging the data measured in these 
rounds, 
excluding excessively scattered data (beyond twice the standard deviation).  
In each optimization, $2.5\times 10^5$ samples are collected, so that 
substantially about $4\times 10^6$ measurements are averaged. 
Only for $\Psi_{\rm SF}$ with $L=16$ and $\delta=0$, the sample number is 
reduced to $2.5\times 10^4$ to save CPU time. 
Typical statistical errors are $10^{-4}t$ in the total energy and 
$10^{-4}$--$2\times 10^{-3}$ in the parameters, except near the 
Mott transition points. 
We use systems of $N_{\rm s}=L\times L$ sites with $L=10$--$18$ under 
periodic-antiperiodic boundary conditions. 
\par

\section{Staggered Flux State at Half Filling\label{sec:square}}
First, we study the unfrustrated cases ($t'=0$) in order to grasp the 
global features of the SF state because most of them do not change 
even if $t'$ is introduced. 
In this section, we focus on the half-filled case.
\par

\subsection{Variational energy}
%
\begin{figure}[htb] 
\begin{center} 
\includegraphics[width=8cm,clip]{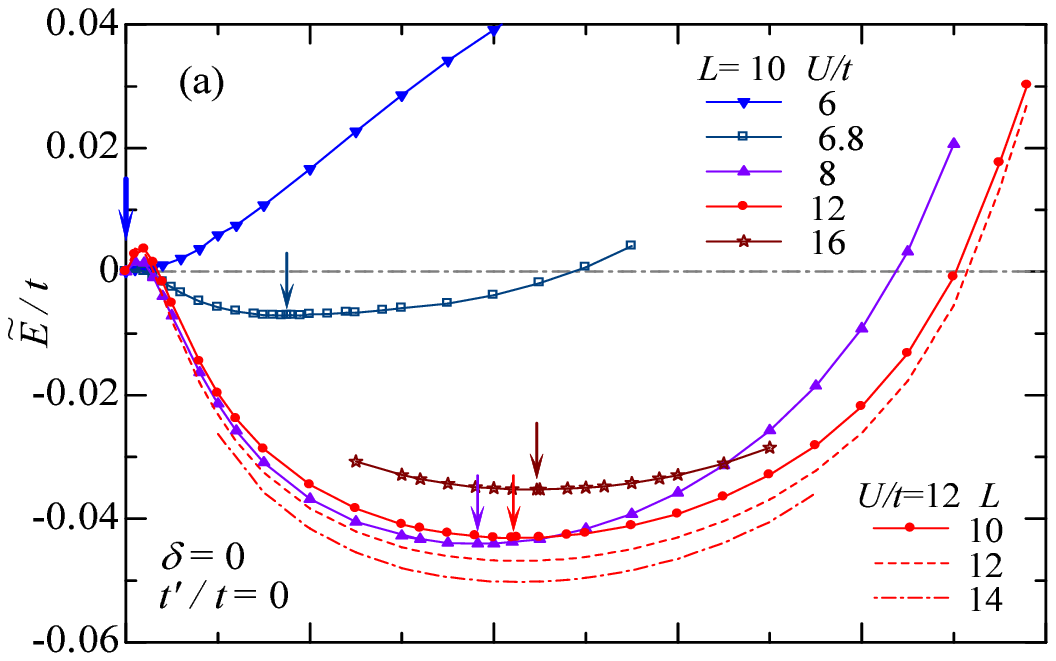} 
\vskip -3mm
\hskip 1mm
\includegraphics[width=8.3cm,clip]{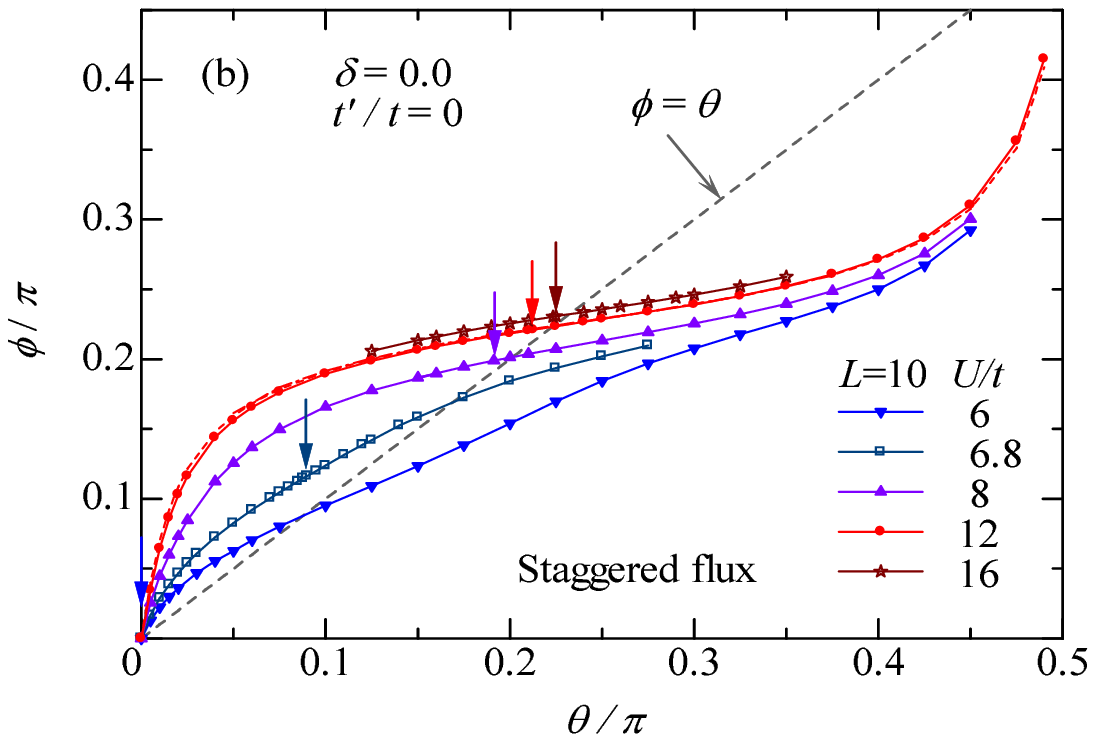} 
\end{center} 
\vskip -3mm  
\caption{(Color online) 
(a) Variational energy per site of the staggered flux state $\Psi_{\rm SF}$ 
(including ${\cal P}_\phi$) measured from that of $\Psi_{\rm N}$ 
[$E(\theta=0)$] as a function of $\theta$ for several values of $U/t$ 
($L=10$) in the Hubbard model at half filling. 
Data for $L=12$ and $14$ are added by dashed lines for $U/t=12$. 
(b) Optimized phase parameter $\phi$ for the same values of $U/t$ as in (a). 
The line of $\phi=\theta$ is added for comparison. 
The size dependence in (b) is small. 
In both panels, the arrows indicate the optimal values of $\theta$ 
when $\tilde E/t$ is minimum. 
} 
\vskip -3mm 
\label{fig:etot} 
\end{figure} 
Figure \ref{fig:etot}(a) shows the variational energy per site of 
$\Psi_{\rm SF}$ measured from that of $\Psi_{\rm N}$,
\begin{equation}
\tilde E=E^{\rm SF}(\theta)-E^{\rm N},
\label{eq:tildeE-H}
\end{equation}
as a function of $\theta$ for five values of $U/t$.
Here, the variational parameters other than $\theta$ are optimized for both 
$\Psi_{\rm SF}$ and $\Psi_{\rm N}$. 
The size dependence in the case of $U/t=12$ is also shown to see the 
convergence of the values.
For $U/t=6$, $\tilde E$ monotonically increases as a function of $\theta$. 
This behavior is the same for $U/t=0$ shown in Fig.~\ref{fig:etot-nonint}(b) 
in Appendix\ref{sec:nonintSF}. 
Hence, $\Psi_{\rm SF}$ is not stabilized for small values of $U/t$. 
The situation changes for $U/t>6$; $\tilde E/t$ becomes considerably 
negative for finite $\theta$ and has a minimum at 
$\theta/\pi\sim 0.2$ for large values of $U/t$ ($=8$--$16$). 
This behavior is qualitatively consistent with that of the $t$-$J$ model, 
the results of which are summarized in Appendix B for comparison. 
In Fig.~\ref{fig:etot}(b), we plot the optimized values of the 
configuration-dependent phase factor $\phi$ as a function of $\theta$. 
At the optimized points indicated by arrows, $\phi$ is very close to 
$\theta$, especially for large values of $U/t$.  
As discussed in Sect.~\ref{sec:phase}, the Peierls phase $\theta$ in the 
hopping process is mostly canceled by $\phi$. 
Although $\theta$ is canceled out, the state $\Psi_{\rm SF}$ preserves the 
nature of the original flux state, as shown shortly in 
Sects.~\ref{sec:spin-gap} and \ref{sec:current}. 
That is, a local staggered current flows and the momentum distribution 
function has a typical ${\bf k}$-dependence. 
\par

\begin{figure}[htb]
\begin{center}
\vskip -27mm
\includegraphics[width=9.0cm,clip]{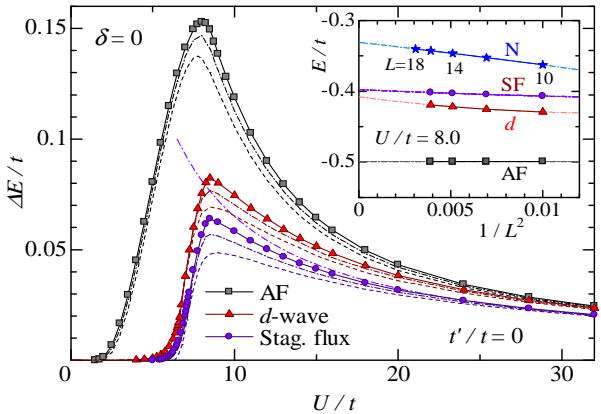} 
\end{center} 
\vskip -20mm 
\caption{(Color online) 
Energy gain of AF, $d$-SC, and SF states with respect to projected 
Fermi sea ($\Psi_{\rm N}$) at half filling as functions of $U/t$. 
Data for $L=14$, $12$, and $10$ for each state are plotted as solid lines with 
symbols, dash-dotted lines, and dashed lines, respectively. 
A guide curve proportional to $t/U$ is drawn for $\Delta E({\rm SF})$ 
with $L=14$ (dash-dotted line). 
In the inset, the system-size dependence is shown for $U/t=8.0$ and fitted by 
second-order polynomials. 
}
\vskip -3mm 
\label{fig:delEtot-jpsj} 
\end{figure}
%
Next, we discuss the $U/t$ dependence of the energy gain of the fully 
optimized 
$\Psi_{\rm SF}$ with respect to the reference state $\Psi_{\rm N}$, 
\begin{equation}
\Delta E({\rm SF})=E({\rm N})-E({\rm SF}), 
\label{eq:gain-N-SF}
\end{equation}
where $E({\rm N})$ and $E({\rm SF})$ are the optimized (including $\theta$) 
energies per site of $\Psi_{\rm N}$ and $\Psi_{\rm SF}$, respectively. 
If $\Delta E({\rm SF})$ is positive, the SF state is stabilized with respect 
to $\Psi_{\rm N}$. 
In Fig.~\ref{fig:delEtot-jpsj}, we show $\Delta E({\rm SF})$ compared with 
other ordered states, i.e., the AF state, 
$\Psi_{\rm AF}={\cal P}\Phi_{\rm AF}$, and the $d$-SC (projected BCS) state, 
$\Psi_d={\cal P}\Phi_d$.
We use the same $\Phi_{\rm AF}$ and $\Phi_d$ as in the preceding study 
(Ref.~\citen{YOTKT})\cite{note-FSRen,Himeda} but we adopt 
Eq.~(\ref{eq:D-H}) for ${\cal P}_Q$. 
For $\delta=0$ and $U>U_{\rm c}$, $\Psi_d$ is not SC but Mott insulating. 
In Fig.~\ref{fig:delEtot-jpsj}, each state exhibits a maximum at 
$U\sim W$ ($=8t$). 
The system-size dependence of $\Delta E$ for each state is large near the 
maximum but, as shown in the inset of Fig.~\ref{fig:delEtot-jpsj}, 
$\Delta E$ remains finite and the order of the variational energy
will not change as $L\rightarrow\infty$. 
$\Delta E(\mbox{AF})$ is largest, i.e., the AF state has the lowest energy 
for any $U/t$.\cite{YTOT} 
For the SF state, $\Delta E({\rm SF})\sim 0$ for small values of $U/t$ 
($\lesssim 5$).
Although $\Delta E({\rm SF})$ is always smaller than $\Delta E(d{\rm-SC})$, 
it is close to $\Delta E(d{\rm-SC})$. 
At $U/t\sim 5$, $\Delta E({\rm SF})$ starts to increase abruptly. 
The range of $U/t$ where $\Psi_{\rm SF}$ is stabilized ($U/t\gtrsim 5$) is 
similar to that of $\Psi_d$. 
In addition, the behavior of physical quantities such as the momentum 
distribution function is similar between $\Psi_{\rm SF}$ and $\Psi_d$ 
as shown shortly. 
As mentioned, in the Heisenberg model, $\Psi_{\rm SF}(g=0)$ and $\Psi_d(g=0)$ 
are equivalent due to the SU(2) symmetry, but in the Hubbard model, 
the two states are not equivalent, probably due to the difference in the
distribution of doublons and holons. 
\par

\subsection{SF transition and Mott transition\label{sec:transitions}}
\begin{figure}[htb]
\begin{center} 
\vskip 1mm
\includegraphics[width=8.5cm,clip]{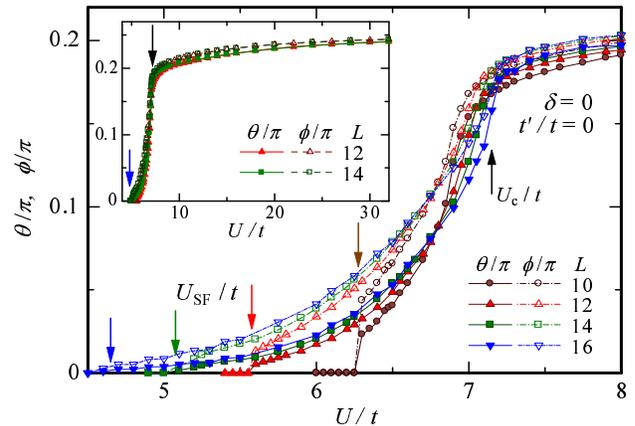}
\end{center} 
\vskip -3mm 
\caption{(Color online) 
Optimized phase parameters ($\theta$ and $\phi$) in $\Psi_{\rm SF}$ 
at half filling.
The arrows indicate the Mott transition ($U_{\rm c}/t$) and 
SF transition ($U_{\rm SF}/t$) for four system sizes. 
The inset shows the same quantities for a wider range of $U/t$,  
with the arrows denoting $U_{\rm c}/t$ and $U_{\rm SF}/t$ for $L=16$.
}
\vskip -3mm 
\label{fig:para-t-ph-Lm} 
\end{figure}

%
Figure \ref{fig:para-t-ph-Lm} shows the optimized $\theta$ and $\phi$ 
in $\Psi_{\rm SF}$ as a function of $U/t$. 
We find two transition points: $U_{\rm SF}/t$ at $\sim4$ -- $5$ and 
$U_{\rm c}/t$ at $\sim 7$. 
The former corresponds to the SF transition at which $\Psi_{\rm SF}$ 
starts to have finite $\theta$ and $\phi$ and its variational energy becomes 
lower than that of $\Psi_{\rm N}$. 
The latter corresponds to a Mott transition at which the system starts 
to have a gap in the charge degree of freedom. 
The symmetry does not change at $U_{\rm c}/t$. 
\par

%

At $U_{\rm SF}/t$, $\theta$ and $\phi$ exhibit first-order-transition-like 
discontinuities, for example, at $U_{\rm SF}/t=6.28$ for $L=10$. 
However, as $L$ increases, $U_{\rm SF}/t$ shifts to lower values  and 
the discontinuities become small and unclear, suggesting that the SF 
transition is continuous and occurs at a small $U/t$. 
Because an appropriate scaling function is not known, 
we simply perform a polynomial fit of $U_{\rm SF}/t$ up to the square of 
$1/L^2$ as a rough estimate.
This yields $U_{\rm SF}/t=2.93$ for $L=\infty$ with a small error. 
In any case, since $\theta$ and $\phi$ are tiny for $U/t\lesssim 5$, we 
consider that $\Psi_{\rm SF}$ is substantially not stable in a weakly 
correlated regime. 
\par

\begin{figure}[htb]
\begin{center}
\includegraphics[width=8.0cm,clip]{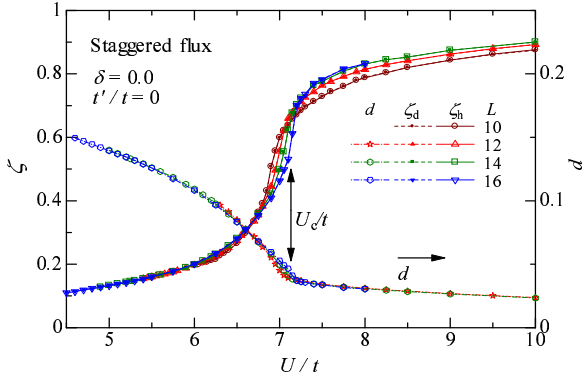} 
\end{center} 
\vskip -3mm 
\caption{(Color online) 
$U/t$ dependences of optimized D-H binding parameter $\zeta$ 
($=\zeta_{\rm d}=\zeta_{\rm h}$) in 
$\Psi_{\rm SF}$ and density of doublons $d$ shown at half filling 
for some system sizes. 
The arrow indicates the Mott transition point.
}
\vskip -3mm 
\label{fig:paramu-L} 
\end{figure}
%
Next, we examine the Mott transition at $U_{\rm c}/t$, where 
the behaviors of $\theta$ and $\phi$ change as shown in Fig.~\ref{fig:para-t-ph-Lm}. 
$U_{\rm c}/t=7.1$ for $L=16$\cite{note-Uc,Sorella,Miyagawa}.
(Note that in the $t$-$J$ model, the Mott transition cannot be discussed.) 
To confirm that $U_{\rm c}/t$ is a Mott transition, we plot the 
$U/t$-dependences 
of the optimized D-H binding parameter $\zeta$ 
($=\zeta_{\rm d}=\zeta_{\rm h}$) and the doublon density $d=E_U/U$ in 
Fig.~\ref{fig:paramu-L}. 
These quantities are sensitive indicators of Mott transitions.
In Fig.~\ref{fig:paramu-L}, we find abrupt changes in both $\zeta$ and $d$ 
at $U_{\rm c}/t$, similarly to those in the Mott transitions in $\Psi_d$ and 
$\Psi_{\rm N}$.\cite{YOT} 
In $\Psi_{\rm SF}$, discontinuities in $\zeta$ and $d$ at $U_{\rm c}/t$ are 
not found even for the largest system we treat ($L=16$). 
However, because the behavior of both $\zeta$ and $d$ becomes more singular 
as $L$ increases, we consider that this transition is first-order, 
similarly 
to those in $\Psi_d$ and $\Psi_{\rm N}$.\cite{YOT} 
\par

\subsection{Spin-gap metal\label{sec:spin-gap}}

\begin{figure*}[t!]
\begin{center}
\includegraphics[width=5.9cm,clip]{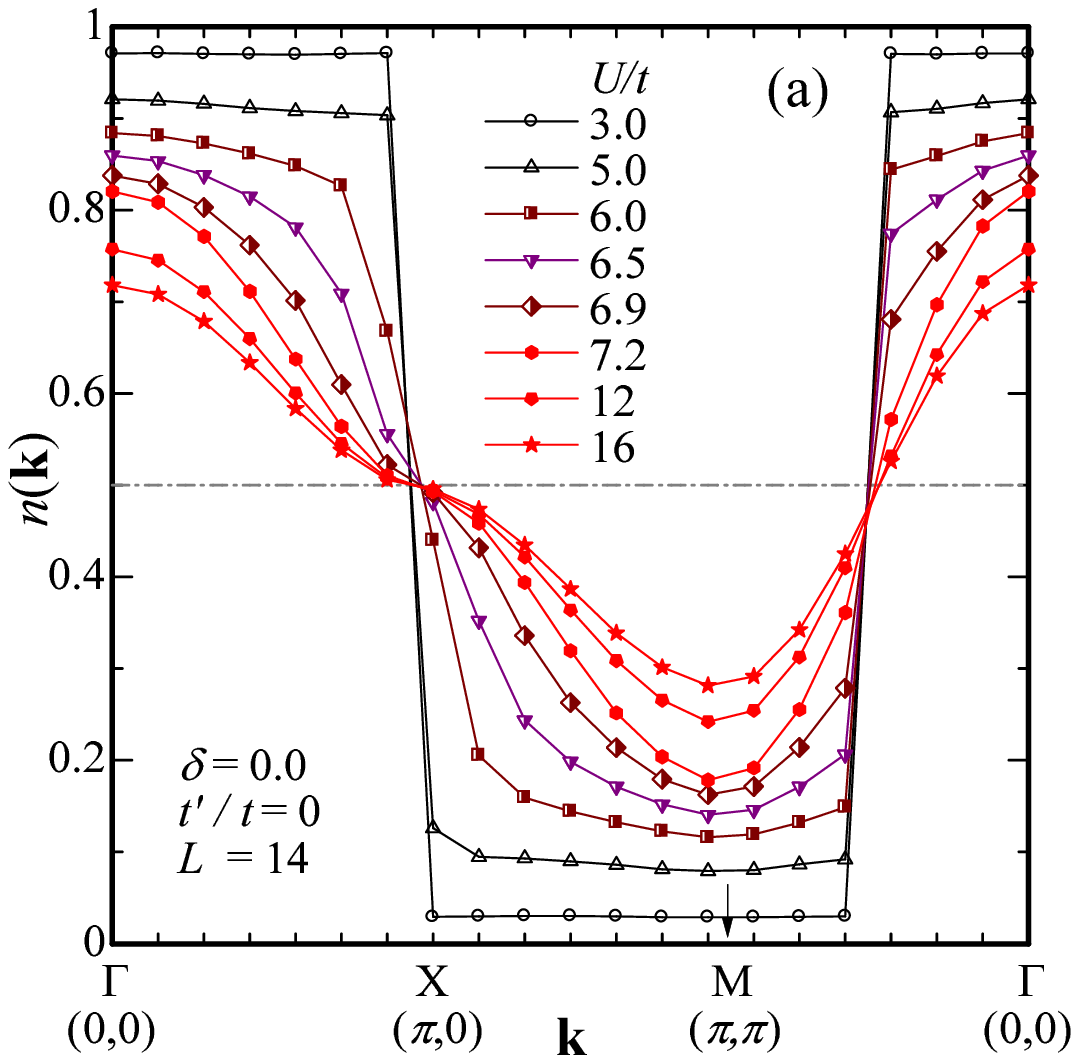}
\hskip -3mm
\includegraphics[width=6.1cm,clip]{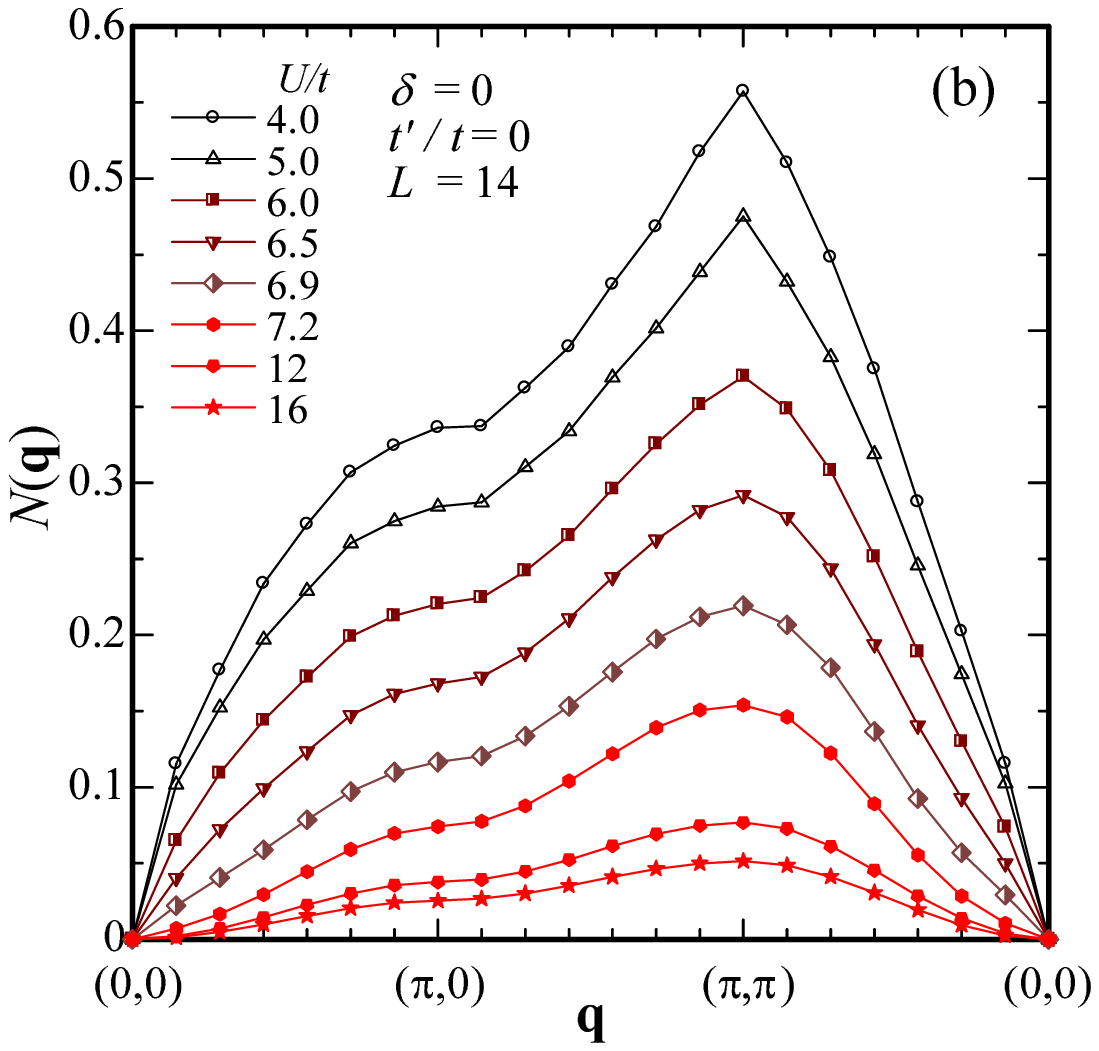}
\hskip -3mm
\includegraphics[width=6.1cm,clip]{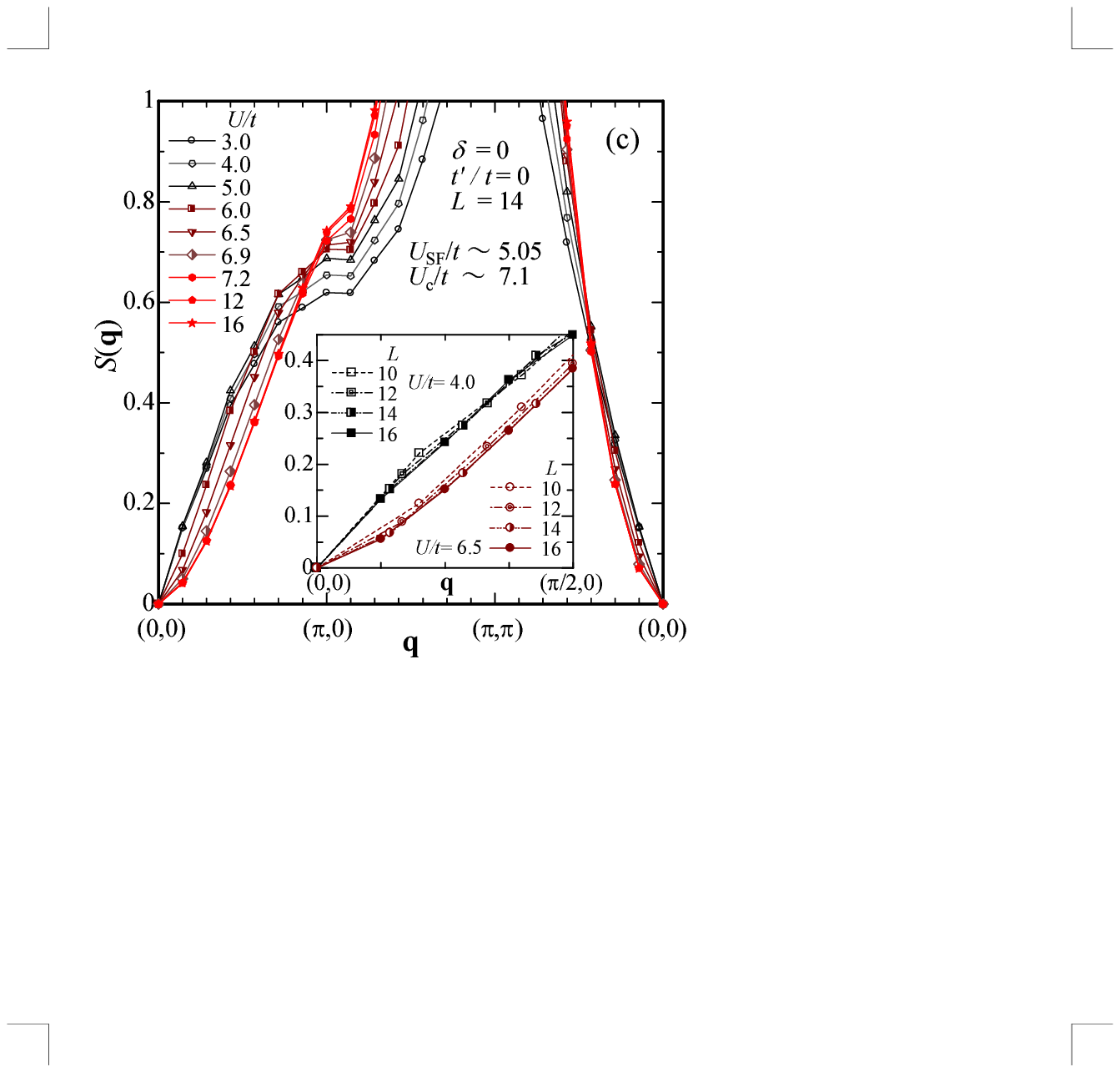}
\end{center} 
\vskip -3mm 
\caption{(Color online) 
Correlation functions in momentum space of 
$\Psi_{\rm SF}$ at half filling for various values of $U/t$. 
(a) Momentum distribution function, 
(b) charge density structure factor, and 
(c) spin structure factor. 
In the inset in (c), the system-size dependence of $S({\bf q})$ ($L=10$--$16$) 
for typical cases of a Fermi liquid ($U/t=4$) and spin-gap metal ($U/t=6.5$) 
is shown for small $|{\bf q}|$ in the direction of $(0,0)$--$(\pi,0)$. 
See also Ref.~\citen{note-sq}. 
Open (black) symbols are for $U<U_{\rm SF}$ (i.e., $\theta=0$), 
half-solid (brown) symbols for $U_{\rm SF}<U<U_{\rm c}$, and 
solid (red) symbols for $U>U_{\rm c}$. 
For this system size ($L=14$), $U_{\rm SF}/t\sim 5.05$ and 
$U_{\rm c}/t\sim 7.1$. 
}
\vskip -3mm 
\label{fig:nk} 
\end{figure*}
%
In the intermediate regime $U_{\rm SF}<U<U_{\rm c}$, 
the present SF state is expected to be metallic. 
In order to clarify the nature of $\Psi_{\rm SF}$, we calculate 
the momentum distribution function 
\begin{equation}
n({\bf k})=
\frac{1}{2}\sum_\sigma\langle c^\dag_{\bf k\sigma}c_{\bf k\sigma}\rangle 
\label{eq:nk}
\end{equation}
for the optimized $\Psi_{\rm SF}$.
Figure \ref{fig:nk}(a) shows $n({\bf k})$ along the path 
$(0,0)$-$(\pi,0)$-$(\pi,\pi)$-$(0,0)$ in the original Brillouin zone 
for various values of $U/t$. 
In the region of $U<U_{\rm SF}$ (i.e., $\theta=0$), 
we find two discontinuities (crossings of the Fermi surface) at 
${\bf k}\sim (\pi,0)$ and $(\pi/2,\pi/2)$, indicating a typical Fermi liquid. 
For $U_{\rm SF}<U<U_{\rm c}$ (half-solid symbols), the discontinuity at 
${\bf k}\sim (\pi,0)$ disappears, while the discontinuity at 
${\bf k}\sim(\pi/2,\pi/2)$ remains. 
This is qualitatively identical to that of the noninteracting SF state 
$\Phi_{\rm SF}$, in which there is a Dirac point at $(\pi/2,\pi/2)$ 
and a certain gap opens near the antinodal points. 
On the other hand, for $U>U_{\rm c}$ (solid symbols), both discontinuities 
disappear, indicating that a gap opens in the whole Brillouin zone. 
This is consistent with a Mott insulator. 
\par

We can reveal the characters of the gaps to some extent by analyzing the 
charge 
density and spin structure factors, 
\begin{eqnarray}
N({\bf q})&=&\frac{1}{N_{\rm s}} 
\sum_{i,j}e^{i{\bf q}\cdot({\bf R}_i-{\bf R}_j)} 
\left\langle{n_{i} n_{j}}\right\rangle - n^2, 
\label{eq:nq} \\
S({\bf q})&=&\frac{1}{N_{\rm s}}\sum_{ij}{e^{i{\bf q}
\cdot({\bf R}_i-{\bf R}_j)} 
\left\langle{S_{i}^zS_{j}^z}\right\rangle}. 
\label{eq:sq}
\end{eqnarray} 
On the basis of the single-mode approximation,
\cite{SMA,SMA-example,Tamura-AHM} 
excitations in the charge sector are gapless when 
$N({\bf q})\propto|{\bf q}|$ for $|{\bf q}|\rightarrow 0$, 
whereas a gap opens in the charge sector when 
$N({\bf q})\propto |{\bf q}|^2$. 
For $S({\bf q})$, a similar relation holds for the spin sector, although 
excitations cannot be sharply divided into the charge and spin 
sectors except for in one-dimensional systems. 
In Figs.~\ref{fig:nk}(b) and \ref{fig:nk}(c), $N({\bf q})$ and $S({\bf q})$ 
are respectively shown for various values of $U/t$. 
For $U<U_{\rm SF}$, both $N({\bf q})$ and $S({\bf q})$ behave linearly 
for $|{\bf q}|\rightarrow 0$ as expected for a Fermi liquid. 
For $U>U_{\rm c}$, the behaviors of both $N({\bf q})$ and $S({\bf q})$ 
appear to be quadratic and consistent with a Mott insulator. 
For $U_{\rm SF}<U<U_{\rm c}$, $N({\bf q})$ is linear near $(0,0)$, whereas 
$S({\bf q})$ is quadratic-like,\cite{note-sq} indicating that the state is a 
spin-gap metal.  
Namely, the charge and spin sectors show different tendencies in excitation. 
This feature is distinct from that of the noninteracting SF state 
$\Phi_{\rm SF}$, which has a gap common to both sectors, and thus  
$N({\bf q})=S({\bf q})$ holds (see Appendix\ref{sec:nonintSF}). 
Therefore, the metallic SF state stable for $U_{\rm SF}<U<U_{\rm c}$ 
is not perturbatively connected to $\Phi_{\rm SF}$. 
We will show in the next section that this state is connected to 
the metallic SF state in the doped case.
\par

\subsection{Circular current\label{sec:current}}

\begin{figure}[htb]
\begin{center}
\includegraphics[width=8.5cm,clip]{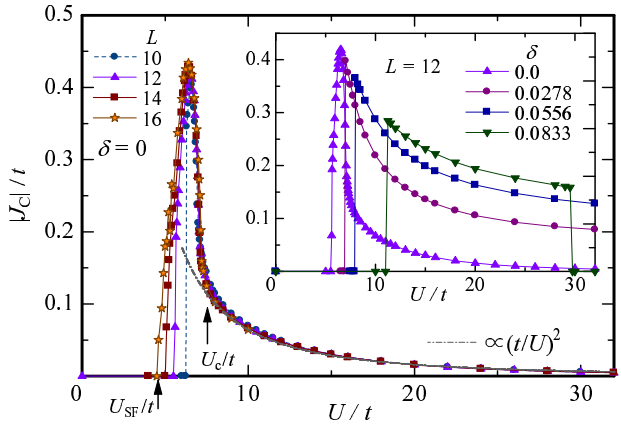}
\end{center} 
\vskip -3mm 
\caption{(Color online) 
Absolute values of local circular current at half filling as 
a function of $U/t$ for some system sizes. 
The SF and Mott transition points are shown by arrows. 
A curve proportional to $(t/U)^2$ is shown by a gray dash-dotted line. 
The inset shows the same quantity for some doping rates for $L=12$ 
(discussed in Sect.~\ref{sec:doped cases}). 
}
\vskip -3mm 
\label{fig:curr-jpsj} 
\end{figure}

Now, we turn to the local circular current in a plaquette defined as 
\begin{equation}
J_{\rm C}/t =\frac{1}{N_{\rm s}}\sum_{\ell\in{\rm A},\sigma}
\sum_\tau(-1)^{\tilde\ell+\tilde\tau}{\rm Im}
\langle c^\dag_{\ell+\tau,\sigma}c_{\ell,\sigma}
       -c^\dag_{\ell,\sigma}c_{\ell+\tau,\sigma}\rangle, 
\label{eq:current}
\end{equation}
where $\ell$ runs over all the A sublattice sites, 
$\tilde\ell=\ell_x+\ell_y$, $\tau$ indicates the nearest-neighbor directions, 
and $\tilde\tau=1$ [$-1$] for $\tau=(\pm 1,0)$ [$(0,\pm 1)$]. 
$J_{\rm C}$ is regarded as the order parameter of the SF phase. 
In the main panel of Fig.~\ref{fig:curr-jpsj}, we show $|J_{\rm C}|/t$ 
at half filling as a function of $U/t$. 
In the metallic SF phase ($U_{\rm SF}<U<U_{\rm c}$), a relatively large 
current flows.
In the insulating SF phase ($U>U_{\rm c}$), the local current is reduced but 
still finite. 
At $U/t=12$, however, $|J_{\rm C}|/t$ is $1/20$ of that in 
$\Phi_{\rm SF}$ with the same $\theta$. 
The $U/t$-dependence of $J_{\rm C}/t$ in this regime is fitted by a curve proportional 
to $(t/U)^2$ and the system-size dependence is small, as shown in 
Fig.~\ref{fig:curr-jpsj}. 
This suggests that $J_{\rm C}$ in this range of $U/t$ has a localized nature. 
More specifically, $|J_{\rm c}|$ will be related to the local four-site ring 
exchange interaction, which appears in the fourth-order perturbation with 
respect to $t/U$ in the large-$U$ expansion of the Hubbard model. 
\par

\section{Staggered Flux State at Finite Doping\label{sec:doped cases}}

\subsection{Energy gain and optimized phase parameters}
\begin{figure}[htb]
\begin{center}
\includegraphics[width=7.45cm,clip]{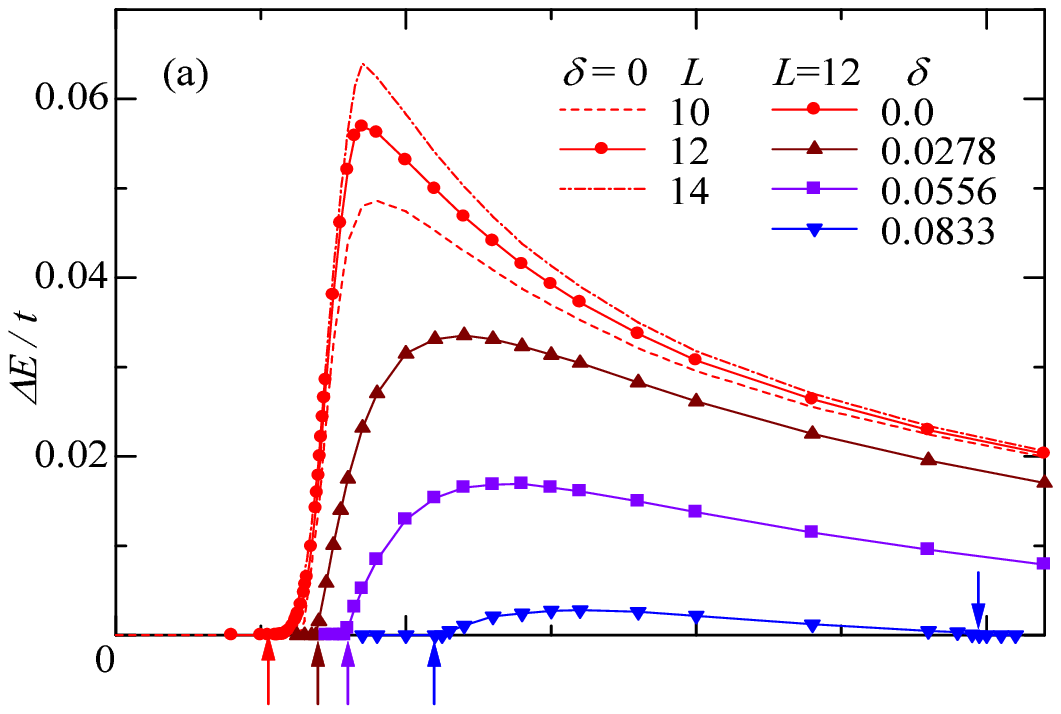}
\vskip -4mm 
\includegraphics[width=7.5cm,clip]{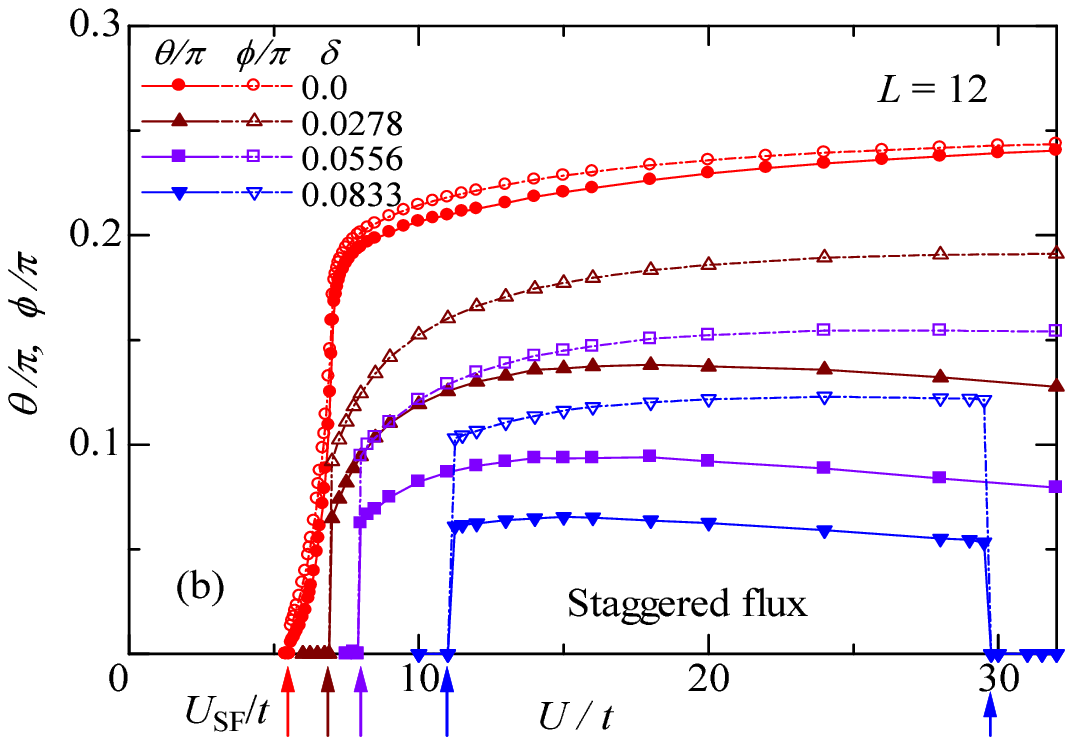} 
\end{center} 
\vskip -3mm 
\caption{(Color online) 
(a) Energy gain of SF state with respect to $\Psi_{\rm N}$ 
as a function of $U/t$ for four values of doping rate $\delta$ ($L=12$).
For $\delta=0$, data for three system sizes are shown. 
(b) Optimized phase parameters $\theta$ and $\phi$. 
In both panels, the SF transition is indicated by arrows.
}
\vskip -3mm 
\label{fig:delEtotvsn.eps} 
\end{figure}

First, we show the energy gain of $\Psi_{\rm SF}$ with respect to the 
reference state $\Psi_{\rm N}$ [Eq.~(\ref{eq:gain-N-SF})]
in Fig.~\ref{fig:delEtotvsn.eps}(a) for four values of the doping rate 
$\delta$. 
Similarly to the half-filled case (Fig.~\ref{fig:delEtot-jpsj}), $\Delta E$ is 
zero for the weakly correlated regime ($U<U_{\rm SF}$); 
the value of the SF transition,  
$U_{\rm SF}/t$, increases as $\delta$ increases. 
The sharp peak of $\Delta E$ for $\delta=0$ changes to a broader peak  
with a maximum at $U/t\sim 12$ -- $16$, and finally vanishes at 
$\delta\equiv\delta_{\rm SF}\sim 0.1$. 
\par

In Fig.~\ref{fig:delEtotvsn.eps}(b), optimized values of the phase parameters 
are plotted. 
Both parameters decrease as $\delta$ increases. 
Although $\theta$ and $\phi$ have a discontinuity at $U_{\rm SF}/t$ at this 
system size, this behavior is owing to a finite-size effect.\cite{note-USF} 
The SF transition for $L\rightarrow\infty$ will be continuous, similarly 
to the half-filled case. 
When we compare with the results at $\delta=0$, we see that 
$\Phi_{\rm SF}$ is realized in the strongly correlated region ($U>W$), and 
it is 
smoothly connected to the Mott insulating state at half filling.
It is also interesting that $\phi$ becomes larger than $\theta$ as $\delta$ increases, 
while they are close to each other when $\delta=0$. 
This suggests that $\phi$ overscreens the phase $\theta$ in the D-H processes 
owing to the increasing number of free-holon processes.
\par

%
\begin{figure}[htb]
\begin{center}
\includegraphics[width=7.55cm,clip]{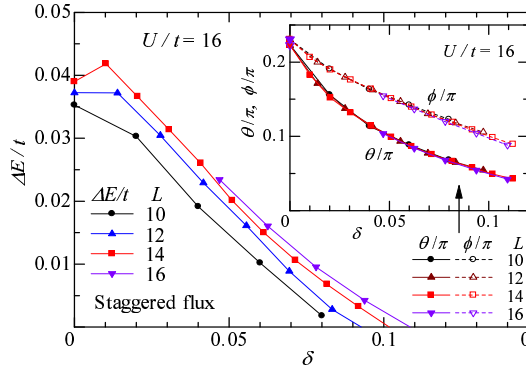} 
\end{center} 
\vskip -3mm 
\caption{(Color online) 
Energy gain of SF state with respect to $\Psi_{\rm N}$ as a function of 
doping rate at $U/t=16$. Data for four system sizes are shown. 
In the inset, the optimized phase parameters $\theta$ and $\phi$ are shown.
}
\vskip -3mm 
\label{fig:delE-para-vsn} 
\end{figure}
%
Figure \ref{fig:delE-para-vsn} shows the $\delta$-dependence of $\Delta E/t$  
for the case with $U/t=16$. 
Except for the case with $\delta=0$, $\Delta E/t$ monotonically decreases 
as a function of $\delta$. 
Because the $L$ dependence is appreciable, $\delta_{\rm SF}$ should be 
somewhat larger in the $L\rightarrow\infty$ limit. 
The behavior of $\Delta E$ is consistent with that for the $t$-$J$ model 
shown in Appendix\ref{sec:t-J}.\cite{note-t-J} 
In the inset of Fig.~\ref{fig:delE-para-vsn}, the $\delta$-dependences of the 
optimized $\theta$ and $\phi$ are plotted. 
Their system-size dependences are very small. 
\par 

\subsection{Various properties\label{sec:properties}}
\begin{figure*}[t!]
\begin{center}
\includegraphics[width=5.9cm,clip]{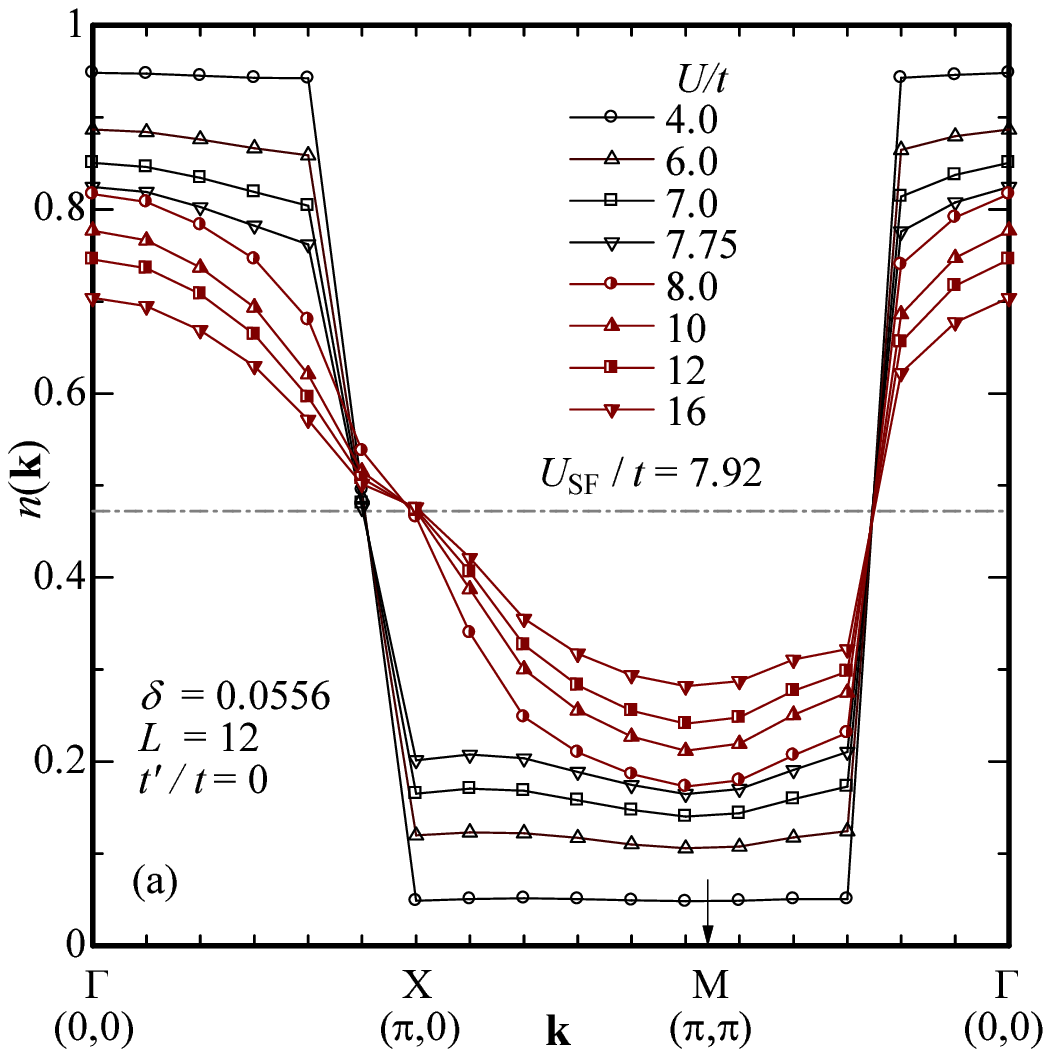}
\hskip -2mm
\includegraphics[width=5.9cm,clip]{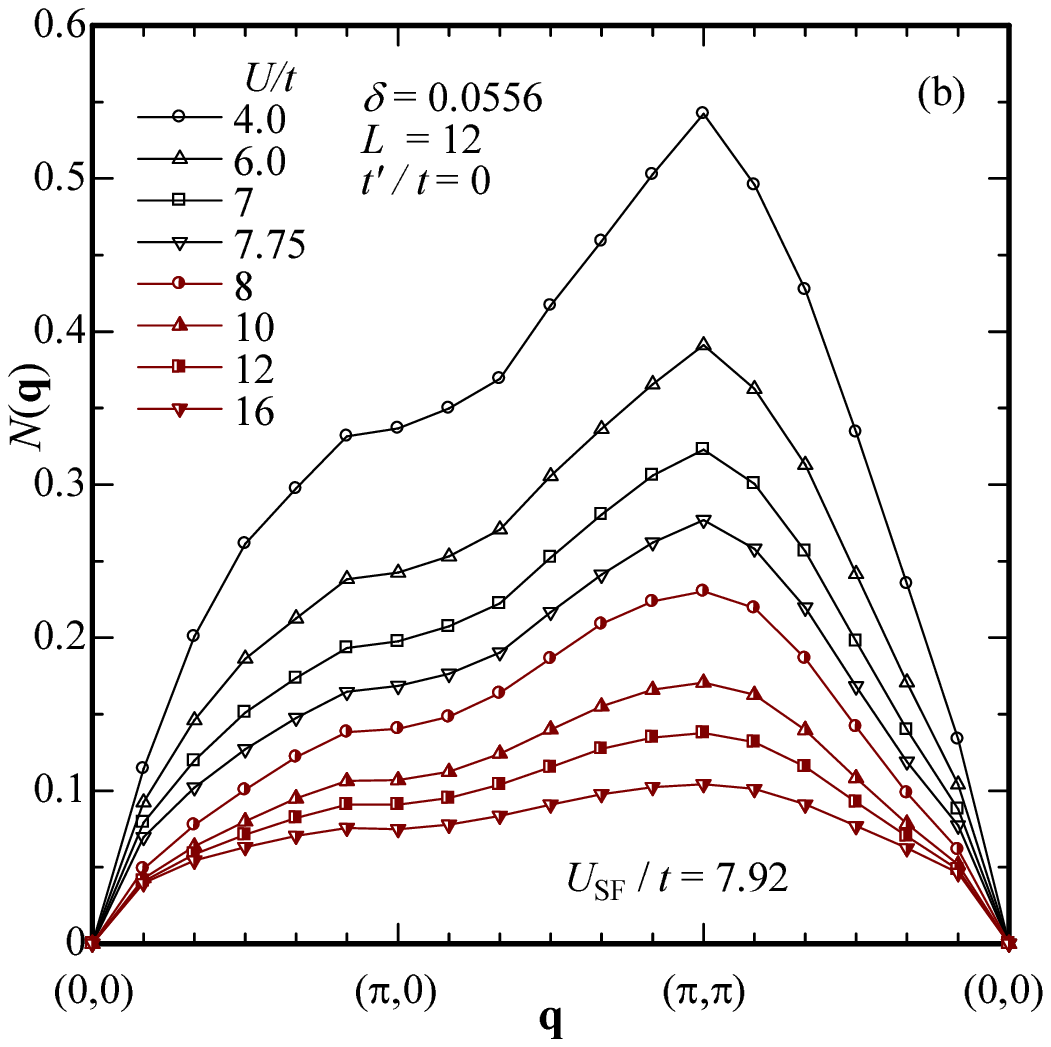}
\hskip -2mm
\includegraphics[width=5.9cm,clip]{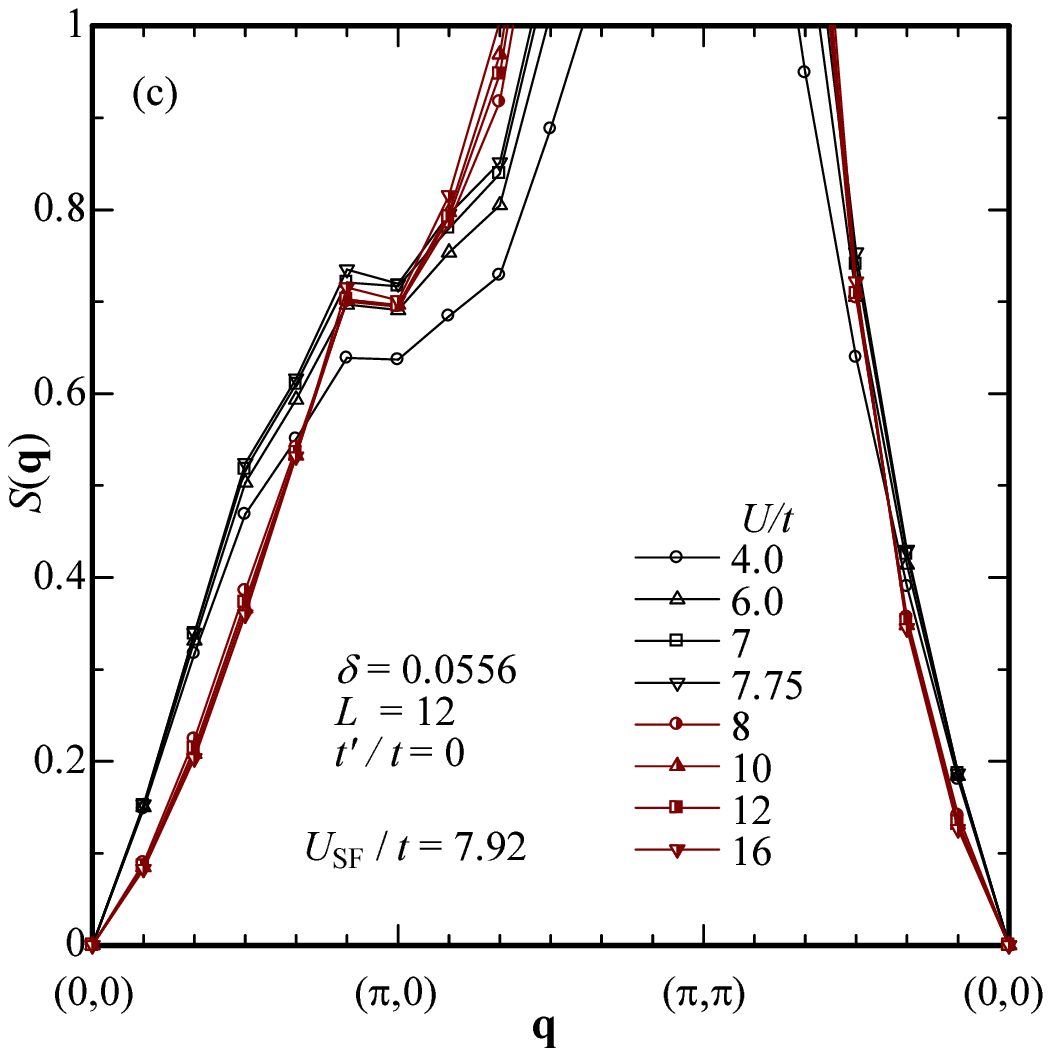}
\end{center} 
\vskip -3mm 
\caption{(Color online) 
Behavior of
(a) momentum distribution function, 
(b) charge density structure factor, and 
(c) spin structure factor 
of $\Psi_{\rm SF}$ for a finite doping ($\delta=0.0556$)
and various values of $U/t$. 
Open (black) symbols are for $U<U_{\rm SF}$ and half-solid (brown) symbols 
are for $U>U_{\rm SF}$. 
For this system size ($L=12$) and doping, the SF transition is at 
$U_{\rm SF}/t=7.92$.
}
\vskip -3mm 
\label{fig:nk-n} 
\end{figure*}
%
(i) {\it Spin-gap metal}: 
In Fig.~\ref{fig:nk-n}, we show the behavior of correlation functions 
in the momentum space $n({\bf k})$, $N({\bf q})$, and $S({\bf q})$ 
for $\delta=0.0556$ ($L=12$). 
The $U/t$-dependences of these quantities are basically 
similar to those at half filling discussed in Fig.~\ref{fig:nk}. 
In the region of $U>U_{\rm SF}$, $n({\bf k})$ preserves a discontinuity 
near $(\pi/2,\pi/2)$, indicating that $\Psi_{\rm SF}$ is always metallic 
and there is no Mott transition.  
Furthermore, $N({\bf q})$ is linear in $|{\bf q}|$ for 
$|{\bf q}|\rightarrow 0$, indicating that the charge degree of 
freedom is gapless. 
On the other hand, $S({\bf q})$ appears to be approximately quadratic at 
small $|{\bf q}|$
for $U>U_{\rm SF}$, suggesting that the SF state in the doped region has 
a gap in the spin sector. 
\par

\begin{figure}[htb]
\begin{center}
\hskip -5mm
\includegraphics[width=9.5cm,clip]{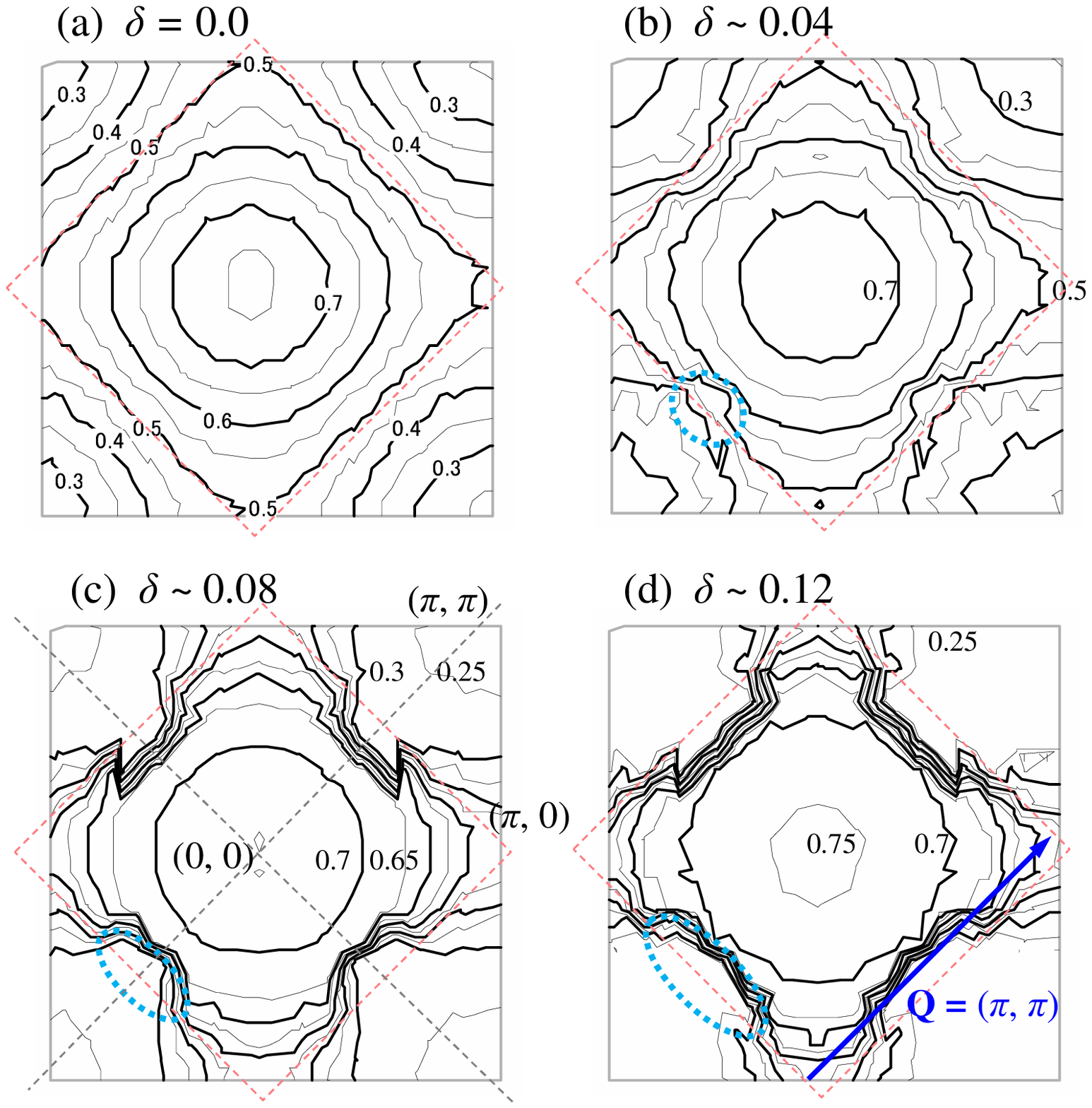} 
\end{center} 
\vskip -48mm 
\caption{(Color online) 
Contour maps of momentum distribution function $n({\bf k})$ of the optimized
$\Psi_{\rm SF}$ in the original Brillouin zone for $t'/t=-0.3$, $U/t=12$, 
$L=10$ -- $14$, and four values of $\delta$ in (a)-(d). 
The wiggles of lines are simply due to the small number of {\bf k} points and 
are not important. 
The AF Brillouin zone boundary is indicated by pink dotted lines, and 
zone-diagonal lines are shown with gray dotted lines in (c). 
Fermi surfaces are indicated with blue ovals in the third quadrants.
In (d), the scattering vector of ${\bf q}={\bf Q}$ connecting the antinodes, 
discussed in Sect.~\ref{sec:related}, is shown with a blue arrow. 
}
\vskip -3mm 
\label{fig:cfnk3d-jpsj} 
\end{figure}
%
(ii) {\it Segmented Fermi surface}: 
The bare SF state, $\Phi_{\rm SF}$, has a segmented (or small) Fermi surface 
around ${\bf k}=(\pi/2,\pi/2)$ as shown in Appendix A 
(see Fig.~\ref{fig:fstha}). 
Here we show that this feature is preserved for strongly correlated cases. 
Shown in Fig.~\ref{fig:cfnk3d-jpsj} are contour maps of $n({\bf k})$ for 
$U/t=12$ and four values of $\delta$. 
Here, we show the data for $t'/t=-0.3$ because $\Psi_{\rm SF}$ is stabilized 
in a wide doping range (see Fig.~\ref{fig:transpoint-jpsj} later) and 
the behavior is similar to that for $t'/t=0$. 
At half filling, there is no Fermi surface, as shown in panel (a). 
Upon doping, however, pocket Fermi surfaces appear around $(\pi/2,\pi/2)$ and, 
as $\delta$ increases, they extend to the antinodes along the AF Brillouin 
zone edge. 
These Fermi surfaces are shown by blue dashed ovals in panels (b)-(d). 
A gap remains open near $(\pi, 0)$. 
\par

%
%
%
(iii) {\it Circular currents}: 
The local circular currents $J_{\rm C}$ defined in Eq.~(\ref{eq:current}) 
for the doped cases have already been shown in the inset of 
Fig.~\ref{fig:curr-jpsj}, where the evolution of $J_{\rm C}$ with 
increasing $\delta$ is shown as a function of $U/t$. 
We find that $J_{\rm C}$ increases as $\delta$ increases, although 
the optimized phase parameters $\theta$ and $\phi$ decrease 
[see Fig.~\ref{fig:delEtotvsn.eps}(b)]. 
This is probably because the number of mobile carriers increases 
as $\delta$ increases in the strongly correlated regime, whose 
feature is typical of a doped Mott insulator.
In contrast, as shown in Appendix A, $J_{\rm C}$ decreases as $\delta$ 
increases in the noninteracting $\Phi_{\rm SF}$. 
At the phase transition point $\delta_{\rm SF}$, where $E({\rm SF})$ 
becomes equal to $E(\rm{N})$, the order parameter $|J_{\rm C}|/t$ drops 
suddenly from $0.25$--$0.3$ (almost the maximum value) to zero. This 
indicates that this transition is first-order, in contrast to the 
corresponding AF and $d$-SC transitions, as a function of $\delta$.
\par

\section{Effect of Diagonal Hopping $t'$\label{sec:t'}}
In this section, we study the effect of diagonal hopping $t'$ in the two 
cases shown in Fig.~\ref{fig:model}. 
\par

\subsection{Frustrated square lattice\label{sec:FSQ}}
\begin{figure*}[t!]
\begin{center}
\includegraphics[width=7.5cm,clip]{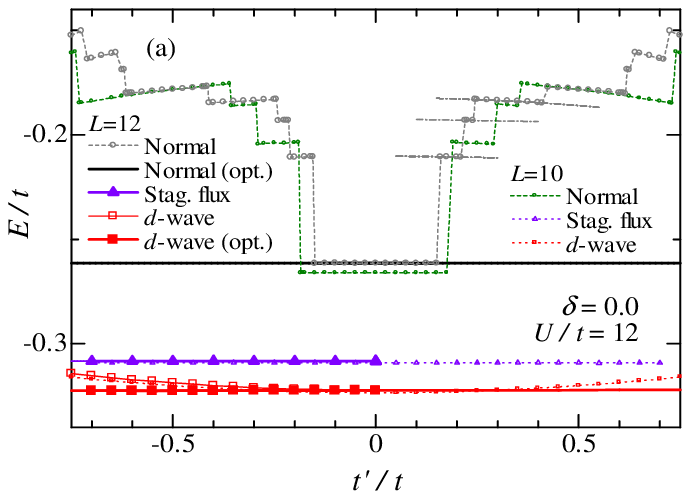} 
\hskip 5mm
\includegraphics[width=7.5cm,clip]{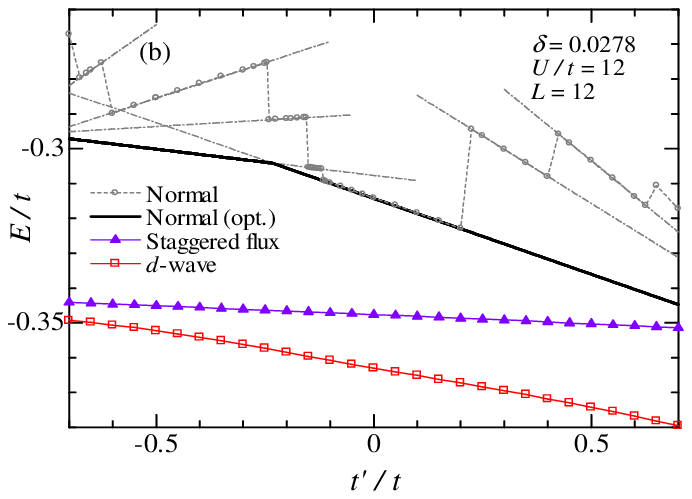} 
\vskip -1mm
\includegraphics[width=7.5cm,clip]{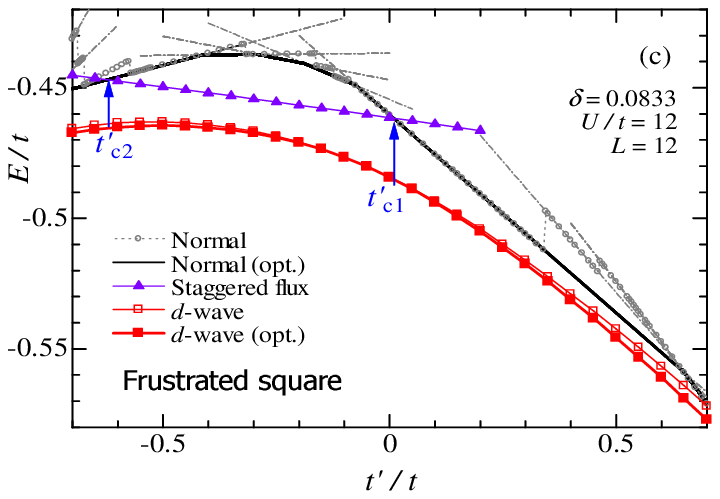} 
\hskip 5mm 
\includegraphics[width=7.5cm,clip]{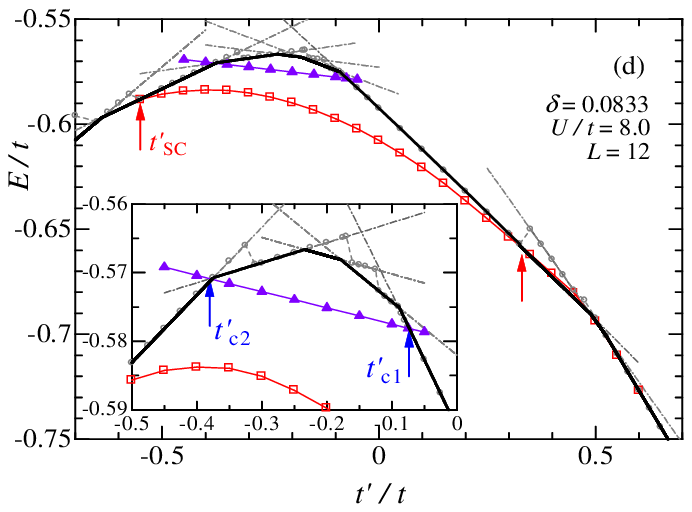} 
\end{center} 
\vskip -3mm 
\caption{(Color online) 
Comparison of total energies among $\Psi_{\rm N}$, $\Psi_{\rm SF}$, and 
$\Psi_d$ as functions of $t'/t$ in frustrated square lattice of $L=12$. 
Panels (a)-(c) display the cases of different doping rates for $U/t=12$ 
and (d) shows the case of $U/t=8$ and $\delta=0.0833$. 
In (a), data for $L=10$ are also shown. 
Symbols are common to all panels. 
The black and red bold lines indicate the values of $\Psi_{\rm N}$ and 
$\Psi_d$, respectively, when the band-renormalization effect is considered 
(for details, see Ref.~\citen{BR}). 
The arrows indicate the positions of the energy crossings.
The inset in (d) shows a magnification of the area of energy crossings. 
}
\vskip -3mm 
\label{fig:evsa} 
\end{figure*}
%
Figure \ref{fig:evsa} summarizes the total energies of 
$\Psi_{\rm SF}$, $\Psi_{\rm N}$, and $\Psi_d$ as functions of $t'/t$.
Note that the energy for $\Psi_{\rm N}$ without band renormalization 
exhibits complicated behaviors as a function of $t'/t$. 
This is because the occupied ${\bf k}$-points in the Fermi surface change 
discontinuously in $\Phi_{\rm N}$. 
However, if we consider the band-renormalization effect\cite{BR} for 
$\Psi_{\rm N}$ and use the optimized $t_1/t$, the lowest energy for 
$\Psi_{\rm N}$ becomes the black solid line in Fig.~\ref{fig:evsa}. 
We use the solid lines as energies for $\Psi_{\rm N}$. 
Although we expect some size effects in $\Psi_{\rm N}$, 
we can see general trends of the energy differences 
between $\Psi_{\rm SF}$, $\Psi_{\rm N}$, and $\Psi_d$. 
\par

At half filling [Fig.~\ref{fig:evsa}(a)], 
$E$ is symmetric with respect to $t'/t=0$ owing to 
the electron-hole symmetry.\cite{YOTKT} 
$E({\rm SF})$ is always lower than $E({\rm N})$ and does not depend on  
$t'/t$ because $\langle{\cal H}_{t'}\rangle=0$ for any $t'/t$ and $U/t$. 
$E(d{\rm-SC})$ tends to increase as $|t'/t|$ increases. 
(When band renormalization is taken into account, $E(d{\rm-SC})$ also becomes 
constant.\cite{ISS,BR}) 
\par

In a slightly doped case [Fig.~\ref{fig:evsa}(b)], $E$ for every state 
becomes a decreasing function of $t'/t$, but the order of the energies does 
not change. 
$E({\rm SF})$ slightly depends on $t'$ and remains a linear function of $t'$.
However, for large $\delta$, the situation changes 
as shown in Fig.~\ref{fig:evsa}(c).
The range of $E({\rm SF})<E({\rm N})$ is restricted to 
$t'_{\rm c2}<t'<t'_{\rm c1}$, as indicated by arrows. 
This range becomes smaller when $U/t$ decreases [Fig.~\ref{fig:evsa}(d)]. 
We also find that this stable range of $\Psi_{\rm SF}$ becomes smaller 
as $\delta$ increases and finally vanishes at $\delta_{\rm SF}\sim 0.16$ 
($0.12$) for $U/t=12$ and $16$ ($8$). 
\par

\begin{figure*}[t!]
\begin{center}
\includegraphics[width=9cm,clip]{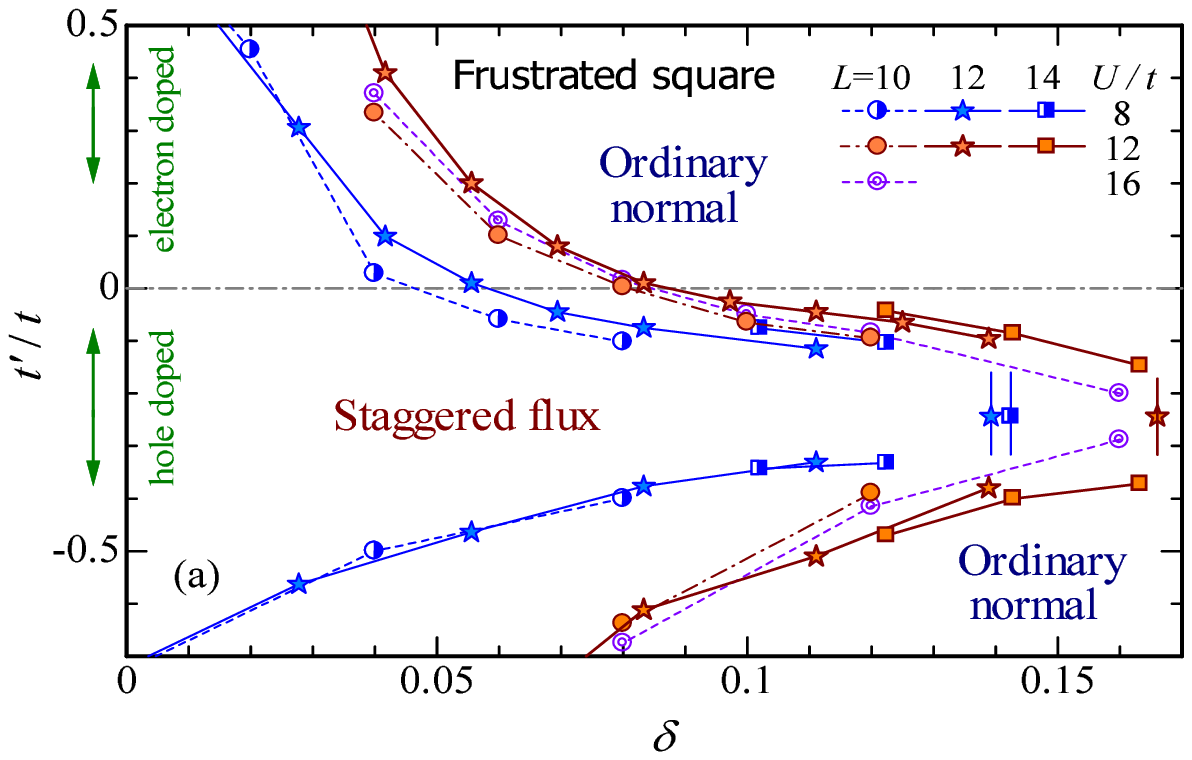}
\hskip 5mm
\includegraphics[width=6.9cm,clip]{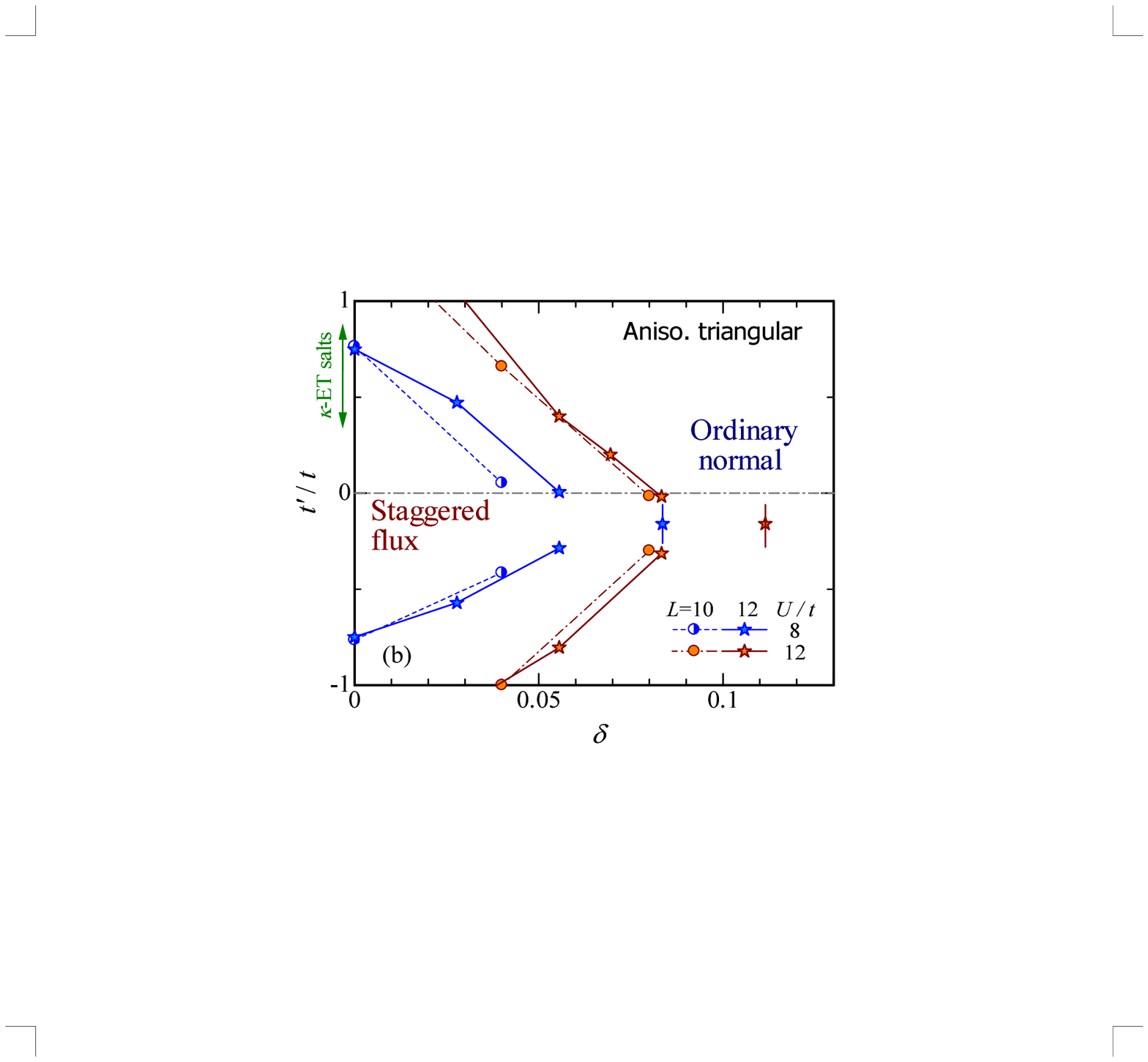}
\end{center} 
\vskip -3mm 
\caption{(Color online) 
Phase diagrams in the $\delta$-$t'$ plane with doping rate ($\delta$) and 
frustration strength ($t'/t$) for (a) the frustrated square lattice and (b) 
the anisotropic triangular lattice for $U/t=8$ and $12$ [and $16$ in (a)]. 
``Ordinary normal" indicates the projected Fermi sea. 
The symbols with vertical bars placed at $t'/t\sim -0.25$ indicate that 
the SF state is not stabilized at these values of $\delta$. 
The scale of the abscissa is identical in the two panels.
}
\vskip -3mm 
\label{fig:transpoint-jpsj} 
\end{figure*}
%
Obtaining similar data for different values of $\delta$, $t'/t$, and $L$, we 
construct a phase diagram, as shown in Fig.~\ref{fig:transpoint-jpsj}(a). 
The stable area of the SF state expands above the optimum doping of cuprates 
($\delta\sim 0.15$) for $-0.4\lesssim t'/t\lesssim -0.1$, which 
corresponds to the hole-doped cuprates. 
As $L$ increases, the area of $\Psi_{\rm SF}$ tends to expand slightly. 
For $t'/t>0$, on the other hand, the area of $\Psi_{\rm SF}$ shrinks to a 
very close vicinity of half filling, especially for $U/t=8$. 
\par 

\begin{figure}[htb]
\begin{center}
\includegraphics[width=7.5cm,clip]{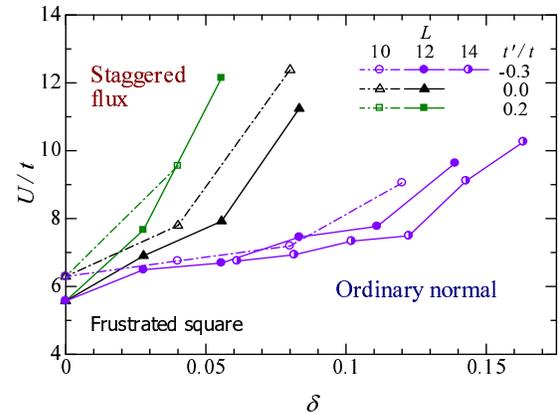}
\end{center} 
\vskip -3mm 
\caption{(Color online) 
Phase diagram between SF and normal states in $\delta$-$U$ plane 
for three values of $t'/t$.
The area of the SF phase may be somewhat smaller for $\delta\sim0$ because the 
band renormalization in $\Psi_{\rm N}$ is not considered here. 
}
\vskip -3mm 
\label{fig:tp-U} 
\end{figure}
%
It is useful to draw a phase diagram in the $\delta$-$U$ plane. 
Figure \ref{fig:tp-U} shows the region in which the SF state is stabilized 
for the cases with $t'/t=-0.3$, $0$, and $0.2$.
Irrespective of $t'/t$, the boundary value, $U_{\rm SF}/t$, 
increases as $\delta$ increases.\cite{note-USF-edge,noteOgata}
For $t'/t=-0.3$ (corresponding to the hole-doped case), 
the SF is stable in the whole underdoped regime ($\delta\lesssim 0.16$) 
for $U/t\gtrsim 10$. 
In contrast, for $t'/t=0.2$ (electron-doped case), $U_{\rm SF}/t$ rapidly 
increases as $\delta$ increases. 
\par

\begin{figure}[htb]
\begin{center}
\includegraphics[width=8.0cm,clip]{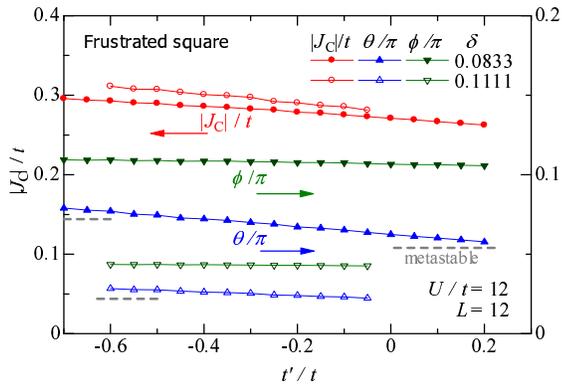} 
\end{center} 
\vskip -3mm 
\caption{(Color online) 
$t'/t$ dependences of the optimized phase parameters $\theta$ and $\phi$ 
and the local circular current $J_{\rm C}$ in $\Psi_{\rm SF}$ 
for two values of $\delta$. 
The unstable regions of $\Psi_{\rm SF}$ are indicated 
by gray dashed lines and ``metastable".
}
\vskip -3mm 
\label{fig:curr-fig} 
\end{figure}

This stability of $\Psi_{\rm SF}$ in a wide range of $\delta$ for 
$-0.4\lesssim t'/t\lesssim -0.1$ originates primarily from the large $t'/t$ 
dependence of $\Psi_{\rm N}$ and the very small $t'/t$ dependence 
of $\Psi_{\rm SF}$. 
This means that  the nature of $\Psi_{\rm SF}$ for $t'=0$ quantitatively 
remains that for $t'/t\ne 0$. 
For example, we show in Fig.~\ref{fig:curr-fig} the $t'/t$ dependences 
of the optimized phase parameters and local circular current, $J_{\rm C}$, 
which do not strongly depend on $t'/t$. 
Furthermore, we confirm that the momentum distribution function $n({\bf k})$ 
is almost the same for $t'/t\ne 0$. 
Note that this is in sharp contrast to the $d$-SC state, in which 
$n({\bf k})$ in the antinodal region markedly changes with $t'/t$ 
(see Fig.~29 in Ref.~\citen{YOTKT}). 
The reason for this difference between $\Psi_{\rm SF}$ and $\Psi_{d}$ 
will be as follows. 
Since $\Psi_{\rm SF}$ is very appropriately defined for the simple square 
lattice, 
$t'$ change the wave function of $\Psi_{\rm SF}$ only slightly.  
On the other hand, $\Psi_{d}$ has a gap opening at the Fermi surface 
near $(\pi,0)$, which is markedly affected by $t'$.
In this context, it is natural to expect that extra current, such as 
diagonal currents in chiral spin states,\cite{chiral} will not be 
favored\cite{SO}. 
\par

\begin{figure}[htb]
\begin{center}
\includegraphics[width=8.0cm,clip]{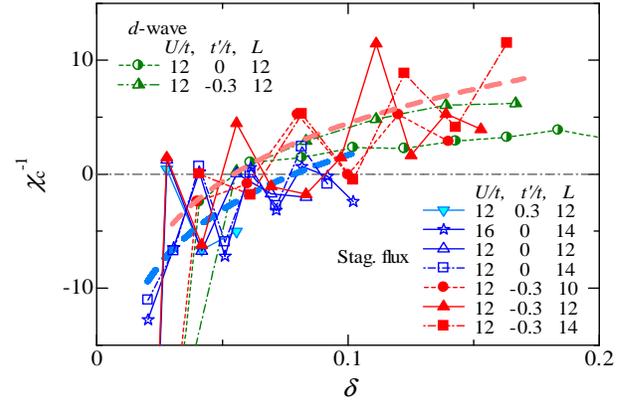} 
\end{center} 
\vskip -20mm 
\caption{(Color online) 
Inverse charge susceptibility of the SF state shown as a function of 
doping rate $\delta$ for some parameter sets for the frustrated square 
lattice [Fig.~\ref{fig:model}(a)]. 
We add guide lines (thick dashed) for $t'/t=0$ and $-0.3$ 
($U/t=12$, $L=14$).
The same quantity of the $d$-SC state is also shown with half-solid green 
symbols for comparison. 
Zigzags of the data for $\Psi_{\rm SF}$ are due to the discontinuous change 
in the occupied {\bf k} points and other finite-size effects. 
}
\vskip -3mm 
\label{fig:chi-n} 
\end{figure}
%
In many studies on the $t$-$J$ and Hubbard models, the instability toward 
phase separation near half filling has been discussed. 
Recently, states with AF long-range orders have been shown to be unstable 
toward phase separation for $t'/t\sim 0$ for the Hubbard model using 
the VMC method.\cite{YOTKT,Kobayashi,Misawa,BR,ISS-2014} 
For the $t$-$J$ model, an SF state has also been shown to be unstable toward 
phase separation in a wide range of $\delta$ for $t'/t=0$.\cite{Ivanov-PS} 
Therefore, we need to check this instability in the present case.
To this end, we consider the charge compressibility $\kappa$ or equivalently 
the charge susceptibility $\chi_{\rm c}$ $(=n^2\kappa)$, the inverse of which 
is given as 
\begin{equation} 
\chi_{\rm c}^{-1}=\frac{\partial^2E(n)}{\partial n^2} 
\sim N_{\rm s}^2\frac{E(N+4)+E(N-4)-2E(N)}{4^2}, 
\end{equation}
with $n=N/N_{\rm s}$. 
If $\chi_{\rm c}^{-1}<0$, the system is unstable toward phase separation. 
In Fig.~\ref{fig:chi-n}, we show the $\delta$ dependence of 
$\chi_{\rm c}^{-1}$ for three values of $t'/t$ and $L$. 
We find that $\chi_{\rm c}^{-1}$ is basically negative for $t'/t=0$ and 
$0.3$, indicating that $\Psi_{\rm SF}$ is unstable toward phase separation 
(data points of $\delta\sim 0$ should be disregarded because they are 
affected by the Mott singularity at $\delta\rightarrow 0$). 
This result is consistent with the previous one for the $t$-$J$ model for 
$t'/t=0$.\cite{Ivanov-PS} 
In contrast, for $t'/t=-0.3$, $\chi_{\rm c}^{-1}$ becomes positive 
for $\delta\gtrsim 0.05$ and comparable to that of $d$-SC (green symbols). 
Therefore, a homogeneous SF state is possible for the parameters of 
hole-doped cuprates. 
\par

\subsection{Anisotropic triangular lattice\label{sec:ATA}}
\begin{figure}[htb]
\begin{center}
\vskip 2mm
\includegraphics[width=7.5cm,clip]{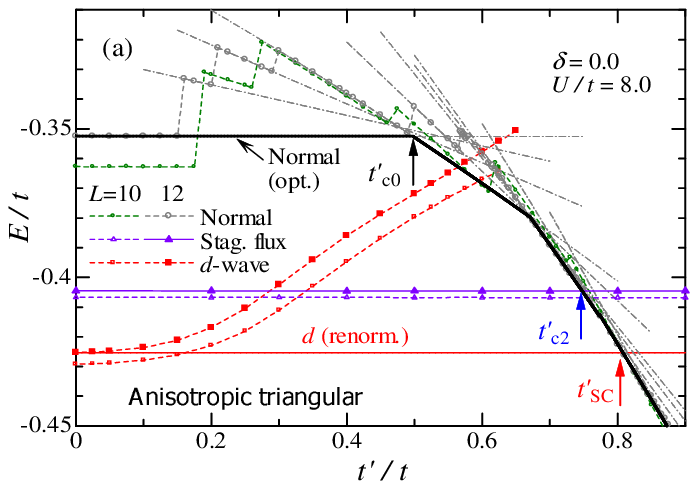} 
\vskip -2mm
\hskip -2mm
\includegraphics[width=7.5cm,clip]{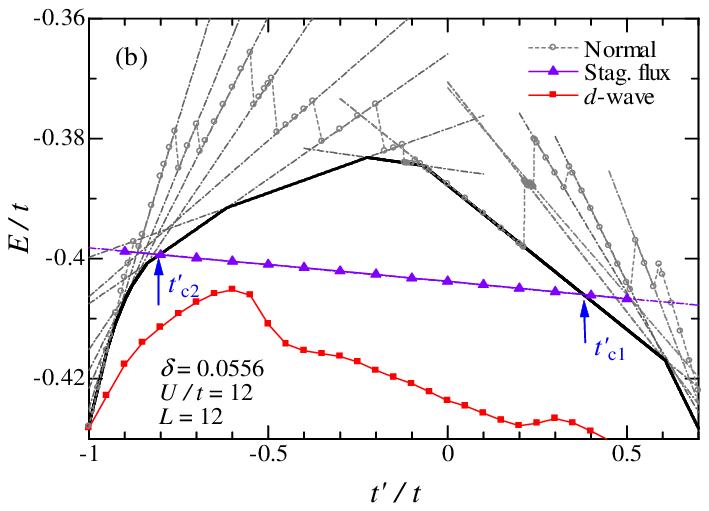} 
\end{center} 
\vskip -5mm 
\caption{(Color online) 
Comparison of total energy among $\Psi_{\rm N}$, $\Psi_{\rm SF}$, and $\Psi_d$
as functions of $t'/t$. 
Panel (a) displays the half-filled case for $U/t=8$ (only the data for 
$t'/t\ge 0$ are shown) and $L=10$ and $12$, and (b) a doped case for $U/t=12$ 
($\delta=0.0556$) and $L=12$. 
We show the optimized values by the band renormalization for 
$\Psi_{\rm N}$ with (black) solid lines. 
The band-renormalized data for $\Psi_d$ are also shown in (a) with a (red) 
solid line. 
}
\vskip -5mm 
\label{fig:evsam} 
\end{figure}

For the anisotropic triangular lattice, Fig.~\ref{fig:evsam} summarizes 
the total energies of $\Psi_{\rm SF}$, $\Psi_{\rm N}$, and $\Psi_d$ 
as a function of $t'/t$. 
Again, $E({\rm N})$, the lowest energy of $\Psi_{\rm N}$ considering 
band renormalization\cite{BR}, is shown by the black solid lines. 
First, let us consider the half-filled case [Fig.~\ref{fig:evsam}(a)]. 
$E$ is symmetric with respect to $t'/t=0$ owing to the electron-hole 
symmetry.\cite{YOTKT} 
Similarly to the case on the frustrated square lattice, $E({\rm FS})$ and 
$E(d{\rm-SC})$ (if the band renormalization is 
considered\cite{Watanabe-PRB,Tocchio-ET2}) are constant. 
For the ground state, we find from Fig.~\ref{fig:evsam}(a) that 
$\Psi_{\rm N}$ has a lower energy than $\Psi_d$ for $t'>t'_{\rm SC}$ with 
$t'_{\rm SC}/t=0.807$ for $L=12$.
However, the $(\pi,\pi)$-AF state or incommensurate AF states including 
the case of a 120$^\circ$ structure has a lower energy at 
half filling.\cite{Watanabe-PRB,Tocchio-ET,Tocchio-ET2} 
\par

\begin{figure}[htb]
\begin{center}
\includegraphics[width=7.5cm,clip]{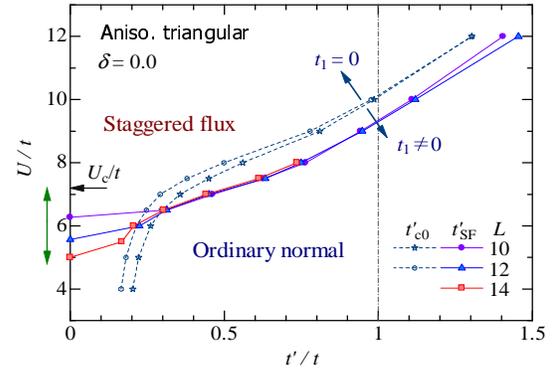} 
\end{center} 
\vskip -27mm 
\caption{(Color online) 
Phase diagram between the SF and normal states in $t'$-$U$ plane at half 
filling on anisotropic triangular lattice. 
Boundaries ($U_{\rm SF}$, solid lines) are determined for three 
system sizes. 
The green arrow near the vertical axis represents the range of the metallic 
SF state. 
The dashed lines show the boundary in $\Psi_{\rm N}$ regarding whether the 
nesting condition is restored ($t_1=0$) or not  ($t_1\ne 0$) in the 
renormalized band (see Sect.~\ref{sec:gain} later). 
}
\vskip -3mm 
\label{fig:tt-U} 
\end{figure}

%
We compare the energies between $\Psi_{\rm N}$ and $\Psi_{\rm SF}$, assuming 
that AF states are not stabilized.
In Fig.~\ref{fig:evsam}(a), $\Psi_{\rm SF}$ is more stable than 
$\Psi_{\rm N}$ for $t'<t'_{\rm c2}$ with $t'_{\rm c2}/t=0.749$ ($0.763$) 
for $U/t=8$ and $L=12$ ($10$). 
We employ the value $U/t=8$ simply because it is frequently used as a 
plausible value for $\kappa$-ET salts.\cite{Organic12} 
Actually, $\Psi_{\rm SF}$ is Mott insulating at $U/t=8$; a value of 
$U/t\lesssim 7.1$ is necessary for a metallic state. 
However, the point here does not change qualitatively irrespective of whether 
$\Psi_{\rm SF}$ is insulating or metallic.  
On the basis of similar calculations for various values of $U/t$ 
and $t'/t$, we construct a phase diagram within the SF and normal state 
at half filling (Fig.~\ref{fig:tt-U}), which is relevant for organic 
conductors. 
The boundary value $U_{\rm SF}/t$ between $\Psi_{\rm SF}$ and 
$\Psi_{\rm N}$ increases as $t'/t$ increases. 
\par

As shown before, the Mott transition occurs at $U_{\rm c}\sim 7.1t$ for 
$t'=0$.
Since the properties of $\Psi_{\rm SF}$ are similar to the case of $t'=0$, 
the SF state is metallic for $U\lesssim U_{\rm c}$ 
and insulating for $U>U_{\rm c}$. 
This phase boundary is also shown in Fig.~\ref{fig:tt-U}.
\par 

%
For a doped case, we show in Fig.~\ref{fig:evsam}(b) the $t'/t$ 
dependence of the total energy for the three states for typical parameters. 
It is noteworthy that $E/t$ for $\Psi_{\rm N}$ and $\Psi_d$ decreases rapidly 
for large values of $|t'/t|$ ($\sim 1$). 
Obtaining similar data for various values of $t'/t$ and $\delta$, 
we construct a phase diagram in the $\delta$-$t'$ 
space [Fig.~\ref{fig:transpoint-jpsj}(b)]. 
Compared with the case of the frustrated square lattice 
[Fig.~\ref{fig:transpoint-jpsj}(a)], the area of $\Psi_{\rm SF}$ is 
restricted to the small doping region.
\par

\section{Discussion\label{sec:discussions}}

\subsection{Phase cancelation mechanism\label{sec:phase-cancel}}

\begin{figure}[htb]
\begin{center}
\includegraphics[width=7.5cm,clip]{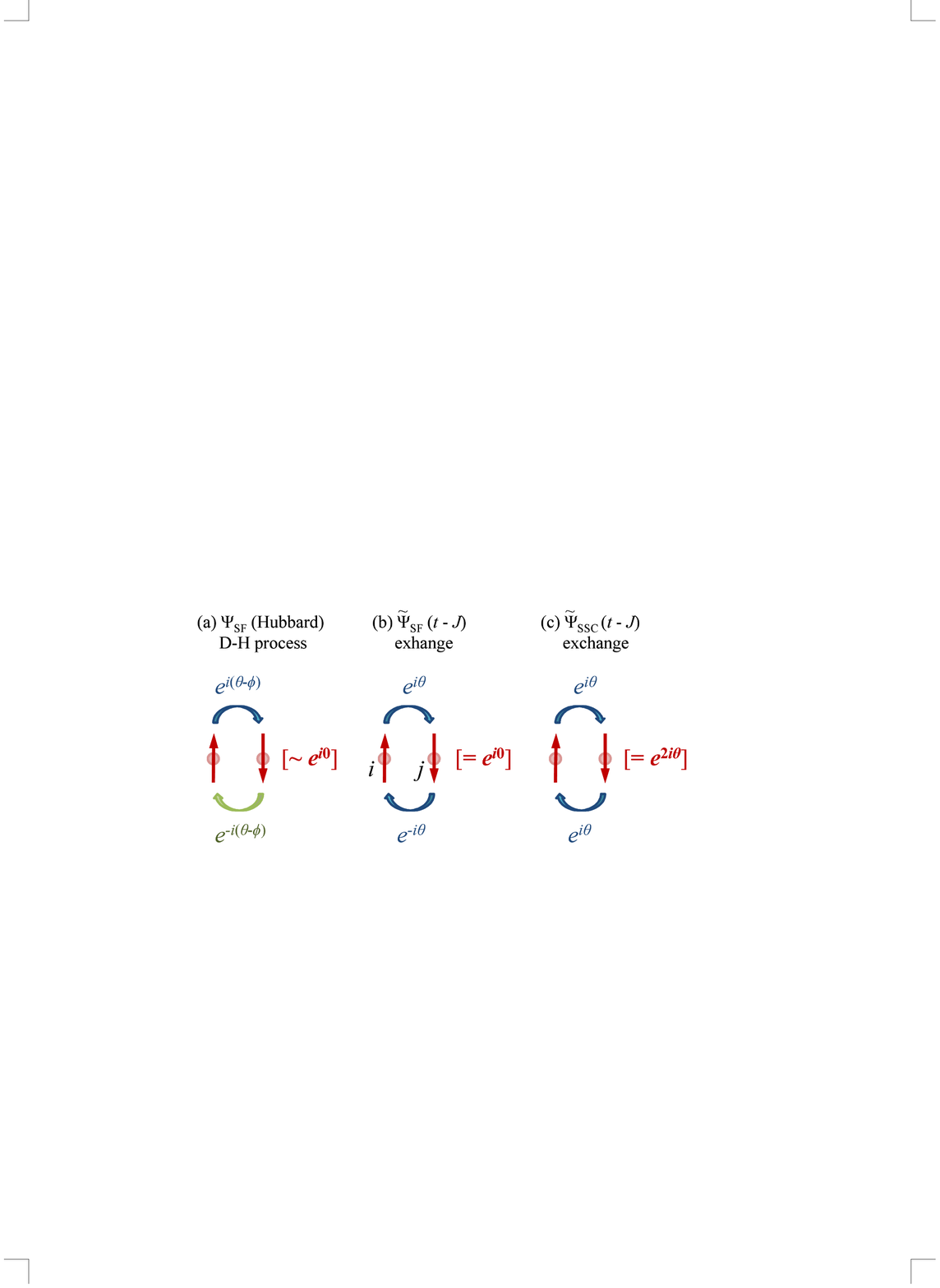}
\end{center} 
\vskip -3mm 
\caption{(Color online) 
Illustration of phase factors added in 
(a) creation or annihilation process of doublon-holon (D-H) pair in 
$\Psi_{\rm SF}$ for large-$U/t$ Hubbard model, 
(b) spin exchange process in $\tilde\Psi_{\rm SF}$ for $t$-$J$ model, and 
(c) spin exchange process in $\tilde\Psi_{\rm SSC}$ for $t$-$J$ model. 
For details, see text.
}
\vskip -3mm 
\label{fig:SSC} 
\end{figure}

First, let us consider intuitively why $\Psi_{\rm SF}$ has a low energy 
in the strongly correlated region of the Hubbard model. 
As discussed in a previous study\cite{YOTKT}, the processes corresponding to 
the $J$ term in the $t$-$J$ model are 
those in which a D-H pair 
is created or annihilated as shown in Fig.~\ref{fig:SSC}(a). 
Generally speaking, the phase yielded in this process causes 
a loss of  kinetic energy. 
In order to reduce this kinetic energy loss, the phase $\theta$ should 
be eliminated by the phase $\phi$ by applying ${\cal P}_\theta$ with 
$\phi\sim\theta$ in the same manner as introduced in Ref.~\citen{Drude}.
\par

We expect a similar phenomenon in the Heisenberg interaction in 
the $t$-$J$ model.
In the $J$ term, 
an $\uparrow$ spin at site $i$ hops to site $j$ ($=i+\tau$) and 
simultaneously a $\downarrow$ spin at site $j$ hops to site $i$. 
As shown in Fig.~\ref{fig:SSC}(b), if the former hopping yields a phase 
$\theta$, the latter yields $-\theta$ in $\Phi_{\rm SF}$; the total phase 
in the exchange process precisely cancels out (shown in the square 
brackets in Fig.~\ref{fig:SSC}).
Since the two processes occur simultaneously, it is unnecessary to 
introduce $\phi$ in the $t$-$J$ model to stabilize the SF state.
On the other hand, in the Hubbard model [Fig.~\ref{fig:SSC}(a)], 
a hopping resulting in D-H-pair annihilation does not necessarily occur 
immediately after a D-H pair is created; these two processes are mutually 
independent events. 
Therefore, it is necessary to introduce the phase $\phi$ 
to eliminate $\pm\theta$ in each process in order to stabilize the SF state. 
\par

Finally, let us apply the present mechanism to the spin-current-carrying 
state. 
Staggered spin current (SSC) states (or sometimes called spin-nematic 
states\cite{Nersesyan})
have been considered to be candidates for hidden orders in various 
systems.\cite{SSC} 
In these states, counter-rotating currents of $\uparrow$ and $\downarrow$ 
spins alternately flow in each plaquette [Fig.~\ref{fig:SF-figA}(b)].
We have carried out similar VMC calculations for the projected SSC state 
$\tilde\Psi_{\rm SSC}={\cal P}_{\rm G}(0)\Phi_{\rm SSC}$.
The results are summarized in Appendix\ref{sec:SSC}.
We conclude that $\tilde\Psi_{\rm SSC}$ is not stabilized for any $J/t$ 
and underdoped $\delta$. 
We can easily see the reason for this by considering the phase cancelation. 
As we can see from Fig.~\ref{fig:SSC}(c), the total phase added 
in an exchange process in $\tilde\Psi_{\rm SSC}$ remains $2\theta$. 
We found that this phase is difficult to eliminate by 
configuration-dependent phase factors such as ${\cal P}_\phi$. 
Therefore, we conclude that the SSC state or spin-nematic state will 
never be stabilized. 
\par

\subsection{Kinetic energy gain\label{sec:gain}}
\begin{figure}[htb]
\begin{center}
\includegraphics[width=7.5cm,clip]{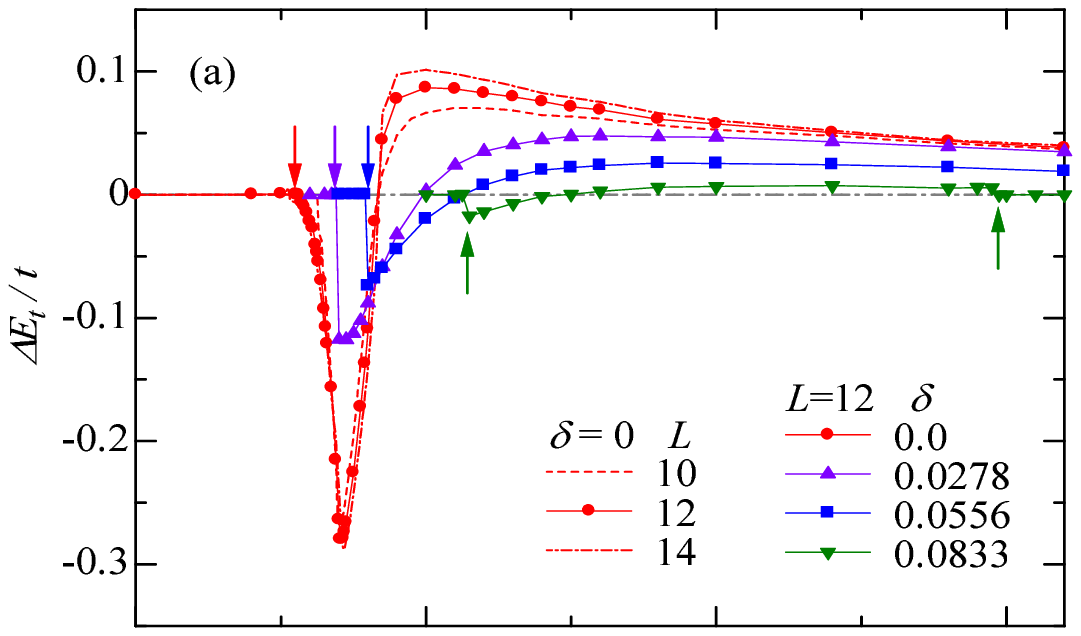} 
\vskip -1mm 
\includegraphics[width=7.4cm,clip]{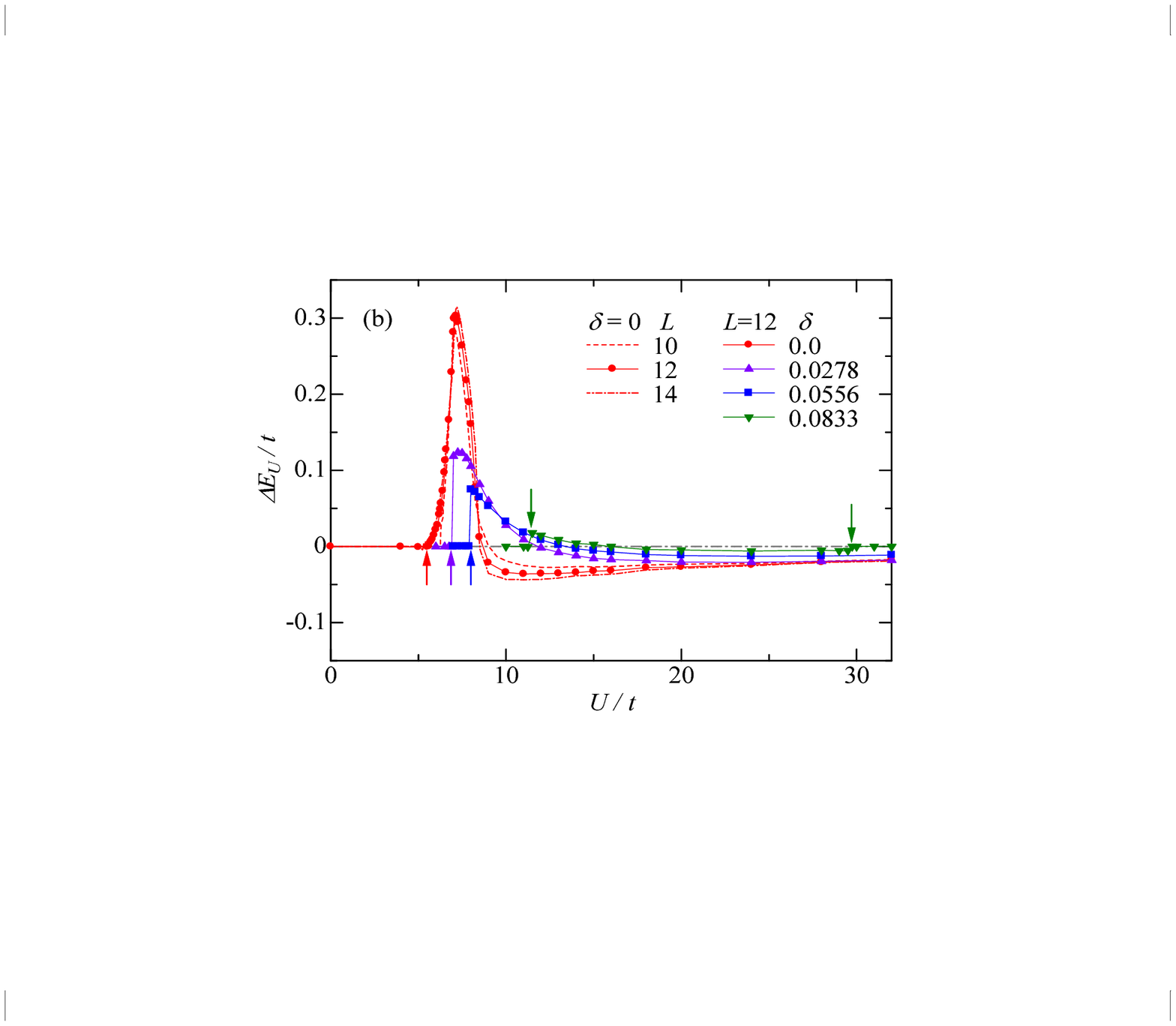} 
\end{center} 
\vskip -3mm 
\caption{(Color online) 
The two components of the energy difference between the SF state and the 
projected Fermi sea are shown for some doping rates ($t'/t=0$, $U/t=12$): 
(a) kinetic energy and 
(b) interaction energy parts. 
For $\delta=0$, we add data for $L=10$ and $14$. 
The arrows indicate $U_{\rm SF}/t$ for $L=12$. 
}
\vskip -5mm 
\label{fig:E-comp-d0} 
\end{figure}

We discuss another physical reason for the stabilization of the SF state.
In Fig.~\ref{fig:E-comp-d0}. we show the difference in the kinetic energy 
$\Delta E_t$ and interaction energy $\Delta E_U$ between 
the optimal SF state and the projected Fermi sea, 
\begin{eqnarray}
\Delta E_t&=&E_t({\rm N})-E_t({\rm SF}),  \cr
\Delta E_U&=&E_U({\rm N})-E_U({\rm SF}), 
\label{eq:gain-n-d}
\end{eqnarray}
for four values of $\delta$. 
In previous papers, we perfomed the same analysis for $\Psi_d$ and 
$\Psi_{\rm N}$\cite{YTOT,YOT}, and showed that the SC transition is driven 
by the kinetic energy gain for $U\gtrsim U_{\rm co}$ with $U_{\rm co}/t$ 
being the crossover value from weakly to strongly correlated regimes. 
In Fig.~\ref{fig:E-comp-d0}, we find that a similar phenomenon emerges 
between $\Psi_{\rm SF}$ and $\Psi_{\rm N}$: 
Kinetic energy gain occurs in the strongly correlated region. 
The physical reason for this will be as follows. 
In the strongly correlated regime, the kinetic energy is dominated by the 
D-H pair creation or annihilation processes (not shown).
Since the phases arising in these processes are canceled out by $\phi$, 
this kinetic energy gain corresponds to that in the $J$-term in the 
$t$-$J$ model.
\par 

%
\begin{figure}[htb]
\begin{center}
\vskip 4mm
\includegraphics[width=7.0cm,clip]{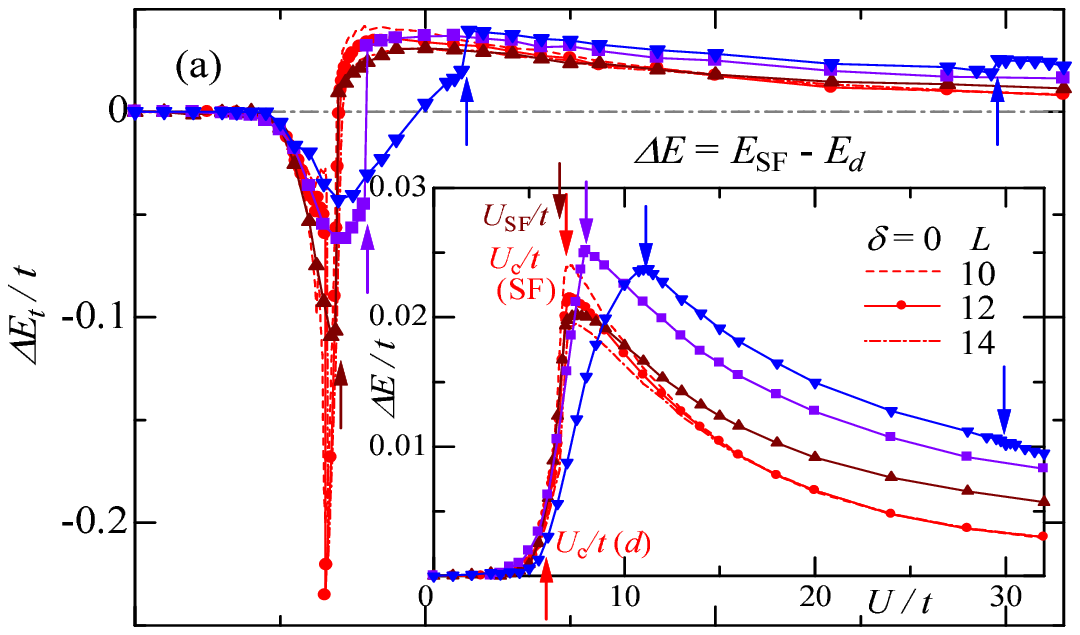}
\vskip -1mm
\includegraphics[width=7.0cm,clip]{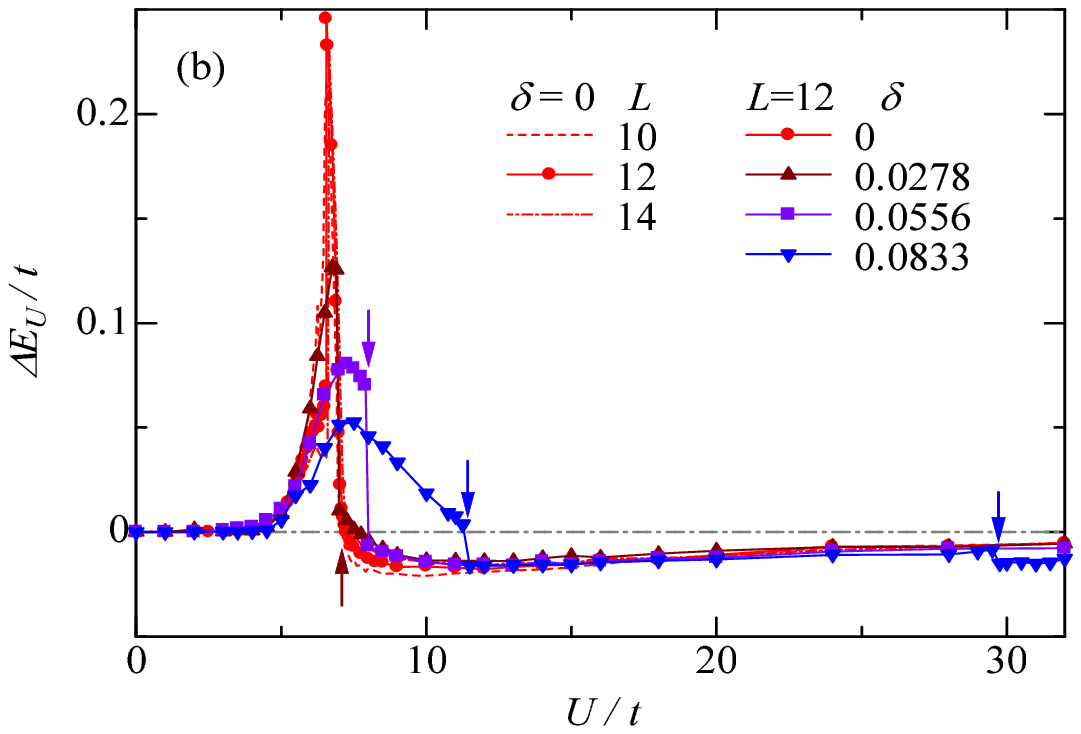} 
\end{center} 
\vskip -3mm 
\caption{(Color online) 
The two components of the energy difference between the $d$-SC and 
SF states are shown for some doping rates ($t'/t=0$, $L=12$): 
(a) kinetic energy and (b) interaction energy parts. 
The symbols are common to all panels. 
For $\delta=0$, data for $L=10$ and $14$ are added. 
The arrows indicate $U_{\rm SF}/t$ in the SF state for $\delta>0$. 
The inset in (a) shows the difference in total energy  
($\Delta E=\Delta E_t+\Delta E_U$) for four values of $\delta$. 
}
\vskip -5mm 
\label{fig:E-comp-d} 
\end{figure}
%
In Fig.~\ref{fig:E-comp-d}. we show a similar comparison between the $d$-SC 
and optimal SF states, i.e.,
\begin{eqnarray}
\Delta E_t&=&E_t({\rm SF})-E_t(d{\rm-SC}), \cr 
\Delta E_U&=&E_U({\rm SF})-E_U(d{\rm-SC}). 
\label{eq:gain-n-d}
\end{eqnarray}
In particular, in the regime of $U>U_{\rm c}({\rm SF})$ at half filling and 
$U>U_{\rm SF}$ 
for $\delta>0$, the energy gain occurs exclusively in the kinetic part 
($\Delta E_t>0$ and $\Delta E_U<0$). 
Thus, the cause of stabilization both in 
$\Psi_{\rm N}\rightarrow \Psi_{\rm SF}$ and in 
$\Psi_{\rm SF}\rightarrow \Psi_{d}$ is the kinetic energy gain for a 
sufficiently large $U/t$.\cite{note-kinetic} 
\par

\subsection{Comparison with experiments}
(i) {\it High-$T_{\rm c}$ cuprates}: 
Here, we discuss the lattice translational symmetry, 
which is broken in the present SF state. 
The peaks arising from local loop currents in the polarized neutron 
scattering spectra are found at 
${\bf k}=(0,0)$,\cite{mag-order1,mag-order2,mag-order3} suggesting that 
the lattice translational symmetry is preserved in the pseudogap phase. 
Some authors have argued that the SF state breaks this symmetry, but physical 
quantities calculated with SF states display a $(0,0)$ peak in addition 
to a ($\pi,\pi$) peak.\cite{Sau}
The above neutron experiments appear to be consistent with more complicated 
circular-current states that do not break this 
symmetry.\cite{Varma,Weber,Weber2} 
Recently, however, one of the authors showed that this type of 
circular-current state is not stabilized with respect to the normal state 
in a wide range of the model parameters on the basis of systematic VMC 
calculations with refined wave functions for $d$-$p$-type models. 
Instead, SF states are stabilized in some parameter ranges.\cite{Thesis} 
On the other hand, the shadow bands observed in the ARPES 
spectra,\cite{SB1,SB2,SB3,SB4} 
which also characterize the pseudogap phase, seem to require the scattering of 
$(\pi,\pi)$ and a folded Brillouin zone. 
Therefore, the issue of translational symmetry breaking is still 
controversial. 
\par

(ii) {\it Organic conductors}:
In Sect.~\ref{sec:ATA}, we studied the anisotropic triangular lattice.
Let us here discuss the relevance of the present results to experiments.
As discussed in Sect.~1, 
deuterated $\kappa$-(ET)$_2$Cu[N(CN)$_2$]Br has $t'/t\sim 0.4$.
The present results show that the SF state is 
stabilized for the case of $t'/t\sim 0.4$.
Therefore, the pseudogap behavior for $T>T_{\rm c}$ observed in deuterated 
$\kappa$-(ET)$_2$Cu[N(CN)$_2$]Br is probably caused by the SF state.
On the other hand, $\kappa$-(ET)$_2$Cu$_2$(CN)$_3$ with $t'/t\sim 0.8$ 
shows Fermi-liquid-like behavior above $T_{\rm c}$.
Since the present result shows that the SF state is not stabilized 
for the case of $t'/t\sim 0.8$, the normal state of 
$\kappa$-(ET)$_2$Cu$_2$(CN)$_3$ is naturally understood on the basis of 
$\Psi_{\rm N}$. 
Although our results are consistent with experiments, quantitative discussions 
will be necessary to determine the effective value of $U/t$ 
as well as $t'/t$ more accurately for each 
compound.\cite{note-flow,Tamura-SSC}. 
\par

For the organic conductors with finite doping, we find that 
$\Psi_{\rm SF}$ is not stabilized at $\delta=0.11$ for both $U/t=8$ and $12$, 
regardless of the value of $t'/t$. 
Thus, concerning the pseudogap phenomena found in a doped $\kappa$-ET 
salt\cite{O9}, we cannot conclude that the SF state is a candidate for the 
pseudogap phase. 
Other factors may be necessary to understand this pseudogap. 
\par

\subsection{Related studies and coexistence with other 
orders\label{sec:related}}
A decade ago, Yang, Rice, and Zhang introduced a phenomenological Green's 
function that can represent various anomalous properties of the pseudogap 
phase.\cite{YangRice}
Their Green's function contains a self-energy that reproduces the 
$d_{x^2-y^2}$-wave RVB state at half filling. 
For finite doping, the Green's function is assumed to have the same 
self-energy but without the features of SC. 
On the other hand, the SF state used in the present paper is also 
connected to the $d_{x^2-y^2}$-wave RVB state due to the SU(2) 
symmetry at half filling. 
For finite doping, however, the SF state does not show SC. 
Therefore, we expect a close relationship between the present SF state 
and the phenomenological Green's function, although the explicit 
correspondence is not known.
\par

As mentioned in Sect.~\ref{sec:intro}, an AF state was recently studied 
by applying a VMC method with a band-renormalization effect to the Hubbard 
model on the frustrated square lattice.\cite{BR} 
It revealed that the AF state is considerably stable and occupies 
a wide range of the ground-state phase diagram. 
In doped metallic cases for $t'/t\lesssim -0.05$, an AF state called 
type-(ii) AF state is stabilized, while for $t'/t\gtrsim -0.05$, a type-(i) 
AF state is stabilized. 
In a type-(ii) AF state, a pocket Fermi surface arises around 
$(\pi/2,\pi/2)$ and a gap opens in the antinode [near $(\pi,0)$]. 
As $\delta$ increases, the Fermi surface around $(\pi/2,\pi/2)$ extends 
toward the antinodes along the AF Brillouin zone edge. 
Such behavior resembles the pseudogap phenomena, as the SF state treated 
in this paper does. 
Thus, if such features are preserved when the AF long-range order is 
broken into a short-range order for some reason, as actually observed in 
cuprates,\cite{AF-SRO} a (disordered) type-(ii) AF state becomes another 
candidate for a pseudogap state, although the symmetry breaking is rather 
different from that in the SF state. 
\par

Let us discuss the coexistence with $d$-SC. 
The same study\cite{BR} as discussed above showed that, although type-(ii) 
AF states do not coexist with $d$-SC, metallic AF states for 
$t'/t\gtrsim -0.05$ 
[called type-(i) AF] coexist with $d$-SC; these type-(i) AF states have
pocket Fermi surfaces in the antinodes. 
This corroborates the fact that the electron scattering of ${\bf q}=(\pi,\pi)$ 
that connects two antinodes is crucial for the appearance of $d$-SC. 
From this result, we expect that the SF state is unlikely to coexist with 
$d$-SC because gaps open in the antinodes in the SF state, as shown 
in Fig.~\ref{fig:cfnk3d-jpsj}(d). 
As an exception, coexistence may be possible for 
$\delta\sim\delta_{\rm SF}$, where the Fermi surfaces extend to the antinodes, 
as discussed in Ref.~\citen{BR}. 
Thus, the SF order probably competes with the $d$-SC order rather than 
underlies it.\cite{note-coexist}
We need to directly confirm this by examining a mixed state of the SF and 
$d$-SC orders. 
\par 

Finally, we consider the possibility of the coexistence of AF and SF orders. 
Recently, a Hubbard model with an SF phase, namely, 
${\cal H}={\cal H}^{\rm SF}+{\cal H}_U$ [see Eqs.~(\ref{eq:Hamil}) and 
(\ref{eq:HamilSF})], was studied using a VMC method with a mixed state of 
SF and AF orders, $\Psi_{\rm SF+AF}$.\cite{Toga-Fermion} 
For $\theta=0$ [Eq.~(\ref{eq:Hamil}) with $t'=0$], the optimized 
$\Psi_{\rm SF+AF}$ is reduced to $\Psi_{\rm AF}$, which belongs to 
the type-(i) AF phase. 
Namely, the SF order is excluded by the type-(i) AF order. 
This is probably because the AF order is energetically dominant 
over the SF order, and the loci of Fermi surfaces compete with each other. 
\par

\section{Conclusions\label{sec:conclusions}}
In this paper, we studied the stability and other properties of 
the staggered flux (SF) state in the two-dimensional Hubbard model at and 
near half filling.
We carried out systematic computations for $U/t$, $t'/t$, and $\delta$, 
using a variational Monte Carlo method, which is useful for treating 
correlated systems. 
In the trial SF state, a configuration-dependent phase factor was 
introduced, which is vital to treat a current-carrying state in the 
regime of Mott physics. 
In this SF state, we found a good possibility of explaining the pseudogap 
phenomena in high-$T_{\rm c}$ cuprates and $\kappa$-ET salts. 
The main results are summarized as follows: 
\par

(1) The SF state is not stabilized in a weakly correlated regime 
($U/t\lesssim 5$), but becomes considerably stable in a strongly correlated 
regime [Figs.~\ref{fig:delEtot-jpsj} and \ref{fig:delEtotvsn.eps}(a)]. 
The physical properties in the latter regime are consistent with those of 
the $t$-$J$ model.\cite{Liang,TKLee,YO,Ivanov} 
The transition from $\Psi_{\rm N}$ to $\Psi_{\rm SF}$ 
at $U_{\rm SF}/t$ is probably continuous. 
\par

(2) At half filling ($\delta=0$), the SF state becomes Mott insulating for 
$U>U_{\rm c}\sim 7t$. 
A metallic SF state is realized for $U_{\rm SF}<U<U_{\rm c}$, which is 
gapless in the charge degree of freedom but gapped in the spin sector.
This gap behavior of the metallic SF state at $\delta=0$ continues to the 
doped cases of $U>U_{\rm SF}$. 
However, it is distinct from the case of the noninteracting SF state 
$\Phi_{\rm SF}$ in the sense that spin-charge separation occurs. 
In doped cases, $\Psi_{\rm SF}$ has a segmentary Fermi surface near the 
nodal point $(\pi/2,\pi/2)$ but is gapped near the antinodal $(\pi,0)$ 
(Fig.~\ref{fig:cfnk3d-jpsj}). 
By analyzing the kinetic energies ($E_t$), we found that 
$E_t(\Psi_d)<E_t(\Psi_{\rm SF})<E_t(\Psi_{\rm N})$ for a large $U/t$, 
meaning that a kinetic-energy-driven SC takes place even if we assume that 
the SF state is realized above $T_{\rm c}$. 
\par

(3) Although $\Psi_{\rm SF}$ is unstable toward phase separation for 
$t'/t\sim 0$ in accordance with the feature in the $t$-$J$ 
model,\cite{Ivanov-PS} $\Psi_{\rm SF}$ restores stability against 
inhomogeneity for $t'/t\sim -0.3$. 
This aspect is similar to that of AF states.\cite{YOTKT,Misawa,BR} 
\par

(4) For the simple square lattice ($t'/t=0$), the stable SF area is 
$\delta\lesssim 0.1$. 
In the anisotropic triangular lattice ($|t'/t|>0$), this area does not 
expand. 
In the frustrated square lattice, however, the $t'$ term makes this 
area expand to $\delta\lesssim 0.16$ for $-0.4\lesssim t'/t\lesssim -0.1$ 
(hole-doped cases) but shrink to a very close vicinity of half filling 
for $t'/t>0$ (electron-doped cases) (Fig.~\ref{fig:transpoint-jpsj}). 
This change is mostly caused by the sensitivity of $\Psi_{\rm N}$ to $t'$, 
while $\Psi_{\rm SF}$ is insensitive to $t'$ because it is defined 
suitably for the square-lattice plaquettes. 
This result may be related to the fact that pseudogap behavior is not 
clearly observed in electron-doped 
cuprates.\cite{note-ED,Armitage,Onose,Matsui0,Matsui,Horio} 
\par

(5) On the basis of this study and another study,\cite{BR} we argue that 
the SF state does not coexist with $d$-SC as a homogeneous state and 
is not an underlying normal state from which $d$-SC arises. 
This is because the SF state has no Fermi surface in the antinodes necessary 
for generating $d$-SC. 
However, when the optimized $\theta$ becomes small (for 
$\delta\sim\delta_{\rm SF}$), coexistence is possible. 
The coexistence of SF and AF orders also does not occur for 
$t'/t\gtrsim -0.05$;\cite{Toga-Fermion} further study is needed to clarify 
the cases of $t'/t\lesssim -0.05$. 
\par

(6) The local circular current in a plaquette, which is an order parameter 
of the SF phase, is strongly suppressed in the large-$U/t$ region, but it 
does not vanish even in the insulating phase. 
A so-called chiral Mott insulator is realized.
\par

(7) We showed that the spin current state (or spin-nematic state) is not
stabilized for the $t$-$J$ and Hubbard models. 
\par

Because these results are mostly consistent with the behaviors in the 
pseudogap phase of cuprates, the SF state should be reconsidered as a 
candidate for the anomalous `normal state' competing with $d$-SC in the 
underdoped regime. 
Note that the AF state is considerably stabilized in a wide region of the 
Hubbard model.\cite{BR}
Therefore, possible disordered AF [type-(ii)] is another candidate for the
pseudogap phase, although the symmetry breaking is different. 
Besides this claim, there are relevant subjects left for future studies. 
(i) What will be the phase transition between the SF and $d$-SC states 
if the SF state is the state above $T_{\rm c}$? 
(ii) In this study, we treated the SF and $d$-SC states independently. 
However, it is important to check directly whether the two orders 
coexist,\cite{coexist1,Uykur,coexist3} and how the coexistent state 
behaves, if it exists.\cite{TKLee,Laughlin}
(iii) In this study, we introduced a phase factor for the doublon-holon 
processes. 
It will be worthwhile to search a useful phase factor that controls isolated 
(doped) holons for $\delta>0$. 
(iv) It will be intriguing to search for a low-lying circular-current state 
other than the SF state and the state proposed by 
Varma\cite{Varma,Weber,Weber2}. 
\par

\begin{acknowledgment}
We thank Yuta Toga, Ryo Sato, Tsutomu Watanabe, Hiroki Tsuchiura, and 
Yukio Tanaka for useful discussions and information. 
This work was supported in part by Grants-in-Aid from the Ministry of 
Education, Culture, Sports, Science and Technology. 
\par
\end{acknowledgment}

\appendix
%
\section{Details of Bare Staggered Flux State\label{sec:nonintSF}}
%
\begin{figure}[htb]
\begin{center}
\includegraphics[width=8.5cm,clip]{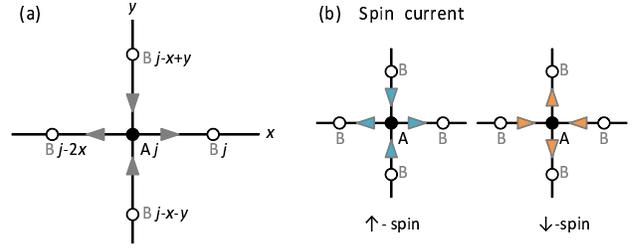}
\end{center} 
\vskip -3mm 
\caption{(Color online) 
(a) Coordinates in the extended unit cell used in ${\cal H}_{\rm SF}$. 
(b) Currents in staggered spin current state for up and down spins 
discussed in Appendix\ref{sec:SSC}. 
The arrows denote the directions of easy flow. 
}
\vskip -3mm 
\label{fig:SF-figA} 
\end{figure}
%
In this Appendix, we give a definition of the one-body SF state 
$\Phi_{\rm SF}$ used in this study (see Sect.~\ref{sec:wf}) and summarize 
its characteristic properties. 
$\Phi_{\rm SF}$ is the ground state of a noninteracting SF Hamiltonian 
${\cal H}^{\rm SF}(\theta)$ written as 
\begin{eqnarray}
&&{\cal H}^{\rm SF}=-t\sum_{j\in A,\sigma}
\Bigl[
 e^{i\theta}\left(c^\dag_{{\rm A}j,\sigma}c_{{\rm B}j,\sigma}
+c^\dag_{{\rm A}j,\sigma}c_{{\rm B}j-2{\bf x},\sigma}\right)
\nonumber\\
&&
+e^{-i\theta}\left(c^\dag_{{\rm A}j,\sigma}c_{{\rm B}j-{\bf x}+{\bf y},\sigma}
+c^\dag_{{\rm A}j,\sigma}c_{{\rm B}j-{\bf x}-{\bf y},\sigma}\right)
+{\rm H.c.}
\Bigr] \qquad\qquad
\label{eq:HamilSF}
\end{eqnarray}
in the sublattice (A,B) representation [see Fig.~\ref{fig:SF-figA}(a)]. 
Here, we abbreviate ${\bf r}_j$ (the position of site $j$) as $j$, and 
${\bf x}$ (${\bf y}$) is the unit vector in the $x$ ($y$) direction. 
For $\theta=0$, ${\cal H}^{\rm SF}$ is reduced to ${\cal H}_t$ 
in Eq.~(\ref{eq:Hamil}).
${\cal H}^{\rm SF}$ is diagonalized as 
\begin{equation}
{\cal H}^{\rm SF}=\sum_{{\bf k},\sigma}
\left[
 E^{\rm SF}_-({\bf k})\ \alpha^\dag_{{\bf k}\sigma}\alpha_{{\bf k}\sigma}
+E^{\rm SF}_+({\bf k})\ \beta^\dag_{{\bf k}\sigma}\beta_{{\bf k}\sigma}
\right], 
\end{equation}
with the band dispersions given as
\begin{equation}
E^{\rm SF}_\pm({\bf k})=\pm 2t\ {\cal S}_{\theta,{\bf k}}, 
\label{eq:EkSF}
\end{equation}
by applying the unitary transformation 
\begin{equation}
\left(
\begin{array}{c}
c_{{\rm A}{\bf k}\sigma} \\
c_{{\rm B}{\bf k}\sigma}
\end{array}
\right)
=\frac{1}{\sqrt{2}}
\left(
\begin{array}{cc}
\gamma_{\theta,{\bf k}} & \gamma_{\theta,{\bf k}} \\
1 & -1
\end{array}
\right)
\left(
\begin{array}{c}
\alpha_{{\bf k}\sigma} \\
\beta_{{\bf k}\sigma}
\end{array}
\right), 
\end{equation}
where
\begin{equation}
\gamma_{\theta,{\bf k}}=
\frac{e^{-ik_x}\left(e^{i\theta}\cos{k_x}+e^{-i\theta}\cos{k_y}\right)}
{{\cal S}_{\theta,{\bf k}}}
\end{equation}
with ${\cal S}_{\theta,{\bf k}}$ given in Eq.~(\ref{eq:sk}). 
The lower band dispersion can be transformed to the form of 
Eq.~(\ref{eq:SF-disp}). 
The one-body SF state for $n\le 1$ is given by filling the lower band as 
\begin{equation}
\Phi_{\rm SF}=\prod_{{\bf k}\in{\bf k}_{\rm F}(\theta),\sigma}
\alpha^\dag_{{\bf k}\sigma}|0\rangle
=\prod_{{\bf k}\in{\bf k}_{\rm F}(\theta),\sigma}
\frac{1}{\sqrt{2}}\left(\gamma_{\theta,{\bf k}}\ 
c^\dag_{A{\bf k}\sigma}+c^\dag_{B{\bf k}\sigma}
\right)|0\rangle, 
\end{equation}
which leads to Eq.~(\ref{eq:SF}) by applying the Fourier transformation,
\begin{equation}
c_{\Lambda j\sigma}=\sqrt{\frac{2}{N_{\rm s}}}\sum_{\bf k}
e^{i{\bf k}\cdot{\bf r}_j}c_{\Lambda{\bf k}\sigma}. 
\qquad (\Lambda={\rm A,B}) 
\end{equation}
Because $\Phi_{\rm SF}$ is a current-carrying state, $\Phi_{\rm SF}$ is 
essentially complex except when $4\theta=0$ and $\pi$. 
\par

\begin{figure}[htb]
\begin{center}
\hskip -1.5mm
\includegraphics[width=7.5cm,clip]{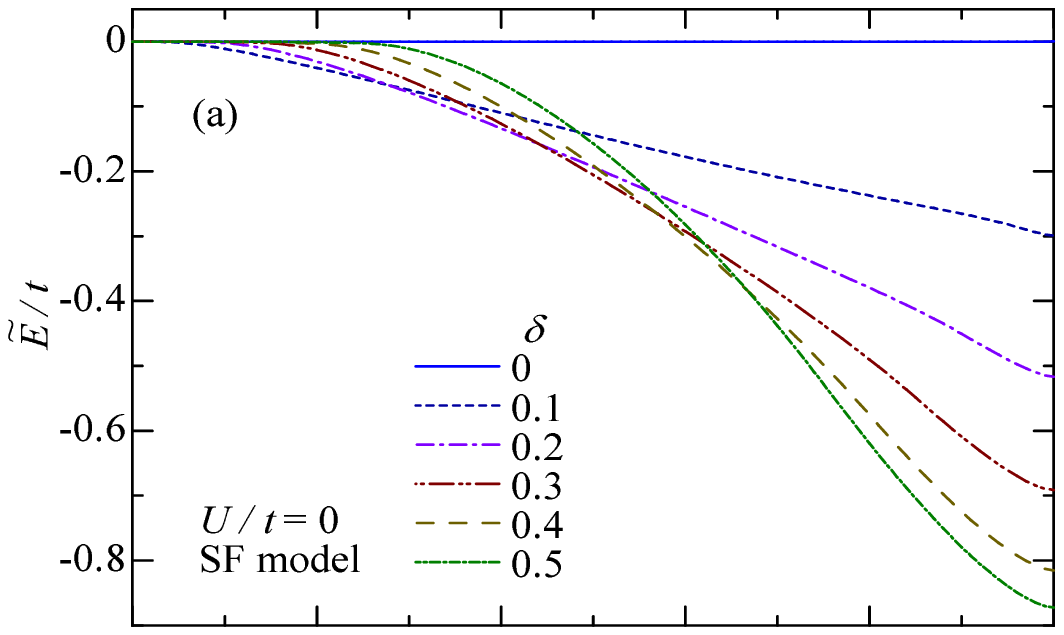}
\includegraphics[width=7.6cm,clip]{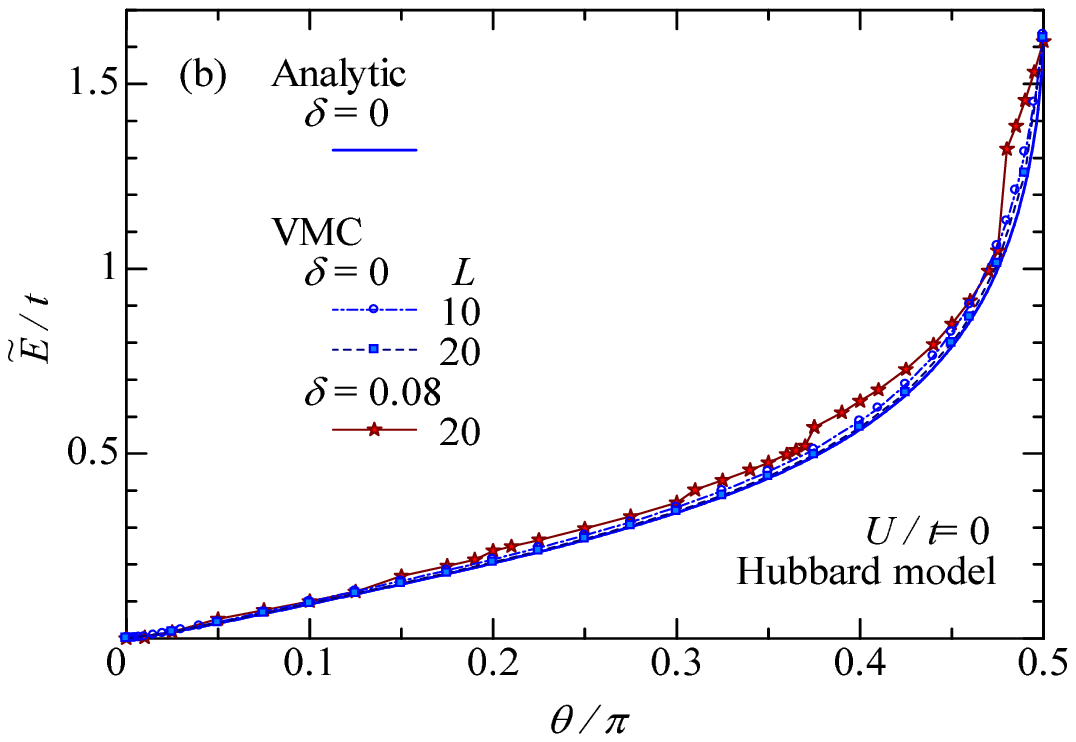}
\end{center} 
\vskip -3mm 
\caption{(Color online) 
Total energies per site of $\Phi_{\rm SF}$ measured from that of 
$\Phi_{\rm N}$ [Eq.~(\ref{eq:tildeE})] are drawn as functions of $\theta$ 
for (a) the SF Hamiltonian [Eq.~(\ref{eq:HamilSF})] and 
(b) the Hubbard Hamiltonian [Eq.~(\ref{eq:Hamil})]. 
In (a), Eqs.~(\ref{eq:Enoni-N-SF}) and (\ref{eq:Enoni-SF}) are used with a 
common $\theta$ in $\Phi_{\rm SF}$ and ${\cal H}^{\rm SF}$ as the ground 
state. 
In (b), Eqs.~(\ref{eq:Enoni-N-Hub}) and (\ref{eq:Enoni-Hub}) are used with 
$\theta$ being varied in $\Phi_{\rm SF}$; we plot VMC data for some cases 
in addition to the analytic result at half filling. 
We confirmed that ${\cal P}$ has almost no effect for $U=0$. 
$t'$ is fixed at $0$. 
} 
\vskip -3mm 
\label{fig:etot-nonint} 
\end{figure}
%
The total energy per site of $\Phi_{\rm SF}$ measured from that of the bare 
Fermi sea $\Phi_{\rm N}$ is written as 
\begin{equation}
\tilde E=E_{\rm SF}-E_{\rm N}. 
\label{eq:tildeE}
\end{equation}
Here, $E_{\rm N}$ for $\Psi_{\rm N}$ is obtained for the Hubbard model 
[Eq.~(\ref{eq:Hamil}) with $U=0$] through
\begin{equation}
E_{\rm N}=\frac{1}{N_{\rm s}}
\frac{\langle\Phi_{\rm N}|{\cal H}_t|\Phi_{\rm N}\rangle}
{\langle\Phi_{\rm N}|\Phi_{\rm N}\rangle}
=\frac{1}{N_{\rm s}}
\sum_{{\bf k}\in{\bf k}_{\rm F}(\theta=0),\sigma}\varepsilon_{\bf k},
\label{eq:Enoni-N-Hub}
\end{equation}
where $\varepsilon_{\bf k}$ [Eq.~(\ref{eq:dispersion})] 
is the bare dispersion of an ordinary Fermi sea, and $E_{\rm N}$ is obtained 
for the SF model [Eq.~(\ref{eq:HamilSF})] through 
\begin{equation}
E_{\rm N}=\frac{1}{N_{\rm s}}
\frac{\langle\Phi_{\rm N}|{\cal H}^{\rm SF}|\Phi_{\rm N}\rangle}
{\langle\Phi_{\rm N}|\Phi_{\rm N}\rangle}
=\frac{1}{N_{\rm s}}
\sum_{{\bf k}\in{\bf k}_{\rm F}(\theta=0),\sigma}E_-^{\rm SF}({\bf k}). 
\label{eq:Enoni-N-SF}
\end{equation}
The energy of $\Phi_{\rm SF}$ for the Hubbard model is given by
\begin{equation}
E_{\rm SF}= 
\frac{\langle\Phi_{\rm SF}|{\cal H}_t|\Phi_{\rm SF}\rangle}
{N_{\rm s}\langle\Phi_{\rm SF}|\Phi_{\rm SF}\rangle}
=-\frac{2t\cos\theta}
{N_{\rm s}}\sum_{{\bf k}\in{\bf k}_{\rm F}(\theta),\sigma}
\frac{\left(\cos{k_x}+\cos{k_y}\right)^2}{{\cal S}_{\theta,{\bf k}}}, 
\label{eq:Enoni-Hub}
\end{equation}
and that for the SF Hamiltonian is given by 
\begin{equation}
E_{\rm SF}=\frac{1}{N_{\rm s}}
\frac{\langle\Phi_{\rm SF}|{\cal H}^{\rm SF}|\Phi_{\rm SF}\rangle}
{\langle\Phi_{\rm SF}|\Phi_{\rm SF}\rangle}=\frac{1}{N_{\rm s}}
 \sum_{{\bf k}\in{\bf k}_{\rm F}(\theta),\sigma}E_-^{\rm SF}({\bf k}). 
\label{eq:Enoni-SF}
\end{equation}
In Fig.~\ref{fig:etot-nonint}(a), we show $\tilde E/t$ for 
${\cal H}^{\rm SF}$ as a function of $\theta$. 
Because $\Phi_{\rm SF}$ is the exact ground state of ${\cal H}^{\rm SF}$, 
$\tilde E/t\le 0$ holds; at half filling, $E_{\rm SF}$ and $E_{\rm N}$ 
are identical because the Fermi surfaces of $\Phi_{\rm N}$ and 
$\Phi_{\rm SF}$ are identical, but the energy of $\Phi_{\rm SF}$ is sizably 
reduced as $\delta$ or $\theta$ increases. 
In contrast, for the Hubbard model with $U=0$ (${\cal H}_t$), 
$\tilde E/t$ is positive because the exact ground state of ${\cal H}_t$ is 
$\Phi_{\rm N}$ [$\Phi_{\rm SF}(\theta=0)$].
$\tilde E/t$ monotonically increases as $\theta$ increases, as shown in 
Fig.~\ref{fig:etot-nonint}(b). 
For $\theta\sim 0$, $E_{\rm SF}$ in Eq.~(\ref{eq:Enoni-Hub}) increases 
quadratically as 
\begin{equation}
E_{\rm SF}=E_{\rm N}
+\theta^2 t\sum_{{\bf k}\in{\bf k}_F(\theta),\sigma}
\frac{\left(\cos{k_x}-\cos{k_y}\right)^2}{\left|\cos{k_x}+\cos{k_y}\right|}
+\cdots,
\label{eq:E-expansion}
\end{equation}
at least for $\delta=0$.
Hence, the SF state is unlikely to be stabilized even if $U/t$ is added as 
a perturbation; this feature is in agreement with that for $U<U_{\rm SF}$ 
discussed in Sects.~\ref{sec:square} and \ref{sec:doped cases}. 
\par

\begin{figure}[htb]
\begin{center}
\includegraphics[width=7.5cm,clip]{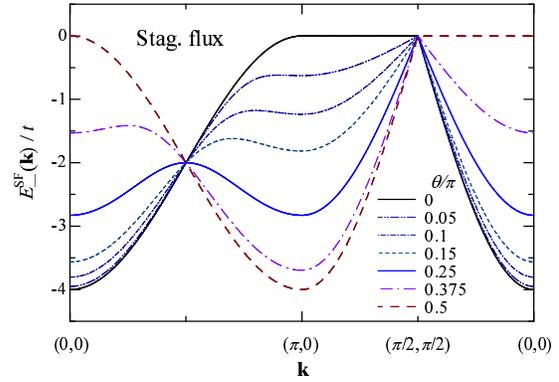}
\end{center} 
\vskip -3mm 
\caption{(Color online) 
Energy dispersion of the bare SF state $\Phi_{\rm SF}$ for several values of 
$\theta$ along $(0,0)$-$(\pi,0)$-$(\pi/2,\pi/2)$-$(0,0)$. 
} 
\vskip -3mm 
\label{fig:ek-jpsj} 
\end{figure}
%
\begin{figure*}[t!]
\begin{center}
\includegraphics[width=5.5cm,clip]{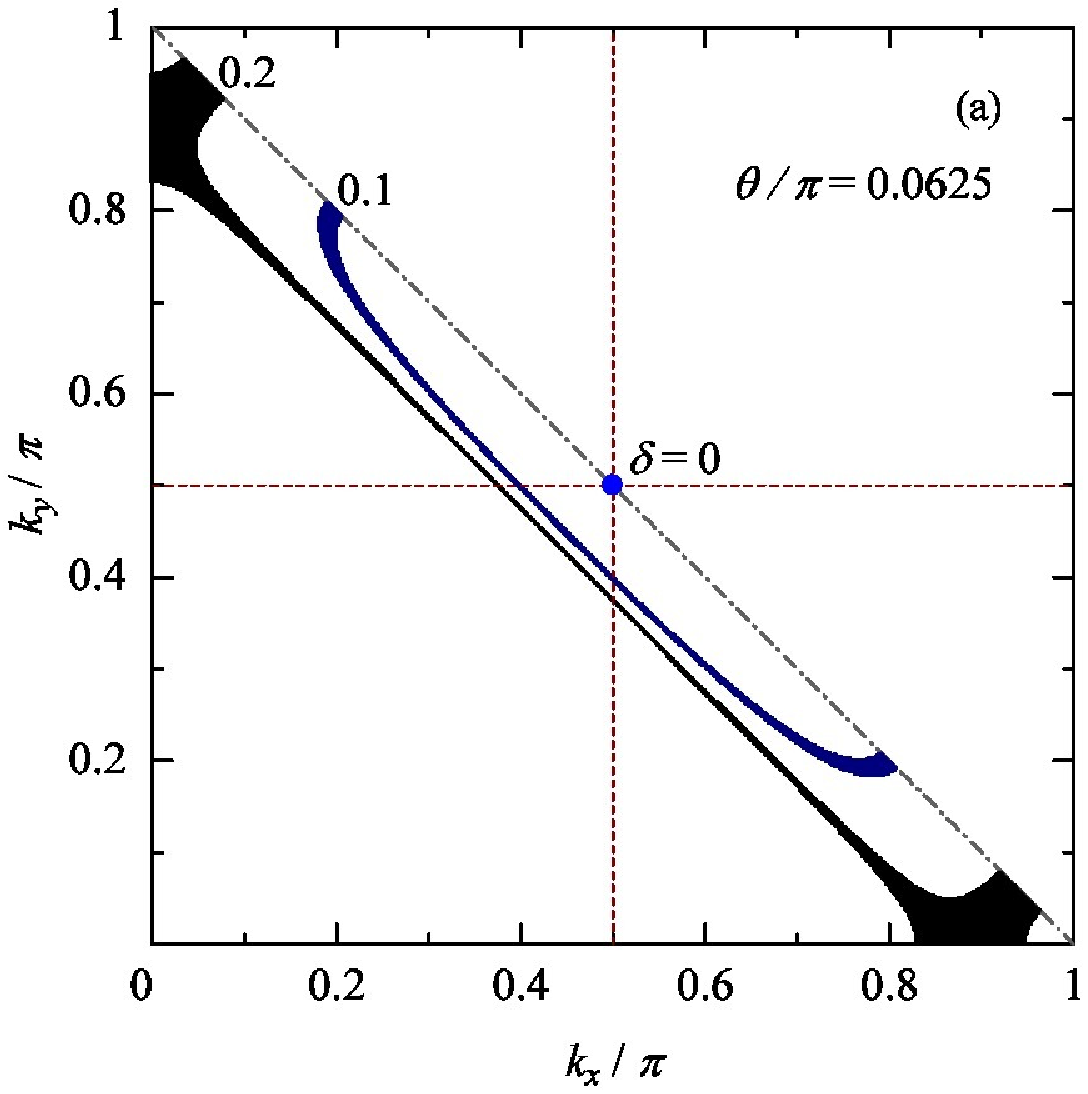}
\includegraphics[width=5.5cm,clip]{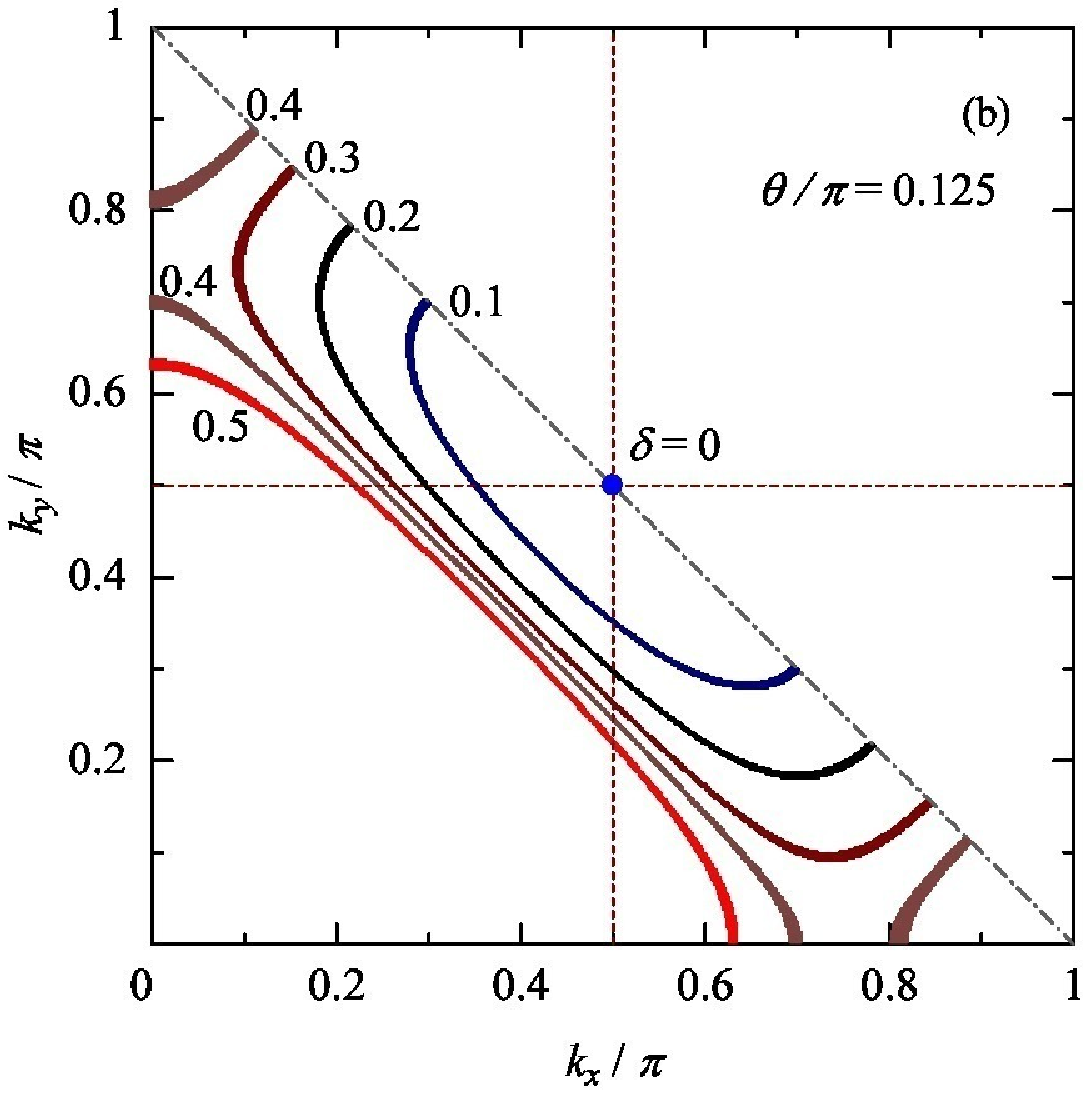}
\includegraphics[width=5.5cm,clip]{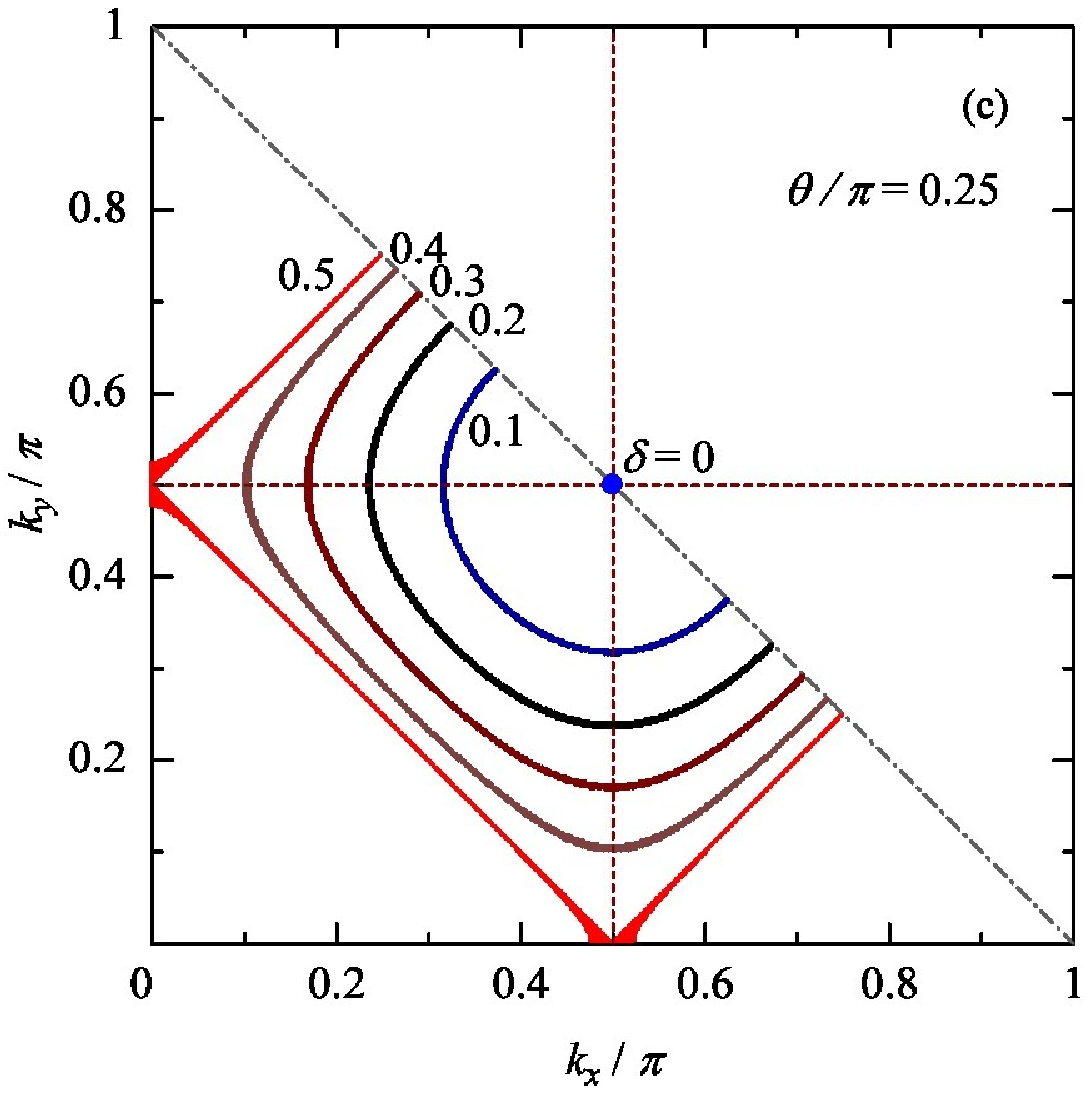}
\end{center} 
\vskip -3mm 
\caption{(Color online) 
Fermi surfaces of the bare SF state $\Phi_{\rm SF}$ shown for 
(a) $\theta=0.0625$, 
(b) $\theta=0.125$, and 
(c) $\theta=0.25$ ($\pi$ flux) 
for some doping rates.
${\bf k}$ points near ${\bf k}_{\rm F}$ 
(${\bf k}\in{\bf k}_{\rm F}$) for $L=1000$ are plotted. 
The $(\pi,\pi)$-folded Brillouin zone edge is shown by a gray dash-dotted 
line. 
} 
\vskip -3mm 
\label{fig:fstha} 
\end{figure*} 
%
The (lower) band structure of the bare SF state, 
$E_{-}^{\rm SF}({\bf k})$ [Eq.~(\ref{eq:EkSF})], is shown in 
Fig.~\ref{fig:ek-jpsj} for several values of $\theta$. 
In the ordinary Fermi sea ($\theta=0$), the band top is degenerate 
along the AF Brillouin zone edge $(\pi,0)$-$(0,\pi)$-$(-\pi,0)$-$(0,-\pi)$, 
namely, the nesting condition is completely satisfied at half filling.  
By introducing $\theta$, this degeneracy is lifted and the band top becomes 
located at $(\pi/2,\pi/2)$ and the three other equivalent points. 
In particular, for the $\pi$-flux state ($\theta=\pi/4$), the band top forms 
an isotropic Dirac cone centered at $(\pi/2,\pi/2)$. 
This cone becomes elongated in the $(\pi,0)$-$(0,\pi)$ direction as 
$\theta$ decreases from $\pi/4$. 
\par

This peculiar band structure brings about anomalous properties in 
$\Phi_{\rm SF}$. 
At half filling, the state for $\theta>0$ is not a conventional metal. 
Although it is not explicitly shown here, $n({\bf k})$ is a smooth continuous 
function except for a discontinuity at ${\bf k}=(\pi/2,\pi/2)$, and 
$N({\bf q})=S({\bf q})$ becomes a quadratic function of $|{\bf q}|$ for 
$|{\bf q}|\rightarrow 0$. 
In a doped case, a Fermi surface appears that is made of a cross section of 
the elongated Dirac cone near ${\bf k}=(\pi/2,\pi/2)$, which is shown in 
Fig.~\ref{fig:fstha} for some values of $\theta$ and $\delta$ and is 
reminiscent of a Fermi arc or hole pocket observed in cuprates by ARPES and 
so forth. 
This is in contrast to the $d$-SC state $\Phi_d$, in which the Fermi 
surface is a point on the nodal line irrespective of the value of $\delta$. 
As $\delta$ increases or $\theta$ decreases, this segmentary Fermi surface 
of $\Phi_{\rm SF}$ becomes longer, and the gap region shrinks to the 
vicinity of the antinodal points. 
However, the behavior of $N({\bf q})$ [$=S({\bf q})$] for 
$|{\bf q}|\rightarrow 0$ in doped $\Phi_{\rm SF}$ is basically unchanged 
from that at half filling. 
This gap behavior differs from the case of the {\it metallic} SF state 
$\Psi_{\rm SF}$ for $U>U_{\rm SF}$, as discussed in 
Sects.~\ref{sec:spin-gap} and \ref{sec:properties}. 
\par

\begin{figure}[htb]
\begin{center}
\includegraphics[width=7.5cm,clip]{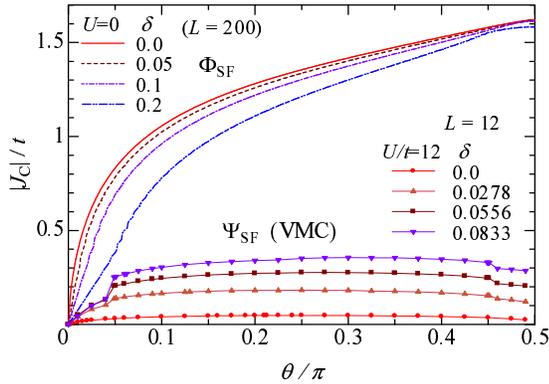} 
\end{center} 
\vskip -3mm 
\caption{(Color online) 
Local circular currents of one-body SF state ($\Phi_{\rm SF}$) are shown 
as functions of Peierls phase for four doping rates. 
The data are numerically computed using Eq.~(\ref{eq:J0-Hub}) for $L=200$.
For comparison, VMC data of $\Psi_{\rm SF}$ for $U/t=12$ and $L=12$ are 
shown for four values of $\delta$.
} 
\vskip -3mm 
\label{fig:curr-u0} 
\end{figure}
%
The local circular current defined by Eq.~(\ref{eq:current}) is calculated 
for $\Phi_{\rm SF}$ as 
\begin{equation}
J_{\rm C}=-\frac{2t\sin\theta}
{N_{\rm s}}\sum_{{\bf k}\in{\bf k}_{\rm F}(\theta),\sigma}
\frac{\left(\cos{k_x}-\cos{k_y}\right)^2}{{\cal S}_{\theta,{\bf k}}}, 
\label{eq:J0-Hub}
\end{equation}
and is shown in Fig.~\ref{fig:curr-u0}; the optimal $\theta$ is always $0$ 
for the Hubbard model [Fig.~\ref{fig:etot-nonint}(a)]. 
For comparison, data for strongly correlated cases are also plotted. 
Here, we only point out two notable features. 
(i) As $\delta$ increases, $|J_{\rm C}|$ decreases for the noninteracting 
$\Phi_{\rm SF}$, while $|J_{\rm C}|$ increases for the strongly correlated 
$\Psi_{\rm SF}$ for the Hubbard model with $U/t=12$. 
(ii) As the interaction increases, $|J_{\rm C}|$ is markedly reduced. 
\par

\section{Staggered Flux State for $t$-$J$ Model\label{sec:t-J}}
In this Appendix, we summarize the stability of the SF state in 
$t$-$J$-type models with calculations of 
reliable accuracy for a comparison with the Hubbard model treated in 
the main text. 
For this purpose, we include the following three-site (or pair-hopping) term 
${\cal H}_{\rm 3site}$, which is the same order as ${\cal H}_J$ 
[$t^2/U$ ($=J/4$)] in the strong-coupling expansion: 
\begin{equation}
{\cal H}_{t{\mbox -}J}={\cal H}_t+{\cal H}_J, \qquad
{\cal H}_3={\cal H}_{t{\mbox -}J}+{\cal H}_{\rm 3site}, 
\label{eq:t-J}
\end{equation}
with
\begin{eqnarray}
{\cal H}_t&=&-t\sum_{\langle i,j\rangle,\sigma}
\left(\tilde c^\dag_{i\sigma}\tilde c_{j\sigma} + {\mbox H.c.}\right), 
\\
{\cal H}_J&=&J\sum_{\langle i,j\rangle}
\left({\bf S}_i\cdot{\bf S}_j-\frac{1}{4}\tilde n_i\tilde n_j\right),
\\
{\cal H}_{\rm 3site}&=&-\frac{J}{4}\sum_{j,\tau\ne\tau',\sigma}
\left(\tilde c^\dag_{j,-\sigma}\tilde c_{j,-\sigma}
      \tilde c^\dag_{j+\tau,\sigma}\tilde c_{j+\tau',\sigma}\right.
\nonumber\\ 
&&\qquad\qquad\ \left.+
      \tilde c^\dag_{j+\tau,-\sigma}\tilde c_{j,-\sigma}
      \tilde c^\dag_{j,\sigma}\tilde c_{j+\tau',\sigma}
\right),
\end{eqnarray}
where $\tilde c_{j\sigma}=c_{j\sigma}(1-n_{j-\sigma})$, 
$\tilde n_{j}=\sum_\sigma\tilde c^\dag_{j\sigma}\tilde c_{j\sigma}$, 
and ${\bf S}_j=\frac{1}{2}\sum_{\alpha,\beta}c_{j\alpha}^\dag
\sigma_{\alpha\beta}c_{j\beta}$ with $\sigma$ being the Pauli matrix for 
$S=1/2$ spins. 
We call ${\cal H}_{t{\mbox -}J}$ the $t$-$J$ model and ${\cal H}_{\rm 3}$ 
the three-site model. 
In doped cases, the behavior of the Hubbard model should be more similar 
to that of ${\cal H}_{\rm 3}$. 
Here, we disregard diagonal hopping terms for simplicity. 
\par 

To these models, we apply a VMC scheme similar to that for the Hubbard model. 
As a many-body factor, we use only the complete Gutzwiller projector,  
${\cal P}={\cal P}_{\rm G}$ with $g=0$, as in previous 
studies.\cite{Liang,TKLee,YO,Ivanov}
Thus, $\tilde\Psi_{\rm SF}={\cal P}_{\rm G}(0)\Phi_{\rm SF}(\theta)$ 
[$\tilde\Psi_{\rm N}={\cal P}_{\rm G}(0)\Phi_{\rm N}$] has one [no] 
variational parameter. 
Here, we concentrate on the decrease in energy of $\tilde\Psi_{\rm SF}$
from that of $\tilde\Psi_{\rm N}$,
\begin{equation}
\tilde E=E^{\rm SF}(\theta)-E^{\rm N}, 
\label{eq:Edecrement}
\end{equation}
where $E^{\rm N}=E^{\rm SF}(\theta=0)$. 
\par

\begin{figure}[htb]
\begin{center}
\includegraphics[width=7.5cm,clip]{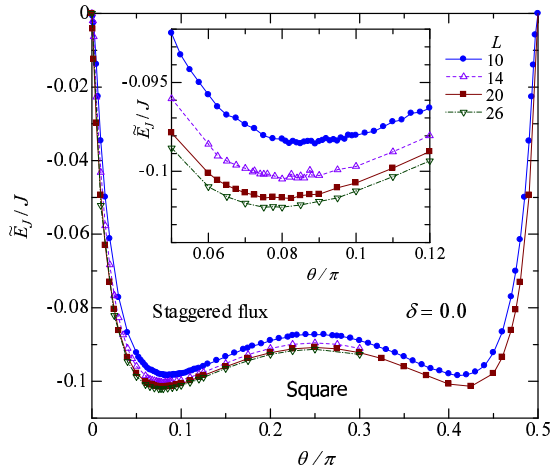}
\end{center} 
\vskip -3mm 
\caption{(Color online) 
Exchange energy per site $\tilde E_J/J$ of $\tilde\Psi_{\rm SF}$ measured 
from that of $\tilde\Psi_{\rm N}$ at half filling as a function of $\theta$. 
The graph is symmetric with respect to $\theta/\pi=1/4$ ($\pi$-flux). 
The inset shows the magnification near the minimum ($\theta_{\rm opt}$) 
of the smaller $\theta$. 
Data for four values of $L$ are compared. 
}
\vskip -3mm 
\label{fig:E2size-JPSJ} 
\end{figure}
%
First, we consider the half-filled case ($\delta=0$), in which 
$E_t$ ($=\langle{\cal H}_t\rangle/N_{\rm s}$) and 
$E_{\rm 3site}$ ($=\langle{\cal H}_{\rm 3site}\rangle/N_{\rm s}$) vanish; 
the total energy is given by the exchange term 
$E=E_J=\langle{\cal H}_J\rangle/N_{\rm s}$. 
In Fig.~\ref{fig:E2size-JPSJ}, we plot $\tilde E_J/J$ ($=\tilde E/J$) 
as a function of $\theta$. 
$\tilde E_J$ has a minimum at $\theta/\pi\sim 0.08$. 
Because $\tilde\Psi_{\rm SF}$ is equivalent to the $d$-SC state 
$\tilde\Psi_d={\cal P}_{\rm G}\Phi_d$ owing to the SU(2) 
symmetry,\cite{Affleck,ZGRS} the minimum energy of $\tilde\Psi_{\rm SF}$
[e.g., $E^{\rm SF}_J/J=-1.1396(4)$ for $L=10$] coincides with that of 
$\tilde\Psi_d$ [$E^d_J/J=-1.1398$]\cite{YO}. 
This value is very low and broadly comparable to the minimum energy of the 
AF state on the same footing, 
$\tilde\Psi_{\rm AF}={\cal P}_{\rm G}\Psi_{\rm AF}$ 
[$E^{\rm AF}/J=-1.1412$]\cite{t-J-AF}. 
Thus, the SF state is very stable at half filling irrespective of the value 
of $J/t$. 
\par

\begin{figure*}[t!] 
\begin{center}
\includegraphics[width=17cm,clip]{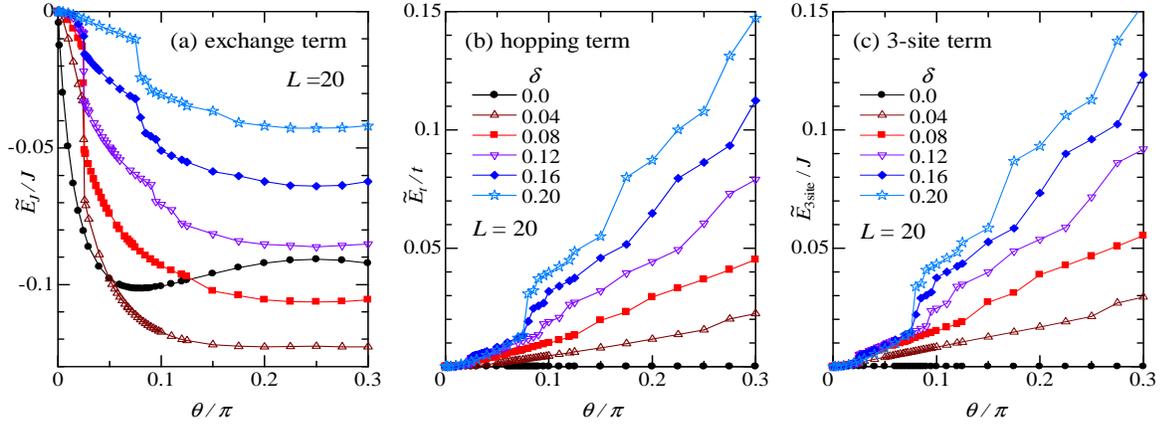} 
\end{center} 
\vskip -20mm 
\caption{(Color online) 
Energy components of SF state measured from those of $\Psi_{\rm N}$ 
as functions of $\theta$: 
(a) exchange term $\langle{\cal H}_J\rangle$, 
(b) hopping term $\langle{\cal H}_t\rangle$, and 
(c) three-site term $\langle{\cal H}_{\rm 3site}\rangle$. 
In each panel, data for several doping rates are plotted. 
The cusplike foldings appearing for $\delta>0$ are caused by the 
discontinuous change in the occupied ${\bf k}$-points in $\Phi_{\rm SF}$ as 
$\theta$ varies owing to a finite $L$. 
}
\vspace{-0.2cm}
\label{fig:DelE-JPSJ} 
\end{figure*}
\begin{figure}[htb] 
\begin{center}
\includegraphics[width=7.5cm,clip]{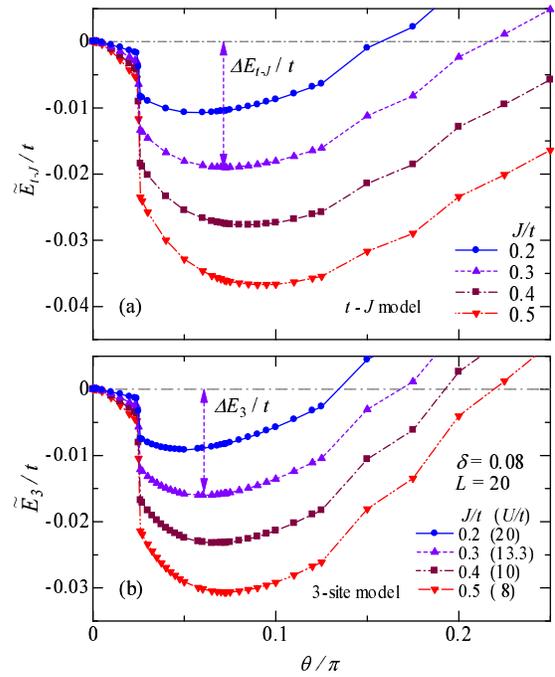} 
\end{center} 
\vskip -3mm 
\caption{(Color online) 
Total energies of the SF state ($\Psi_{\rm SF}$) measured from that of 
$\Psi_{\rm N}$ plotted as a function of $\theta$ for 
(a) the $t$-$J$ model and (b) the three-site model. 
The doping rate is fixed at 0.08 and the values of $J/t$ are chosen 
appropriately for cuprates. 
The values of $U/t$ in (b) are converted using $J=4t^2/U$.  
Arrows in both panels indicate the energy gains $\Delta E/t$ by the SF 
state for $J/t=0.3$. 
}
\vspace{-0.2cm}
\label{fig:detot-tj-e-JPSJ} 
\end{figure}

%
For doped cases ($\delta>0$), $E_t$ and $E_{\rm 3site}$ also make 
contributions. 
Figure \ref{fig:DelE-JPSJ} shows the $\theta$ dependence of the three energy 
components of $\tilde\Psi_{\rm SF}$ measured from those of 
$\tilde\Psi_{\rm N}$. 
By introducing $\theta$, the exchange energy is lowered ($\tilde E_J/J<0$), 
similarly to the case of half filling, in a wide range of $\delta$ 
[Fig.~\ref{fig:DelE-JPSJ}(a)]; $|\tilde E_J|/t$ is large, especially near 
half filling. 
In contrast, $\tilde E_t/t$ and $\tilde E_{\rm 3site}/J$ monotonically 
increase with $\theta$, namely, they destabilize the SF state, and become more 
marked as $\delta$ increases 
[Figs.~\ref{fig:DelE-JPSJ}(b) and \ref{fig:DelE-JPSJ}(c)]. 
For fixed values of $\delta$ and $J/t$, the total energy $\tilde E/t$ is 
the sum of these competing components. 
For example, in Fig.~\ref{fig:detot-tj-e-JPSJ}, we plot $\tilde E/t$ for the 
$t$-$J$ and three-site models for typical values of $\delta$ and $J/t$ of 
underdoped cuprates. 
We find that $\Psi_{\rm SF}$ is stable with respect to $\Psi_{\rm N}$ 
in a wide range of $J/t$. 
Because $\tilde E_{\rm 3site}/J$ is disadvantageous to the SF state 
[Fig.~\ref{fig:DelE-JPSJ}(c)], the decrease in $\tilde E_3/t$ is somewhat 
smaller than that in $\tilde E_{t-J}/t$. 
\par

%
\begin{figure}[htb]
\begin{center}
\includegraphics[width=7.5cm,clip]{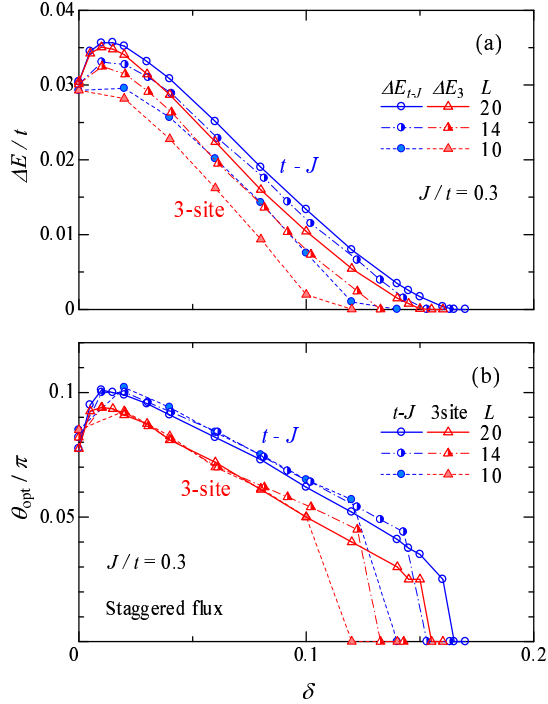}
\end{center} 
\vskip -3mm 
\caption{(Color online) 
(a) Doping rate dependence of energy gain of SF state measured from the 
energy of $\Psi_{\rm N}$ for the $t$-$J$ ($\Delta E_2$) and the three-site 
($\Delta E_3$) models. 
Data for three values of $L$ are plotted for each model. 
(b) Optimized values of $\theta$ in SF state as functions of doping rate. 
}
\vskip -3mm 
\label{fig:t-J-delE-theta} 
\end{figure}
%
Finally, we look at the $\delta$ dependence of the stability of 
$\tilde\Psi_{\rm SF}$. 
In Fig.~\ref{fig:t-J-delE-theta}(a), we plot the energy gain or difference 
of $\tilde\Psi_{\rm SF}$ as compared with $\tilde\Psi_{\rm N}$, defined as 
\begin{equation}
\Delta E({\rm SF})=E({\rm N})-E({\rm SF}),  
\end{equation}
for $J/t=0.3$. 
Note that $\Delta E$ has the inverse sign to $\tilde E$. 
In Fig.~\ref{fig:t-J-delE-theta}(b), we show the optimized $\theta$ 
($\theta_{\rm opt}$). 
The behavior of $\theta_{\rm opt}$ is similar to that of $\tilde E$, 
but $\theta_{\rm opt}$ vanishes abruptly at the boundary $\delta_{\rm SF}$
owing to finite-size effects.
From the system-size dependence, the range of the SF state seems to expand 
to some extent in the thermodynamic limit. 
\par

\section{Staggered Spin Current State\label{sec:SSC}}
In this Appendix, we study the staggered spin current (SSC) state 
$\Psi_{\rm SSC}={\cal P}\Phi_{\rm SSC}$,\cite{Nersesyan,Schulz,Ozaki} 
as illustrated in Fig.~\ref{fig:SF-figA}(b). 
The one-body state, $\Phi_{\rm SSC}$, is obtained as the ground state 
of the noninteracting SSC model written as
\begin{eqnarray}
&&{\cal H}^{\rm SSC}=-t\sum_{i\in A,\sigma}
\Bigl[
 e^{i\theta s(\sigma)}\left(c^\dag_{{\rm A}i,\sigma}c_{{\rm B}i,\sigma}
        +c^\dag_{{\rm A}i,\sigma}c_{{\rm B}i-2{\bf x},\sigma}\right)
\nonumber\\
&&
+e^{-i\theta s(\sigma)}
\left(c^\dag_{{\rm A}i,\sigma}c_{{\rm B}i-{\bf x}+{\bf y},\sigma}
+c^\dag_{{\rm A}i,\sigma}c_{{\rm B}i-{\bf x}-{\bf y},\sigma}\right)
+{\rm H.c.}
\Bigr], \qquad
\label{eq:HamilSSC}
\end{eqnarray}
where $s(\sigma)=1$ or $-1$ according to whether $\sigma=\uparrow$ or 
$\downarrow$. 
${\cal H}^{\rm SSC}$ is diagonalized in the same way as ${\cal H}^{\rm SF}$. 
The energy dispersion is identical to $E_\pm^{\rm SF}({\bf k})$ 
[Eq.~(\ref{eq:EkSF})]. 
Consequently, we have
\begin{equation}
\Phi^{\rm SSC}
=\prod_{{\bf k}\in{\bf k}_{\rm F},\sigma}
\frac{1}{\sqrt{2}}\left[\gamma^\sigma_{\bf k}(\theta)\ 
c^\dag_{A{\bf k}\sigma}+c^\dag_{B{\bf k}\sigma}
\right]|0\rangle, 
\end{equation}
\begin{equation}
\gamma^\sigma_{\bf k}(\theta)=
\frac{e^{-ik_x}\left(e^{i\theta s(\sigma)}\cos{k_x}
                    +e^{-i\theta s(\sigma)}\cos{k_y}\right)}
{{\cal S}_{\theta,{\bf k}}}. 
\end{equation}
$\Phi^{\rm SSC}$ has a doubled unit cell but, in contrast to 
$\Psi_{\rm SF}$, it has no magnetic flux and preserves the time-reversal 
symmetry. 
The SU(2) symmetry is broken in $\Phi^{\rm SSC}$ even at half filling.  
\par

\begin{figure*}[t!] 
\begin{center}
\includegraphics[width=6.0cm,clip]{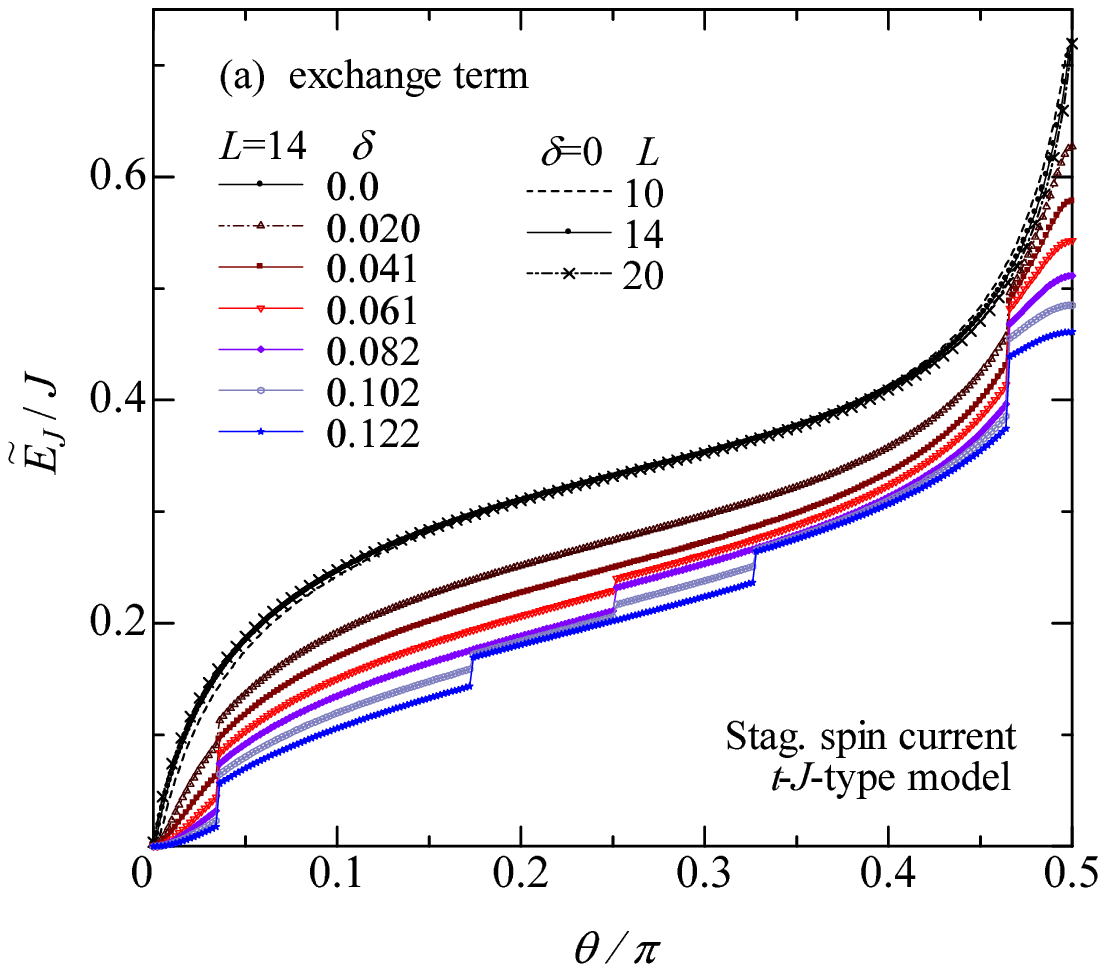} 
\hskip -3mm
\includegraphics[width=6.0cm,clip]{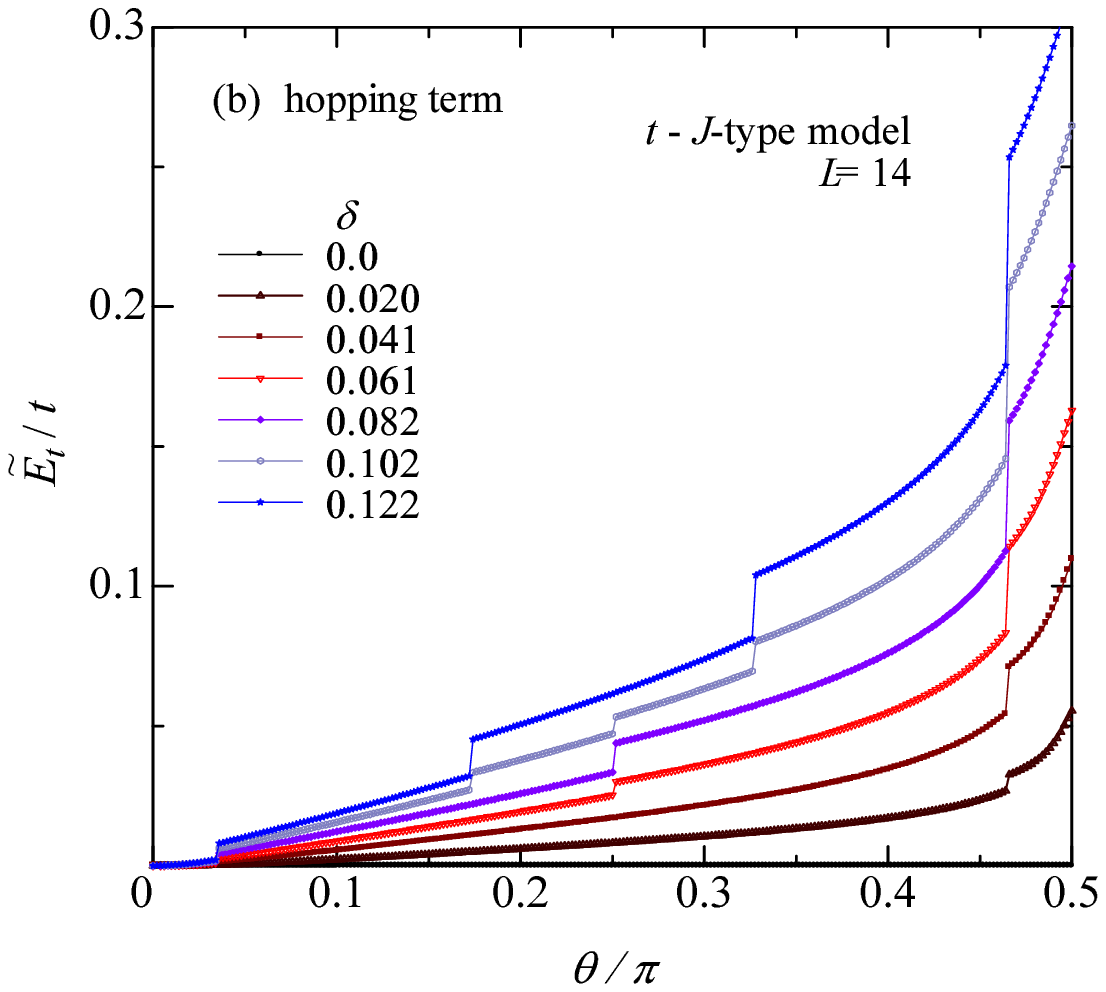} 
\hskip -3mm
\includegraphics[width=6.0cm,clip]{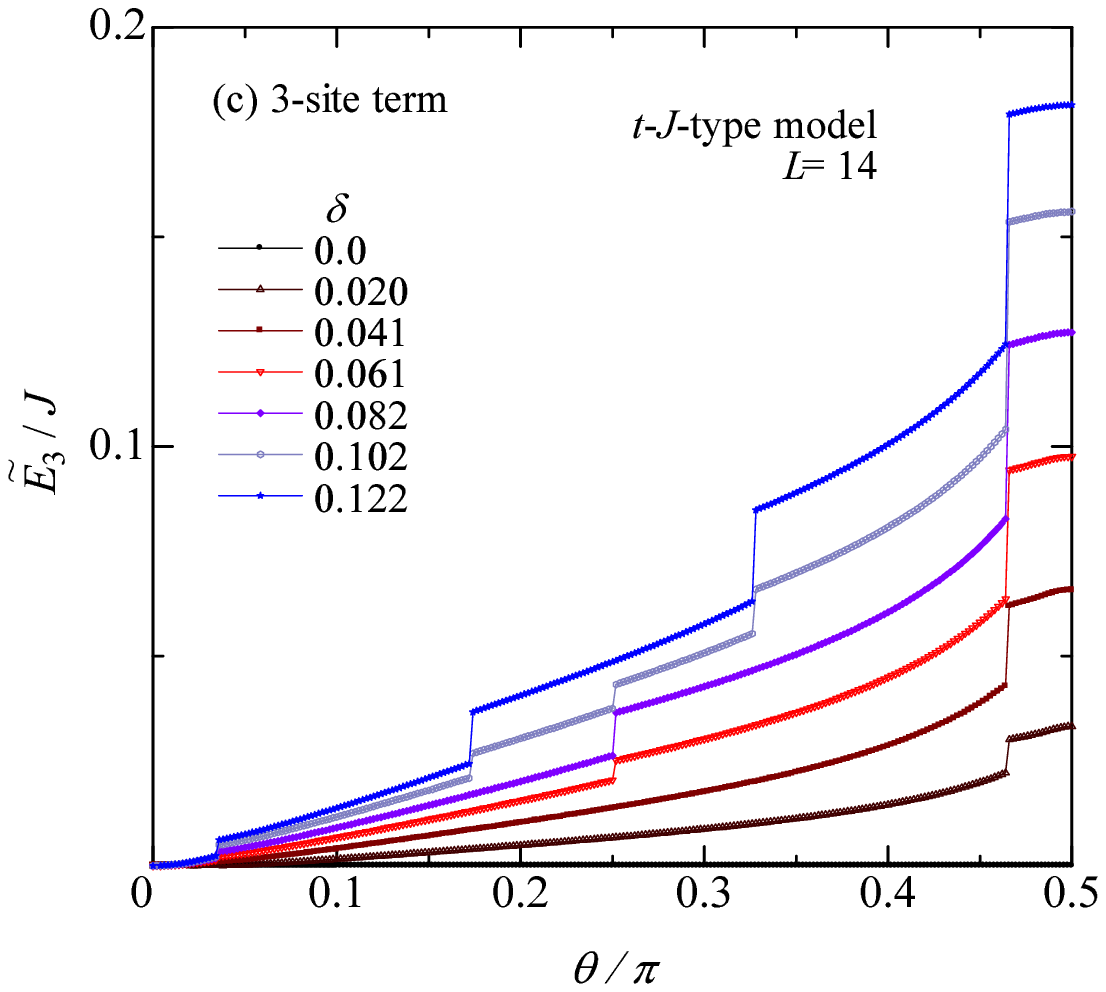} 
\end{center} 
\vskip -3mm 
\caption{(Color online) 
Energy components of staggered spin current state measured from those of 
$\Psi_{\rm N}=\Psi_{\rm SSC}(\theta=0)$, similarly to in 
Fig.~\ref{fig:DelE-JPSJ}.
In (a), data for three values of $L$ are compared at half filling to show 
the small finite-size effect. 
}
\vspace{-0.2cm}
\label{fig:SSC-E} 
\end{figure*}
%
For the noninteracting Hubbard model, it is trivial that $\tilde E$ 
[$=E^{\rm SSC}(\theta)-E^{\rm N}$] increases as $\theta$ increases because 
$\Psi_{\rm N}$ is the exact ground state and $E^{\rm SSC}(\theta)$ is 
equivalent to $E^{\rm SF}(\theta)$ [see Eq.~(\ref{eq:E-expansion})]. 
To consider strongly correlated cases, we apply 
$\tilde\Psi_{\rm SSC}={\cal P}_{\rm G}(0)\Phi_{\rm SSC}$ to the $t$-$J$-type 
model $\langle{\cal H}_3\rangle$ in Eq.~(\ref{eq:t-J}) in the same manner 
as $\tilde\Psi_{\rm SF}$ in Appendix\ref{sec:t-J}. 
Figure \ref{fig:SSC-E} shows the $\theta$ dependence of the three energy
components of ${\cal H}_3$. 
The behavior of $\tilde E_t/t$ and $\tilde E_3/J$ is similar to that of 
$\tilde\Psi_{\rm SF}$ 
[Figs.~\ref{fig:DelE-JPSJ}(b) and \ref{fig:DelE-JPSJ}(c)]. 
In contrast to $\Psi_{\rm SF}$, however, $\tilde E_J/J$ also monotonically 
increases with $\theta$. 
The cause of this difference is discussed in Sect.~\ref{sec:phase-cancel}. 
Because every component of the energy increases as $\theta$ increases, we 
conclude that $\Psi_{\rm SSC}$ never has a lower energy than $\Psi_{\rm N}$ 
for $\delta\sim 0$ and positive $J/t$. 
\par

Although we have argued that $\Psi_{\rm SSC}$ is not stabilized in the 
square-lattice $t$-$J$ and Hubbard models, a recent VMC study\cite{Tamura-SSC} 
showed that an SSC state has a lower energy than the paramagnetic state for 
the Heisenberg model on anisotropic triangular lattices with $J'\sim J$ 
in magnetic fields.\cite{SSC}
\par



\begin{thebibliography}{999}
\bibitem{note-PG}
Studies in this line are broadly classified into two groups: 
(i) SC fluctuation or precursors to SC and 
(ii) resonating valence bond (RVB) theory.
It was found that the SC fluctuation appears below $T_{\rm pair}$ with 
$T_{\rm c}<T_{\rm pair}<T^*$ 
by measurements of the Nernst effect,\cite{Nernst1,Nernst2} ARPES,\cite{Kondo} 
and optical conductivity.\cite{Dubroka,Uykur}
Therefore, the SC fluctuation will not be related to the pseudogap below 
$T^*$. 
In contrast, the RVB mean-field theory successfully explains the pseudogap, 
particularly, in the spin sector, because the spin degrees of freedom form 
RVBs, which represent a singlet pairing.\cite{OF,AtoZ}
In this theory, $T^*$ is not a phase-transition temperature but a crossover 
temperature.
\bibitem{note-phase} 
Precisely speaking, $\theta$ in the wave function slightly deviates from 
the actually averaged value of the phase variation $\langle\theta\rangle$ 
in the hopping processes as in Fig.~\ref{fig:SF-scheme}. 
At half filling, $\langle\theta\rangle$ is slightly larger than $\theta$ 
but $\theta\sim\langle\theta\rangle$ holds as a whole. 
For a sufficiently large $U/t$, $\phi=\langle\theta\rangle$ holds rather 
than $\phi=\theta$.
\bibitem{note-free-holon} 
In this process, a phase assignment similar to Eq.~(\ref{eq:phase}), such as
%
\begin{equation}
{\cal P}_\xi=\exp\left\{
i\xi\sum_j(-1)^jh_j\left[\prod_\tau\left(1-d_{j+\tau}\right)\right]
\left(n_{j+x}+n_{j-x}-n_{j+y}-n_{j-y}\right)
\right\}, 
\nonumber
\end{equation}
%
can be assumed, where $n_j=n_{j\uparrow}+n_{j\downarrow}$ and $\xi$ is 
a variational parameter. 
However, we can easily show that ${\cal P}_\xi=1$, namely, the phase variation 
$\theta$ in the hopping of isolated holons is not controlled by this type 
of phase factor.
\bibitem{note-real} 
If we apply ${\cal P}_\phi$ to a real wave function such as $\Psi_{\rm N}$, 
$\Psi_{\rm AF}$, or $\Psi_d$, the phase factor in ${\cal P}_\phi$ is 
optimized as $\phi=0$, namely, ${\cal P}_\phi=1$. 
Therefore, expectation values associated with the optimized real state do 
not alter regardless of whether ${\cal P}_\phi$ is introduced or not.
\bibitem{note-FSRen} 
Because the band-renormalization effect\cite{Himeda} is small in 
$\Psi_{\rm AF}$ as well as in $\Psi_d$ even for $\delta\sim 0$ 
(See Ref.~\citen{BR}) in the case of $t'=0$, we do not treat it here.
It does not directly affect the present result.
\bibitem{note-Uc} 
To estimate $U_{\rm c}/t$ more accurately, the long-range 
part of the D-H (Jastrow) factor is important.\cite{Sorella,Miyagawa} 
Therefore, here we do not pursue the quantitative accuracy of $U_{\rm c}/t$.
\bibitem{SMA-example} For a procedure of applying SMA to finite-size 
systems, see, for instance, the treatment in constructing Fig.~5 in 
Ref.~\citen{Tamura-AHM}, in which a conductive spin-gap state was similarly 
confirmed in the Bose-Einstein-condensation regime for the attractive 
Hubbard model.
\bibitem{note-t-J} 
In the $t$-$J$ model, not only $\delta_{\rm SF}$ but also 
$\delta_d$ (the transition point from $\Psi_d$ to $\Psi_{\rm N}$) tends to be 
large as compared with in the Hubbard model.
\bibitem{note-USF-edge} In Fig.~\ref{fig:tp-U}, we find that the system-size 
dependence becomes large when $\delta$ approaches $\delta_{\rm SF}$. 
This is because the optimal $\theta$ discontinuously switches from the lower 
edge of the 
second continuous segment ($\theta_{\rm a}$ in Ref.~\citen{note-USF}) to 
zero (minimum in the first segment) at $U=U_{\rm SF}$. 
The value of $\theta_{\rm a}$ depends on $L$; $\theta_{\rm a}\rightarrow 0$ 
as $L\rightarrow\infty$.
\bibitem{noteOgata}
Here, $U_{\rm SF}/t$ is estimated without considering the band 
renormalization in $\Psi_{\rm N}$ for ease of calculation; therefore, the 
area of the SF state may shrink to some extent near half filling.
\bibitem{note-kinetic} Generally, the band renormalization also occurs 
so as to reduce the kinetic energy.\cite{BR} 
In Fig.~\ref{fig:tt-U}, we show the boundary where the effective band in 
$\Psi_{\rm N}$ is renormalized to $t_1=0$ and then the nesting condition, 
which is lost for $t'\ne0$, is restored. 
Note that this boundary for $t'/t\gtrsim0.3$ behaves similarly to the 
phase boundary between $\Psi_{\rm N}$ and $\Psi_{\rm SF}$ of concern 
in the text.
\bibitem{note-flow} 
In this connection, we mention the renormalization flow in $\Psi_{\rm N}$. 
As in Fig.~\ref{fig:evsam}(a), $t_1<t'$ holds and the value of 
$t'_{\rm c2}/t$ at $U/t=8$ decreases as $L$ increases for $t'/t\lesssim 0.8$, 
indicating that the flow is directed to $t'/t\rightarrow 0$ in this regime. 
On the other hand, we find that $t_1>t'$ for $t'/t\gtrsim 1.3$ and 
$t'_{\rm c2}/t$ 
increases with $L$ at $U/t=12$, meaning that the flow is directed to 
$t'/t\rightarrow\infty$. 
Similar analysis has been performed for a paramagnetic state in the Heisenberg 
model ($U/t\rightarrow\infty$) on an anisotropic triangular lattice, and it 
was concluded that the boundary is $J'/J\sim 0.95$.\cite{Tamura-SSC}
\bibitem{note-coexist} We should rectify the argument in Refs.~\citen{SCES} 
and \citen{ISS} that the SF state may be an underlying normal state 
from which $d$-SC arises.
\bibitem{note-ED} 
In electron-doped cuprates, pseudogap behavior has been found at the 
so-called hot spots---the intersections of a quasi-Fermi surface with the 
AF-Brillouin-zone boundary---, which are detached from the antinodal 
points.\cite{Armitage} 
This pseudogap is related to AF correlations\cite{Onose} or AF 
orders.\cite{Matsui0}
The $d$-SC gap also has a maximum at the hot spot.\cite{Matsui} 
However, recent experiments on high-quality samples showed that such an 
AF pseudogap is suppressed.\cite{Horio}
\end{thebibliography}
\end{document}